\documentclass[]{ar2e}

\usepackage{ARAstroBib}
\usepackage{journals}

\begin{document}

\input epsf.tex    %<-If you need EPS figures to be
                   %  called in {figure} environment for PC
\input psfig.sty

\jname{Ann. Rev. Ast. Astro.}
\jyear{2013}
\jvol{51}
\ARinfo{1056-8700/97/0610-00}

\textbf{NOTE}: The following paper will appear in the 2013 volume of
\textsl{Annual Reviews of Astronomy and Astrophysics.}\\

\title{Asteroseismology of Solar-Type and Red-Giant Stars}

\markboth{Asteroseismology of Solar-Type and Red-Giant
  Stars}{Asteroseismology of Solar-Type and Red-Giant Stars}

\author{William J. Chaplin, Andrea Miglio \affiliation{School of
    Physics and Astronomy, University of Birmingham, Edgbaston,
    Birmingham, B15 2TT, UK}}

\begin{abstract}

We are entering a golden era for stellar physics driven by satellite
and telescope observations of unprecedented quality and scope. New
insights on stellar evolution and stellar interiors physics are being
made possible by asteroseismology, the study of stars by the
observation of natural, resonant oscillations. Asteroseismology is
proving to be particularly significant for the study of solar-type and
red-giant stars. These stars show rich spectra of solar-like
oscillations, which are excited and intrinsically damped by turbulence
in the outermost layers of the convective envelopes. In this review we
discuss the current state of the field, with a particular emphasis on
recent advances provided by the \emph{Kepler} and CoRoT space missions
and the wider significance to astronomy of the results from
asteroseismology, such as stellar populations studies and exoplanet
studies.\\

\end{abstract}

\maketitle

 \section{Introduction}
 \label{sec:intro}

Almost two decades have passed since a general review of
asteroseismology appeared in this journal \citep{Brown1994a,
  gautschy95,gautschy96}. At the time, data on solar-like oscillations
-- pulsations excited and damped by near-surface convection -- were
available for only one star, the Sun, and the case for detections
having been made in other solar-type stars was mixed at best. That
situation has since changed dramatically. There are already several
excellent reviews in the literature which discuss the huge efforts and
considerable ingenuity that went into detecting oscillations in
solar-type and red-giant stars using ground-based telescopes and
Doppler-velocity instrumentation (for a detailed overview, see
\citealt{Bedding2011b} and references therein).  The concerted drive
to reduce noise levels for the detection of exoplanets led to
considerable benefits for asteroseismic studies.  To results from
episodic ground-based campaigns were added a few detections from the
first space-based photometric observations of solar-like oscillations
(e.g., WIRE, MOST and SMEI; again, see \citealt{Bedding2011b} and
references therein). However, it was the launch of the French-led
CoRoT satellite (in late 2006) and the NASA \emph{Kepler} Mission (in
2009) that heralded major breakthroughs for the field: exquisite
quality photometric data are now available on solar-like oscillations
in unprecedented numbers of solar-type and red-giant
stars. \emph{Kepler} has provided multi-year, uninterrupted coverage
on many of these targets, with its asteroseismology programme being
conducted by the \emph{Kepler} Asteroseismic Science Consortium (KASC;
see \citealt{Kjeldsen2010, Gilliland2010}).

This review has two overarching themes: first, how data on solar-like
oscillations can now test theories of stellar structure, stellar
dynamics and evolution, and constrain the physics of stellar
interiors, sometimes in ways that have, hitherto, not been possible;
and second, how precise estimates of the fundamental properties of
stars obtained from asteroseismology impact more widely on
astrophysics. We consider two areas in particular that highlight the
second theme: the study of the structure, dynamics and evolution of
exoplanetary systems, and stellar population and Galactic evolution
studies.

Our review is organised as follows. We begin in
Section~\ref{sec:theory} with a concise description of the key
features of solar-like oscillations. We show how the oscillation
spectrum changes as the star evolves, in ways that allow us to use
seismology to diagnose the stellar properties and internal
structure. Section~\ref{sec:obs} summarizes the observational data
that are available from \emph{Kepler} and CoRoT, and some of the key
data-analysis challenges. In Section~\ref{sec:prop}, we discuss the
use of average or global asteroseismic parameters in estimation of the
fundamental stellar properties. Section~\ref{sec:nuf} begins by
summarizing key issues for stellar properties estimation when
estimates of individual oscillation frequencies are available. We then
explain how the individual frequencies may be used to probe the
internal structure and dynamics, and test stellar interiors
physics. Sections~\ref{sec:exo} and~\ref{sec:pop} present the strong
links that asteroseismology has with studies of exoplanets, stellar
activity and stellar populations. We conclude our review in
Section~\ref{sec:conc} with some remarks on future prospects for
asteroseismology.

It is worth stressing that this is currently a very fast-moving field,
thanks in large part to the data collected by \emph{Kepler} and
CoRoT. Most of the results discussed in the review come from these
missions.  Our aim is to provide the non-expert with a good overview
of recent progress and future opportunities. Many papers have recently
been published, and it is not possible to cite them all. We have
therefore also tried to cite papers that serve as a useful entry to
the subject for those wishing to explore further. \citet{Aerts2010}
provide an excellent overview and introduction to asteroseismology in
general.  Some recent reviews and articles on asteroseismology of
solar-like oscillators include \citet{Cunha2007, jcd2010,
  Bedding2011b, jcd2011a} and \citet{Michel2012}.

 \section{A little background theory: solar-like oscillations}
 \label{sec:theory}

The oscillations have their physical origins in two types of standing
waves, those that are predominantly acoustic in character (commonly
referred to as pressure modes or p modes) where gradients of pressure
act as the restoring force; or internal gravity waves (g modes), where
the effects of buoyancy are relevant. Modes of mixed character may
also exist, displaying g-mode like behaviour in the central region of
a star, and p-mode like behaviour in the envelope.  

Solar-like oscillations are intrinsically stable: they are driven
stochastically and damped intrinsically by vigorous turbulence in the
superficial layers of the sub-surface convection zone \citep[e.g.,
  see][ and references therein]{Houdek1999, Samadi2001}. A necessary
condition for stars to show solar-like oscillations is therefore the
presence of near-surface convection.  The oscillation spectra shown by
solar-type and red-giant stars are very rich, with multiple overtones
excited to observable amplitudes.  Because of geometrical
cancellation, only modes of low angular (spherical) degree, $l$, can
be observed.

In this section we shall consider the oscillation spectra of cool
stars in different evolutionary states, beginning with main-sequence
stars which show spectra of pure p modes. Our purpose is to lay some
of the ground work for detailed discussion of results and future
prospects in following sections, where we also elaborate on some of
the theory (as appropriate).  We begin here by using expressions based
on an asymptotic treatment to illustrate the principal characteristics
of the spectra. The asymptotic formalism has its limitations,
particularly when describing the oscillation characteristics of
evolved stars. Indeed, significant deviations from asymptotic
behaviour can even become apparent in main-sequence stars. We shall
see that these deviations may be used to make key inferences on the
internal structures of the stars.

Detectable p modes in main-sequence stars have high overtone numbers
(radial orders) $n$, meaning asymptotic theory may be applied to
describe the frequencies $\nu_{nl}$ (i.e., as $l/n \rightarrow 0$). An
approximate expression given to second order may be written (e.g., see
\citealt{Gough1986, jcd2010}):
 \begin{equation}
 \nu_{nl} \simeq \Delta\nu \left( n + \frac{l}{2} + \epsilon \right) - 
                \Delta\nu^2 \left[\frac{A[l(l+1)]-B}{\nu_{nl}}\right],
 \label{eq:asspall}
 \end{equation}
where
 \begin{equation}
  \Delta\nu = \left( 2 \int_{0}^{R} \frac{{\rm d}r}{c} \right)^{-1}
 \label{eq:Dnu}
 \end{equation}
is the inverse of the acoustic diameter, i.e., the sound travel time
across a stellar diameter, $c$ being the sound speed and $R$ the
stellar radius, and the coefficient $\epsilon$ depends on the cavity
boundary conditions, the behaviour close to the stellar surface being
most important.  The coefficient $A$ in the second-order term is given
by
 \begin{equation}
 A = \left(4\pi^2\Delta\nu \right)^{-1} \left[ \frac{c(R)}{R} -
 \int_{0}^{R} \frac{{\rm d}c}{{\rm d}r} \frac{{\rm d}r}{r} \right],
 \label{eq:A}
 \end{equation}
while $B$ is a small correction that also depends on the surface
boundary conditions.

The leading term on the right-hand side of Equation~\ref{eq:asspall}
implies a dominant overtone spacing in the spectrum, the so-called
large frequency separation between modes of the same $l$, i.e.,
$\Delta\nu_{nl} = \nu_{nl} - \nu_{n-1\,l} \simeq \Delta\nu$. Modes of
even and odd degree should be separated by $\simeq \Delta\nu/2$. The
final term on the right-hand side describes the departure from the
degeneracy $\nu_{nl} \approx \nu_{n+1\,l-2}$ implied by the first
term, i.e.,
 \begin{equation}
 \delta\nu_{l\,l+2}(n) = \nu_{nl}-\nu_{n-1\,l+2} \simeq -(4l+6) \frac{\Delta\nu}{4\pi^2\nu_{nl}}
                       \int_{0}^{R} \frac{{\rm d}c}{{\rm d}r} \frac{{\rm d}r}{r}. 
 \label{eq:dsmall}
 \end{equation}
This so-called small frequency separation depends on the sound-speed
gradient in the central regions, which in main-sequence stars depends
critically on the evolutionary state. In the asymptotic limit,
Equation~\ref{eq:asspall} implies that $\delta\nu_{13}(n) = 5/3\,
\delta\nu_{02}(n)$. Other useful small frequency separations may be
formed by $l=0$ and $l=1$ modes (e.g., see \citealt{Roxburgh2009} and
references therein). For example, the small separation
$\delta\nu_{01}(n) = \nu_{n\,0} - \frac{1}{2} \left( \nu_{n-1\,1} +
\nu_{n\,1} \right)$ measures deviations of $l=0$ frequencies from the
exact halfway frequencies of the adjacent $l=1$ modes; while
$\delta\nu_{10}(n) = \frac{1}{2} \left( \nu_{n\,0} + \nu_{n+1\,0}
\right) - \nu_{n\,1}$ does the same for $l=1$ frequencies and the
adjacent $l=0$ modes. In the asymptotic limit, $\delta\nu_{01}(n) =
\delta\nu_{10}(n) = 1/3\,\delta\nu_{02}(n)$. Similar separations may
also be formed from combinations of five (or more)
frequencies. Frequency separation ratios \citep{Roxburgh2003} are
formed by taking the ratio of small to large frequency separations,
e.g., $r_{02}(n) = \delta\nu_{02}(n)/\Delta\nu_1(n)$. Use of these
ratios offers the advantage that they are somewhat insensitive to the
structure of the near-surface layers of stars -- which are poorly
described by models -- because the small and large separations are
affected in a similar way by near-surface effects. We discuss the
issue of the near-surface layers later in
Section~\ref{sec:modelling}. As noted previously, it should be borne
in mind that the asymptotic expressions may provide poor descriptions
in more evolved stars.

Rotation lifts the degeneracy in the oscillation frequencies
$\nu_{nl}$, so that the frequencies of non-radial modes ($l>0$) depend
on the azimuthal order, $m$. For the fairly modest rates of rotation
typical of solar-like oscillators we may ignore, to first order, the
effects of the centrifugal distortion. Only in the most rapidly
rotating solar-like oscillators are these effects expected to give
rise to measureable frequency asymmetries of detectable high-$n$,
low-$l$ p modes \citep{Reese2006, Ballot2010}.  The $2l+1$
rotationally split frequencies may be written:
 \begin{equation}
  \nu_{nlm} \equiv \nu_{nl} + \delta\nu_{nlm},
 \end{equation}
with
 \begin{equation}
 \delta\nu_{nlm} \simeq \frac{m}{2\pi}
                \int_0^R \int_0^\pi K_{nlm}(r,\theta) 
                \Omega(r,\theta)r\,{\rm d}r\,{\rm d}\theta.
 \label{eq:rot}
 \end{equation}
Here, $\Omega(r,\theta)$ is the position-dependent internal angular
velocity (in radius $r$, and co-latitude $\theta$), and $K_{nlm}$ is a
weighting kernel that reflects the sensitivity of the mode to the
internal rotation. For the high-order p modes observed in solar-like
oscillators, modest rates of differential rotation (in latitude and
radius) and absolute rotation mean the splittings $\delta\nu_{nlm}$ of
the observable modes will typically take very similar values (hence
tending to the approximation of solid-body rotation). The above
neglects any contributions to the splittings from near-surface
magnetic fields, which give rise to frequency asymmetries of the
observed splittings. We might expect to detect departures from
symmetry in more active stars \citep{Gough1990}, a point we return to
briefly in subsequent sections.

\begin{figure}
 \centerline {\epsfxsize=14.0cm\epsfbox{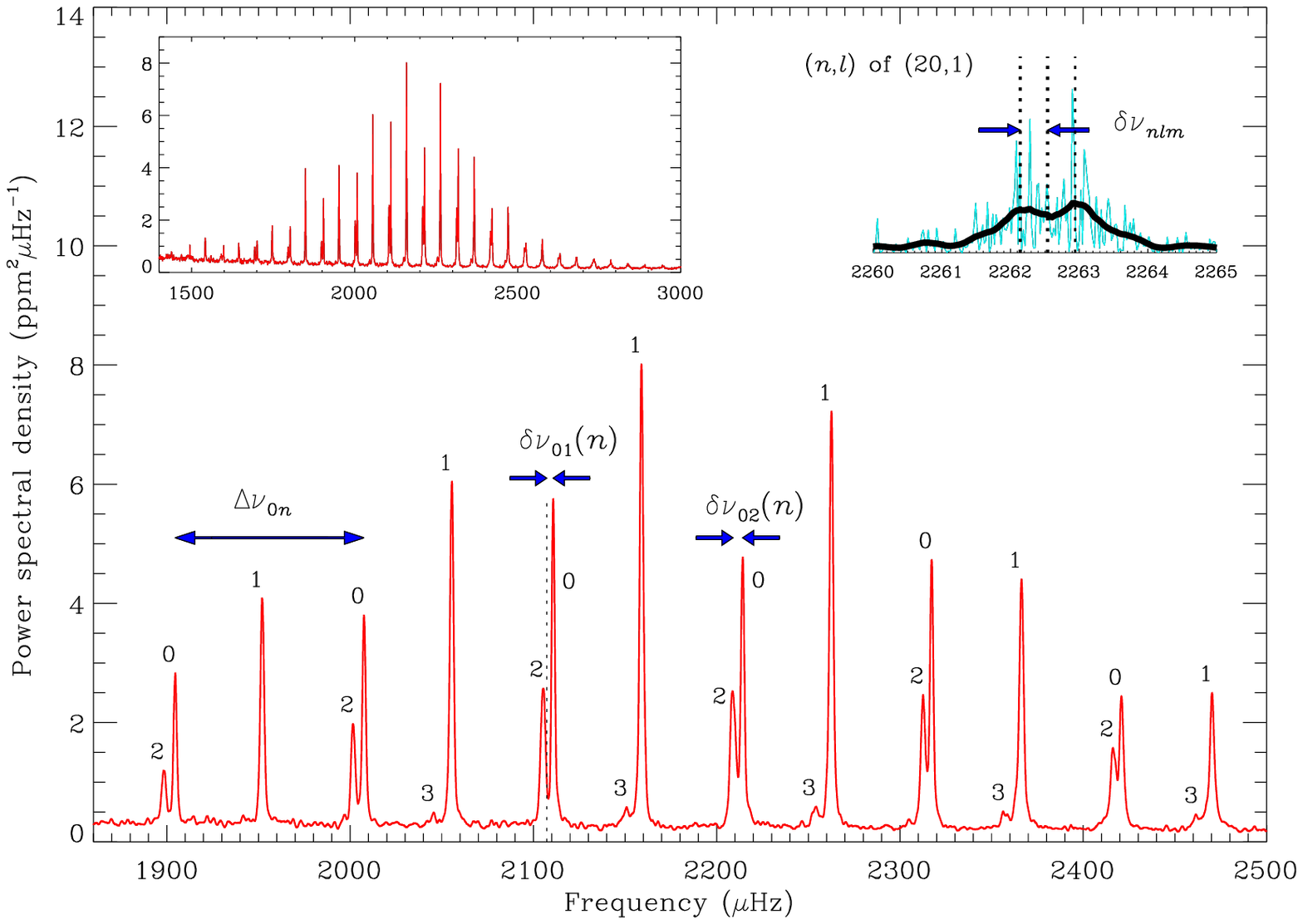}}

 \caption{Oscillation spectrum of the G-type main-sequence star in
   16\,Cyg\,A (KIC\,12069424, HD\,186408), as observed by
   \emph{Kepler}. Main plot: smoothed frequency-power spectrum showing
   the frequency range containing the most prominent modes in the
   spectrum, with annotations marking key frequency separations.  (The
   smoothing filter was a double-boxcar filter of width $0.2\,\rm \mu
   Hz$.) Top left-hand inset: Plot of a wider range in frequency,
   showing the Gaussian-like modulation (in frequency) of the observed
   powers of the modes. The frequency of maximum oscillations power,
   $\nu_{\rm max}$, lies at about $2200\,\rm \mu Hz$. Top right-hand
   insent: zoom in frequency showing rotational frequency splitting of
   the non-radial $l=1$, $n=20$ mode. The raw spectrum is shown in
   light blue, and the smoothed spectrum in black. The rotation axis
   of the star is inclined such that the outer $|m|=1$ components are
   visible (outer vertical dashed lines), but the inner $m=0$
   component is not as prominent (central vertical dashed line).}

 \label{fig:spec} 
\end{figure}

Figure~\ref{fig:spec} shows in detail the observed oscillation
spectrum of the G-type main-sequence star 16\,Cyg\,A (KIC\,12069424,
HD\,186408; see \citealt{Metcalfe2012}), the more massive component of
the solar-type binary system 16\,Cyg, two of the brightest stars with
\emph{Kepler} data.  The observed frequencies and frequency splittings
carry information on the internal structure and dynamics of the star,
and may in turn be used to place tight constraints on the fundamental
stellar properties (including the age). As noted previously, the
dominant spacing visually is the large frequency separation.  The
average large separation $\left< \Delta\nu_{nl} \right>$ is to good
approximation equal to $\Delta\nu$ in Equation~\ref{eq:asspall}, and
hence provides a measure of the acoustic radius of the star. In turn,
it may be shown that the average large separation scales to very good
approximation with the square root of the mean density
\citep{Ulrich1986}, i.e.,
 \begin{equation}
 \left< \Delta\nu_{nl} \right> \propto \left< \rho \right>^{1/2},
 \label{eq:scalednu}
 \end{equation}
since the frequencies and overtone spacings are related to the
dynamical timescale of the star.

\begin{figure}
 \centerline {\epsfxsize=7.5cm\epsfbox{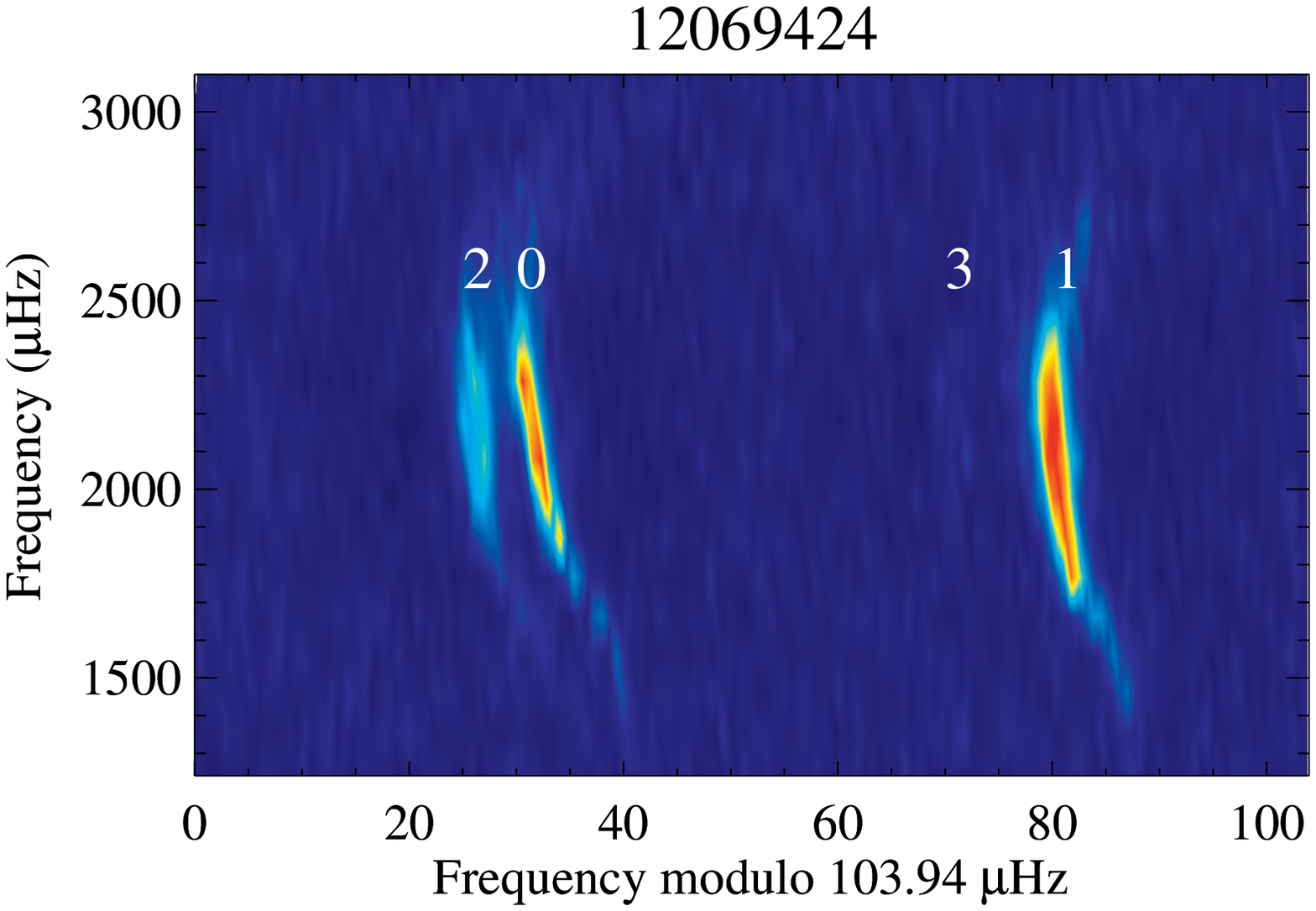}
              \epsfxsize=7.5cm\epsfbox{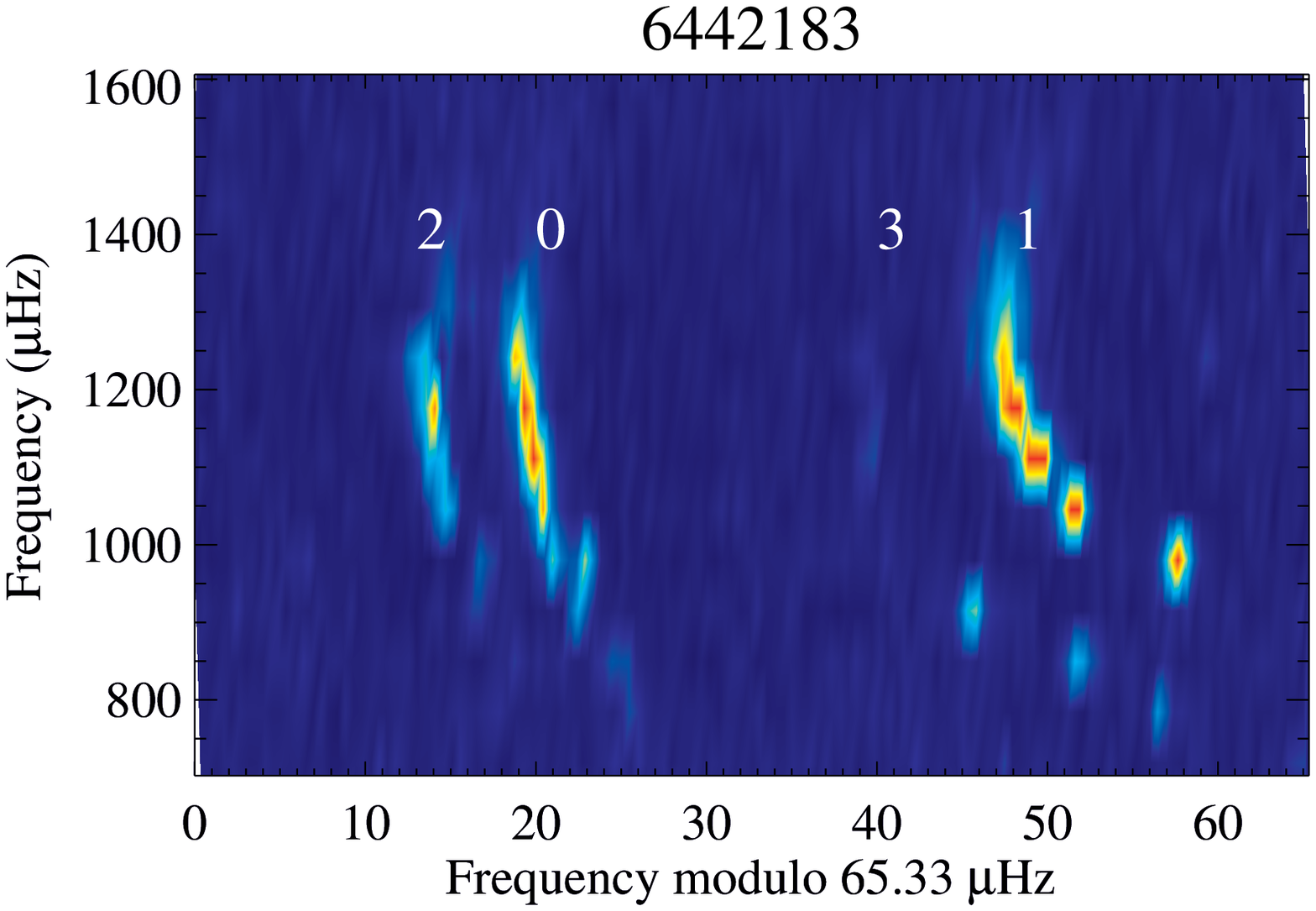}}
 \centerline {\epsfxsize=7.5cm\epsfbox{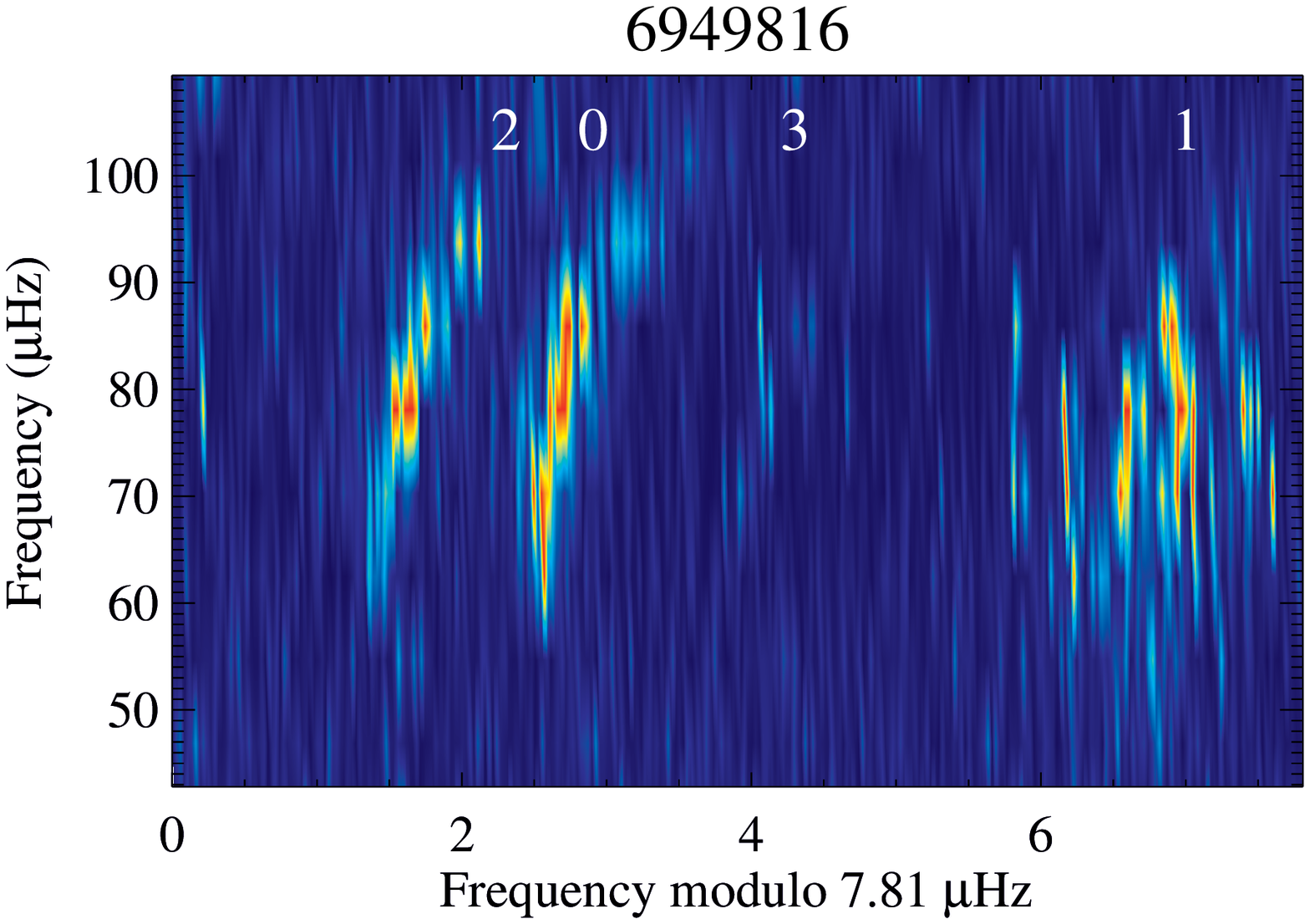}}
 \centerline {\epsfxsize=7.5cm\epsfbox{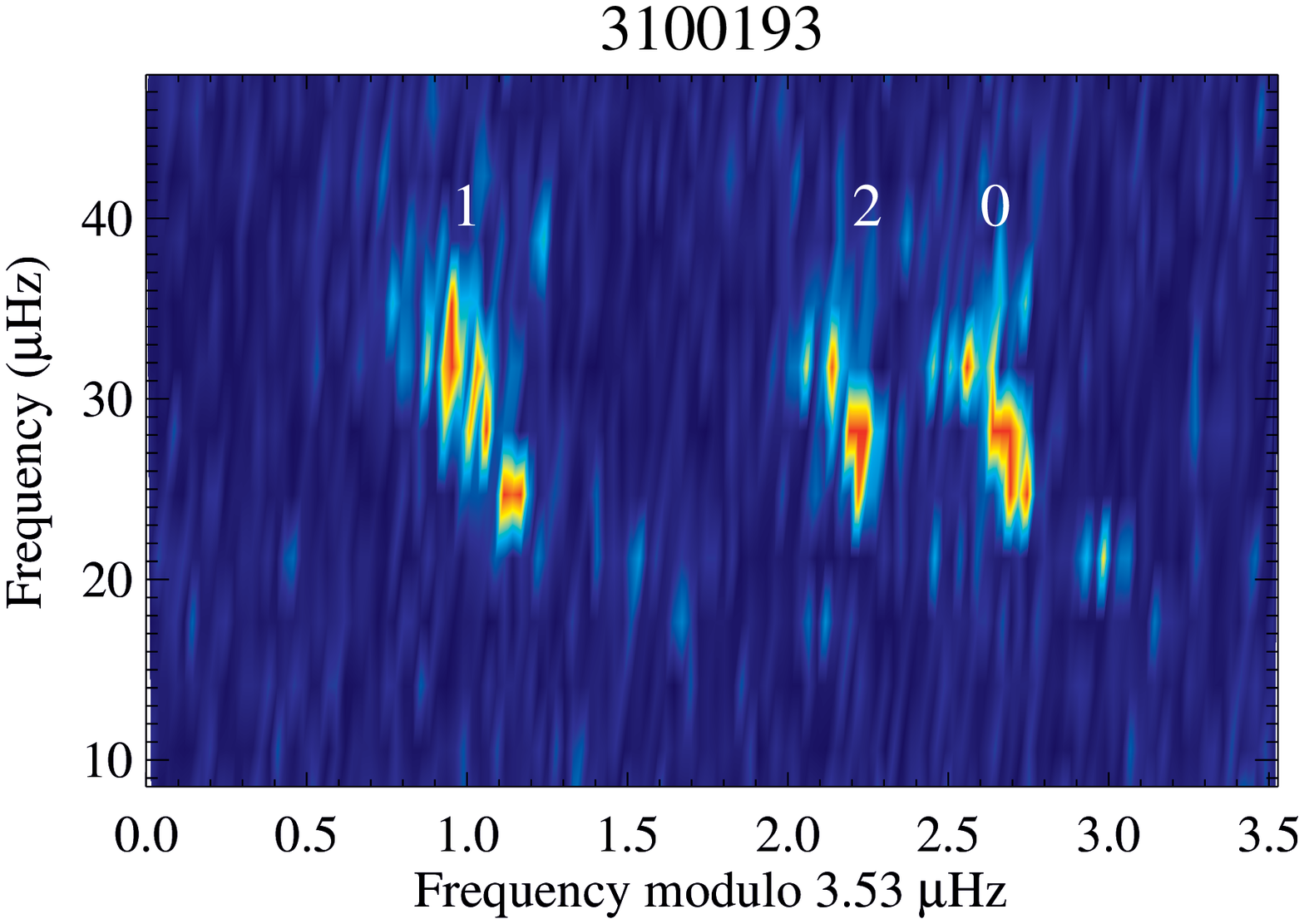}
              \epsfxsize=7.5cm\epsfbox{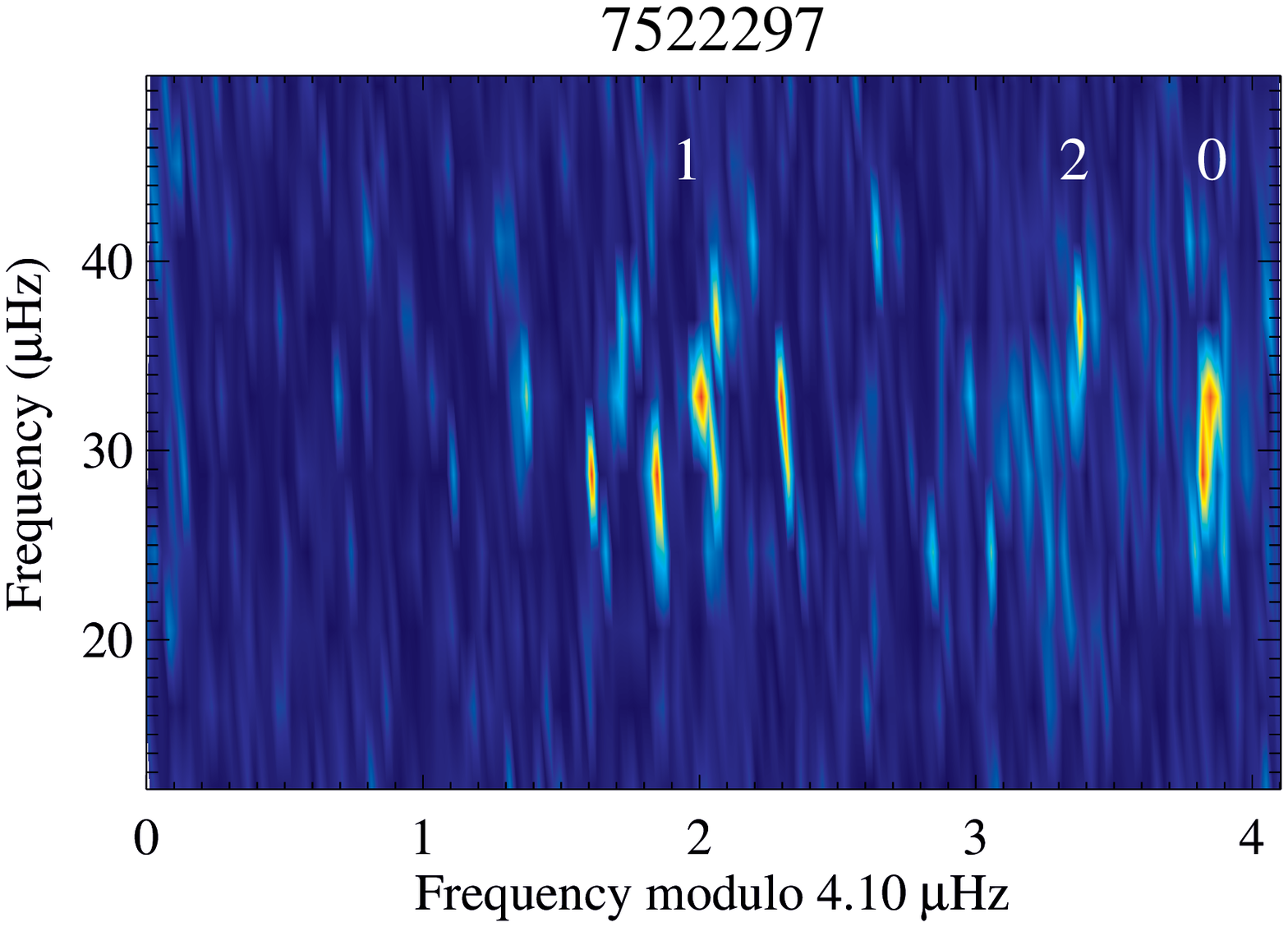}}

 \caption{\small Echelle diagrams (see text) of the oscillation
   spectra of five stars observed by \emph{Kepler}. Annotations mark
   the angular degrees, $l$. Top left-hand panel: diagram for the
   main-sequence star 16\,Cyg~A, showing vertically aligned ridges of
   oscillation power. Note the faint (but significant) power of the
   $l=3$ ridge. Top right-hand panel: diagram for the subgiant
   KIC\,6442183 showing a beautiful avoided crossing of the $l=1$
   modes at a frequency around $1000\,\rm \mu Hz$. Middle panel:
   diagram for the first-ascent RGB star KIC\,6949816, which shows
   clusters of closely spaced $l=1$ mixed modes in its
   spectrum. Bottom panels: diagrams of RGB (KIC\,3100193) and RC
   (KIC\,7522297) stars that have similar surface properties (note the
   complexity of the $l=1$ modes in the spectrum of KIC\,7522297
   compared to KIC\,3100193).}

 \label{fig:echelle} 
\end{figure}

The top-left hand panel of Figure~\ref{fig:echelle} shows a so-called
\'echelle diagram of the oscillation spectrum of 16\,Cyg~A. This was
made by dividing the oscillation spectrum into segments of length
$\left< \Delta\nu_{nl} \right>$ in frequency. The segments are then
arranged above one another, in order of ascending frequency. Were
stars to show a strict correspondence to Equation~\ref{eq:asspall}
vertical, straight ridges would be found in the \'echelle diagram
(assuming use of the correct spacing), since the overtone spacings
would all be exactly equal to $\Delta\nu$. In practice, stars show
departures from this simple description. In main-sequence stars, like
16\,Cyg~A, those departures are modest but clearly detectable. They
carry frequency signatures of regions of abrupt structural change in
the stellar interiors, e.g., the near-surface ionization zones and the
base of the convective envelope, which we discuss in
Section~\ref{sec:abrupt}; and more subtle signatures due to small
convective cores (if present; see Section~\ref{sec:cores}), and the
general background state.

It is worth remarking that $l=3$ modes are detectable in
16\,Cyg~A. This is due to the exceptional quality of its \emph{Kepler}
data. Indeed, the quality is so good that the S/N in the modes is
limited by intrinsic stellar noise, and not by shot noise. Typically,
CoRoT and \emph{Kepler} give useable data for solar-type stars up to
$l=2$, with the very weak $l=3$ modes usually lost in the noise (see
also Section~\ref{sec:obs} below).

The oscillation peaks in the frequency-power spectrum have an
underlying Lorentzian-like appearance, i.e., the form expected for
damped oscillations. The widths of the Lorentzians provide a measure
of the linear damping rates, while the amplitudes -- which increase
with decreasing $\log\,g$, as stars evolve -- are set by the delicate
balance between the excitation and damping.  Small asymmetries of the
Lorentzian peaks are detectable in solar p modes, and are expected in
other stars. These asymmetries arise from the very localized (in
radius) excitation source, and the correlation of the p-mode signal
and granulation signal (the latter excites and damps the former; e.g.,
see \citealt{Roxburgh1997}). Measurement of the excitation and damping
parameters provides the means to infer various important properties of
the still poorly understood near-surface convection (e.g., see
\citealt{Chaplin2005, Samadi2007, Houdek2010}).

The observed powers of modes in the oscillations spectrum are
modulated in frequency by a Gaussian-like envelope (see top left-hand
inset of Figure~\ref{fig:spec}). The frequency of maximum oscillations
power, $\nu_{\rm max}$, carries diagnostic information on the
excitation and damping and hence physical conditions in the
near-surface layers. The behaviour of waves close to the surface is
influenced strongly by the acoustic cut-off frequency, $\nu_{\rm ac}$,
which is given by
 \begin{equation}
 \nu_{\rm ac}^2 = \left( \frac{c}{4\pi H} \right)^2 
                \left(1 - 2 \frac{{\rm d}H}{{\rm d}r} \right),
 \label{eq:nuac}
 \end{equation}
with $c$ the speed of sound and $H = -({\rm d\,ln}\rho/{\rm d}r)^{-1}$
the density scale height. The sharp rise in $\nu_{\rm ac}$ close to
the surface of a star describes an efficient boundary for the
reflection of waves having $\nu < \nu_{\rm ac}$, hence fixing a
notional upper limit in frequency for the trapped oscillations.

\begin{figure}
 \centerline {\epsfxsize=14.0cm\epsfbox{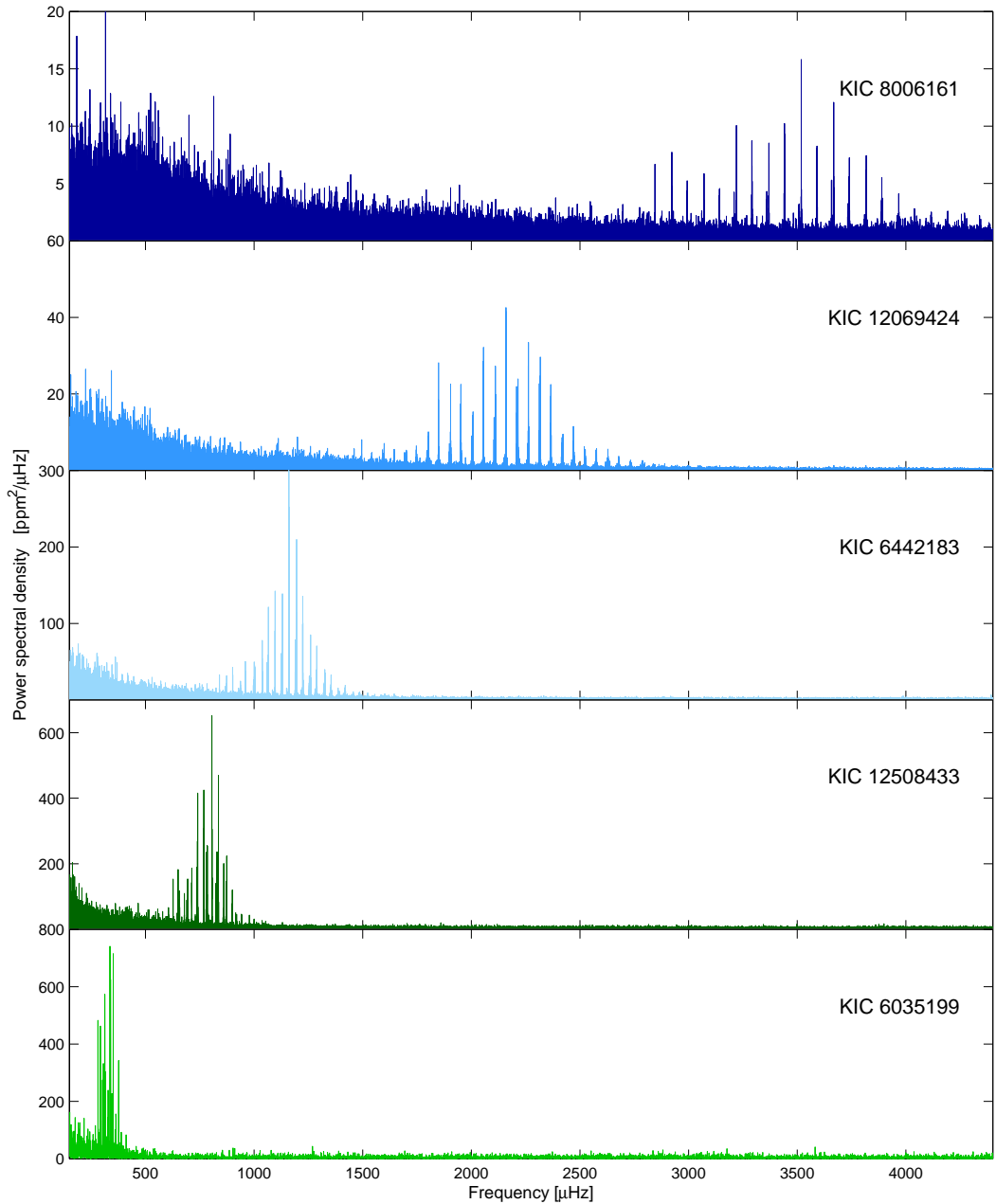}}

 \caption{Solar-like oscillation spectra of five stars observed by
   \emph{Kepler}, using its short-cadence data (see
   Section~\ref{sec:obs}). Each star has a mass around $1\,\rm
   M_{\odot}$. Stars are arranged from top to bottom in order of
   decreasing $\nu_{\rm max}$, i.e., decreasing surface gravity. The
   top two stars -- KIC\,8006161 and KIC\,12069424 (16\,Cyg~A) -- are
   main-sequence stars. The third and fourth stars down --
   KIC\,6442183 (HD\,183159) and KIC\,12508433 -- are subgiants. The
   bottom star (KIC\,6035199) lies at the base of the RGB. Echelle
   diagrams of KIC\,12069424 and KIC\,6442183 may be found in
   Figure~\ref{fig:echelle}.}

 \label{fig:stacked1} 
\end{figure}

\citet{Brown1991} conjectured that $\nu_{\rm max} \propto \nu_{\rm
  ac}$, since both frequencies are determined by the near-surface
properties. We may turn this into a relation linking $\nu_{\rm max}$
to measureable surface properties of a solar-like oscillator by noting
that for the relevant stellar models little accuracy is lost by
applying an isothermal approximation of Equation~\ref{eq:nuac}, where
$\nu_{\rm ac} = c/(4\pi H)$. This suggests a scaling relation of the
form \citep{Kjeldsen1995a}:
 \begin{equation}
 \nu_{\rm max} \propto \nu_{\rm ac} \propto \frac{c}{H} \propto
 g\,T_{\rm eff}^{-1/2},
 \label{eq:scalenumax}
 \end{equation}
where $g \propto M/R^2$ is the surface gravity and $T_{\rm eff}$ is
the effective temperature of the star. As a solar-type star evolves,
the frequencies shown by its most prominent oscillations will
therefore decrease, largely in response to the falling surface
gravity.  Figure~\ref{fig:stacked1} shows the oscillation spectra of
five stars observed by \emph{Kepler} (including 16\,Cyg~A), each
having a mass around $1\,\rm M_{\odot}$. Surface gravity decreases
from top to bottom by about one order of magnitude. The top two stars
lie on the main sequence; the third and fourth stars are subgiants,
whilst the bottom star lies near the base of the red giant branch
(RGB). As we shall see in Section~\ref{sec:scaling},
Equation~\ref{eq:scalenumax} appears to work remarkably well in
practice, although the power envelopes of some of the hottest F-type
stars have a flatter maximum (e.g., Procyon~A; see
\citealt{Arentoft2008}) raising question marks over the diagnostic
potential, and even the definition or meaning, of $\nu_{\rm max}$ in
those stars. More work is clearly needed to understand the observed
$\nu_{\rm max}$ scaling, and theoretical studies have made progress on
the problem (e.g., see \citealt{Belkacem2011}).

After cessation of hydrogen core burning, stars leave the main
sequence and their oscillation spectra become more complicated.  This
is because there is no longer a clear separation of the frequency
ranges that will support p modes and g modes. The behaviour is
controlled largely by the bouyancy or Brunt-V\"ais\"al\"a frequency,
$N$, given by
 \begin{equation}
 N^2 = g \left(
         \frac{1}{\Gamma_1} \frac{{\rm d\, ln}p}{{\rm d}r} - 
                            \frac{{\rm d\,ln}\rho}{{\rm d}r}
         \right).
 \label{eq:N}
 \end{equation}
The g modes have frequencies that are lower than $N$, and high-order g
modes may be described by an asymptotic relation in the periods,
$\Pi_{nl}$, i.e.,
 \begin{equation}
  \Pi_{nl}  = \nu_{nl}^{-1} \simeq \Delta\Pi_{l} \left( n + \epsilon_{g} \right),
 \label{eq:gass}
 \end{equation}
where the period separation $\Delta\Pi_{l}$ (analagous to $\Delta\nu$
for p modes) is given by:
 \begin{equation}
  \Delta\Pi_{l} = \frac{2\pi^2}{\sqrt{l(l+1)}}
                 \left( \int_{r_1}^{r_2} N \frac{{\rm d}r}{r} \right)^{-1},
 \label{eq:gspace}
 \end{equation}
assuming that $N^2 \geq 0$ in the convectively stable region bounded
by $\left[r_1, r_2\right]$, with $N = 0$ at $r_1$ and $r_2$
\citep{Tassoul1980}.

After exhaustion of the central hydrogen, the buoyancy frequency in
the deep stellar interior increases to such an extent that it extends
into the frequency range of the high-order p modes. When the frequency
of a g mode comes close to that of a non-radial p mode of the same
degree, $l$, the modes undergo an ``avoided crossing''
\citep{Aizenman1977}, which is analagous to avoided crossings of
atomic energy states. Interactions between the modes will affect (or
bump) the frequencies and also change the intrinsic properties of the
modes, so that they take on mixed p and g characteristics (having a
g-mode character in the deep interior and a p-mode character in the
envelope).

The first subgiant showing evidence of an avoided crossing in its
oscillation spectrum ($\eta$\,Boo) also happened to be the first case
of a star other than the Sun showing unambiguous evidence of
solar-like oscillations \citep{Kjeldsen1995b, jcd1995}. Ground-based
observations also provided another case of a subgiant showing mixed
modes ($\beta$\,Hyi; see \citealt{Bedding2007}). \emph{Kepler} and
CoRoT have now added a significant number of further cases, with much
more precisely determined oscillation frequencies.  The top right-hand
panel of Figure~\ref{fig:echelle} shows the \'echelle diagram of one
such \emph{Kepler} target, the G-type subgiant KIC\,6442183
(HD\,183159; see also Figure~\ref{fig:stacked1}).  First, we note that
the $l=0$ p modes are unaffected by the phenomenon because
perturbations in buoyancy will not support radial modes, and so there
are no g modes for the radial p modes to couple to. However, several
$l=1$ p modes have been significantly shifted from the putative,
undisturbed $l=1$ ridge (including one mode that has been shifted
close to the $l=0$ ridge). The larger the shift in frequency from the
undisturbed ridge, the stronger is the interaction or coupling between
the p and g modes.  The coupling is much weaker at $l=2$.

The frequency of the avoided crossing -- where the frequency
perturbation changes sign, across the undisturbed ridge -- corresponds
to the pure g-mode frequency that the star would show if it was
comprised only of the central, g-mode cavity, and is hence a sensitive
diagnostic of the core properties and the exact evolutionary state of
the star. The displaced $l=1$ frequencies of KIC\,6442183 are examples
of several p modes coupling to a single g mode. The detectability of
the modes may be partially understood in terms of the mode inertia, a
measure of the fraction of the star's mass that is engaged in
pulsation (e.g., see \citealt{Aerts2010}). The g-dominated mixed modes
have high mode inertias because their eigenfunctions show large
amplitudes in the core, where the density is high; this contributes
to making them undetectable in the observations (see also
Section~\ref{sec:obs} below). The amplitudes of the eigenfunctions of
the p-dominated modes are in contrast highest in the envelope, hence
these modes are clearly detectable.

\begin{figure}
 \centerline {\epsfxsize=14.0cm\epsfbox{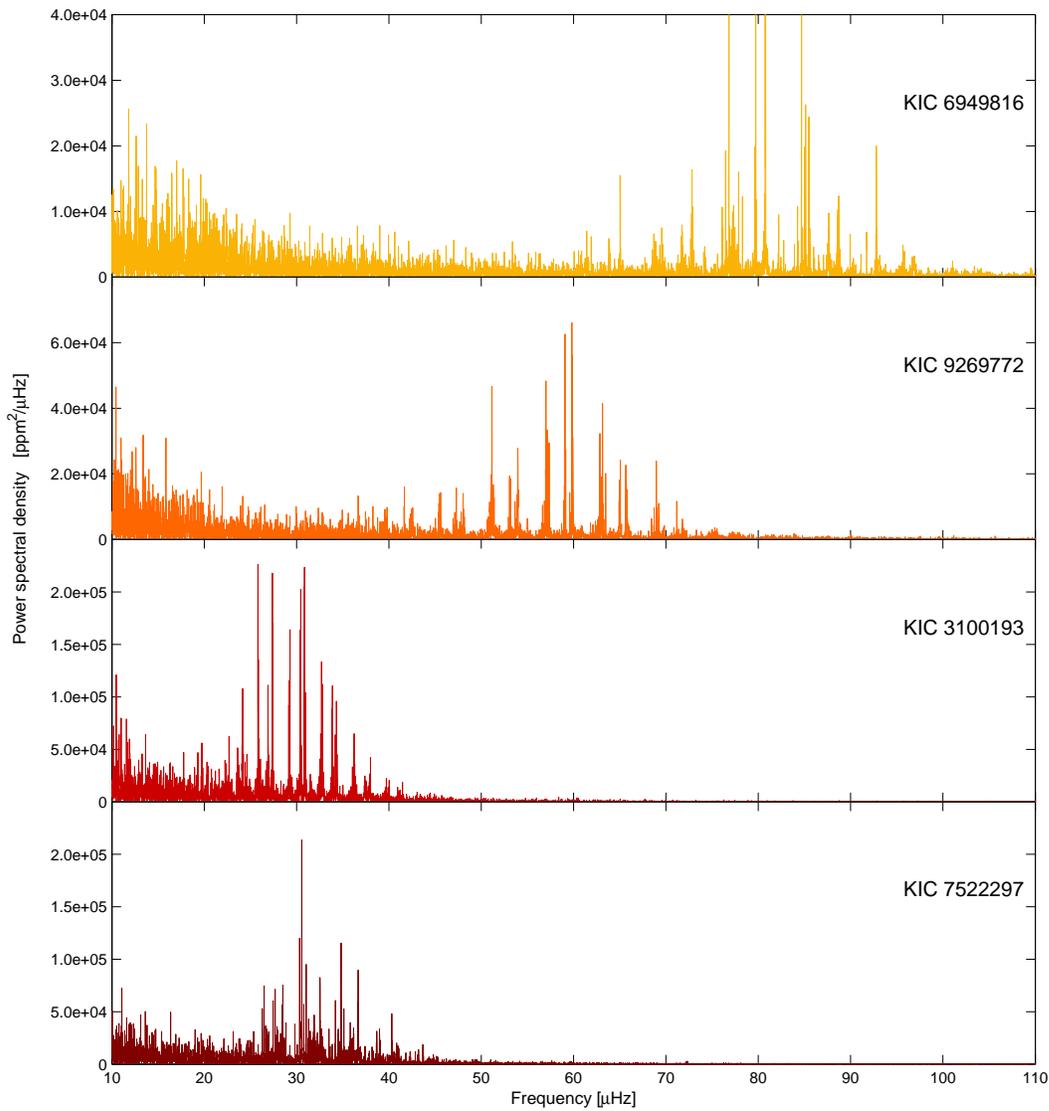}}

 \caption{Solar-like oscillation spectra of five stars observed by
   \emph{Kepler}, using its long-cadence data (see
   Section~\ref{sec:obs}). Each star has a mass around $1\,\rm
   M_{\odot}$. KIC\,6949816 and KIC\,9269772 are both first-ascent RGB
   stars. KIC\,3100193 and KIC\,7522297 are, respectively, RGB and RC
   stars sharing similar surface properties. Echelle diagrams of
   KIC\,6949816, KIC\,3100193 and KIC\,7522297 may be found in
   Figure~\ref{fig:echelle}.}

 \label{fig:stacked2} 
\end{figure}

The impact of mode coupling is much more apparent in red giants,
leading to a much richer ensemble of observable signatures
\citep{Dupret2009}, which, as we shall see in later sections, may be
used to discriminate different advanced phases of
evolution. Figure~\ref{fig:stacked2} shows the oscillation spectra of
four red giant stars observed by \emph{Kepler}, again all having
masses around $1\,\rm M_{\odot}$. The middle panel of
Figure~\ref{fig:echelle} shows the \'echelle diagram of one of these
stars, KIC\,6949816, a star ascending the RGB. The observed
oscillation frequencies are significantly lower than in earlier phases
of evolution. Again, this is largely the result of the marked
expansion of the outer layers in the post main-sequence phase, which
leads to a large reduction in the surface gravity.

In the main-sequence phase we see one $l=1$ mode in each radial
order. In the early post main-sequence phase it may be possible to
observe more than one $l=1$ mode per order, assuming the g-dominated
modes have a low enough inertia to be detectable. In the red-giant
phase there is a much denser spectrum of g modes for the p modes to
couple to, so that there are potentially many observable $l=1$
mixed-modes per order, as is apparent from Figure~\ref{fig:echelle}
(middle panel) and Figure~\ref{fig:stacked2} (top panel, notably at
frequencies around 77 and $85\,\rm \mu Hz$). The p-dominated mixed
modes are most likely to be observed (again, because they have lower
intertia than their g-dominated counterparts). Here, at each order we
see a cluster of p-dominated mixed modes, arranged about the putative,
undisturbed pure p-mode frequency at that order. Because the observed
mixed modes are not pure g modes, the observed period spacings are in
practice smaller than those implied by Equation~\ref{eq:gspace}.

Following ignition of helium (He) in the core, stars enter a
relatively long-lived core-He burning phase. The ignition of He in low
mass stars takes place in a highly degenerate He core of mass $\approx
0.47\,\rm M_\odot$, irrespective of the total mass of the star. This
He flash is followed by a structural readjustment resulting in an
accumulation of stars in the Hertzsprung-Russell diagram known as the
Red Clump (RC). The bottom right-hand panel of Figure~\ref{fig:echelle}
shows the \'echelle diagram of an RC star, KIC\,7522297 (see also
Figure~\ref{fig:stacked2}). Its oscillation spectrum is very
complicated, in contrast to the spectrum of KIC\,3100193 (bottom
left-hand panel, and again Figure~\ref{fig:stacked2}), an RGB star
with similar surface properties and hence a similar $\nu_{\rm
  max}$. In Section~\ref{sec:mixed} we shall discuss at some length
use of the g-mode period spacings as a diagnostic of the evolutionary
state, and see that increased period spacings in RC stars allows them
to be distinguished from RGB stars that lie in close proximity in the
Hertzsprung-Russell diagram.

 \section{Observational data for solar-like oscillators}
 \label{sec:obs}

Asteroseismology is currently an observationally driven field, with
large volumes of exquisite quality long timeseries data now available,
in particular from \emph{Kepler} and CoRoT. Their long datasets give
the frequency resolution needed to extract accurate and precise
estimates of the basic parameters of individual modes covering several
radial orders, such as frequencies, frequency splittings, amplitudes,
and damping rates.  Uninterrupted data offer huge advantages for the
analysis. The diurnal gaps present in ground-based observations cause
well-known frequency-aliasing problems. The frequency separations
associated with the resulting ``sidebands'' can be close to the
characterisic frequency separations presented by solar-like
oscillators, making interpretation of the observed spectra more
difficult. A less well-known problem arises from the fact that
ground-based window functions typically also carry a quasi-random
component (e.g., weather), which introduces quasi-white (i.e.,
broadband) noise in frequency. This may significantly degrade the
underlying signal-to-noise ratio in the modes.

In spite of these complications, ground-based Doppler velocity
observations have an important r\^ole to play. The signatures of
granulation are much less prominent in velocity, relative to the
oscillations, than they are in photometry.  Since the amplitude of the
granulation signal increases with decreasing frequency (i.e., its
spectrum is ``pink''), the intrinsic stellar noise presents less of a
challenge to the detectability of low-frequency modes when
observations are made in velocity. In main-sequence stars, these modes
present the narrowest widths in frequency (being more lightly damped
than their higher-frequency counterparts) and so it is possible to
extract extremely accurate and exquisitely precise frequencies and
rotational frequency splittings for making inference on the internal
structure and dynamics. Doppler velocity observations are also more
sensitive to modes of $l=2$ and 3 than are photometric observations.
Finally, when complementary photometric and Doppler-velocity data are
available on the same target, important inferences can be made on the
interactions of the oscillations with the convection, and the physics
of non-adiabatic processes in the near-surface layers of the star
\citep{Houdek1999, Houdek2010, Huber2011a}. Results of such
comparisons can help to calibrate convection models and the treatment
of radiative transport.

Asteroseismic detections provided by \emph{Kepler} and CoRoT are
considerably more numerous for red giants than for solar-type
stars. Detection of oscillations in solar-type stars requires the
target-limited, short-cadence data\footnote{CoRoT had the capability
  to observe up to 10 targets simulaneously in short cadence, and five
  following the failure of one of its detectors. \emph{Kepler} has the
  capability to observe up to 512 targets simultanously in short
  cadence.} of each mission (58.85\,s and 32\,s cadences,
respectively) since the dominant oscillations have periods of the
order of minutes. These short periods are not accessible to the
long-cadence data of either mission (having 29.4\,min and 8.5\,min
cadences, respectively). The periods and amplitudes of the
oscillations in red giants are significantly longer and higher,
respectively -- the strongest oscillations of stars at the base of the
RGB having periods of the order of 1\,hr -- meaning oscillations may
be detected in fainter (more numerous) targets observed in long
cadence.

CoRoT provided the first multi-month datasets on solar-type stars
\citep{Michel2008} and red giants \citep{DeRidder2009}. At the time of
writing the short-cadence ``seismo'' observations have provided
asteroseismic datasets on 12 solar-type stars in the apparent
magnitude range $5.4 \le m_v \le 9.4$, with dataset lengths ranging
from 20 to 170\,days. The long-cadence ``exo'' observations have
yielded asteroseismic datasets of similar length on more than one
thousand red giants in the range from roughly $11 \la m_v \la 16.5$
\citep[e.g.][]{Hekker2009, Mosser2011a}, and in several different
fields (as we shall discuss later in Section~\ref{sec:pop}). There had
previously been some debate as to whether red giants would show
detectable non-radial modes, but CoRoT resolved the issue with
unambiguous detections in many targets.

\emph{Kepler} has revolutionized asteroseismology of solar-type
stars. During the first 10\,months of science operations an
asteroseismic survey yielded detections of solar-like oscillations in
more than 500 stars observed for one month each in short cadence
\citep{Chaplin2010, Chaplin2011}. These data span spectral types from
early F through to late K, with most targets in the \emph{Kepler}
apparent magnitude range $8 \le K_p \le 12$. There are a small number
of even brighter targets, for which special dedicated photometry masks
have been designed. These include the brightest star falling on
detector pixels, the F-type star $\theta$\,Cyg, and the G-type binary
comprised of the solar-analogues 16\,Cyg\,A and B
\citep{Metcalfe2012}.  These data represent a unique, homogeneous
ensemble for testing stellar evolutionary theory. Around 150
solar-type stars with the highest-quality data were subsequently
selected to be observed for longer durations in short cadence. These
stars now have data ranging in length from several months to years.
Around 60 stars should be observed for the entire duration of the
mission.

The long-cadence \emph{Kepler} data have yielded asteroseismic data of
unprecented quality on stars in the red-giant phase \citep[e.g.,
  see][]{Bedding2010, Huber2010, Kallinger2010, Hekker2011}, including
giants in the open clusters NGC\,6791, NGC\,6819 and NGC\,6811
(\citealt{Stello2011}; the clusters are most likely too faint to allow
detection of oscillations in solar-type stars). High-quality
asteroseismic data are available on around 14,000 red giants having
uninterrupted coverage throughout the first 3.5\,yr of the
mission. The expectation is that many of these targets will have
continued observations throughout the extended mission. These very
long datasets will be required to yield accurate seismic information
on stars near the tip of the RGB, and on the asymptotic giant branch
(AGB) where $\nu_{\rm max}$ and $\left< \Delta\nu_{nl} \right>$ are
both small fractions of a $\rm \mu Hz$ in size.

We come back in Section~\ref{sec:exo} to the approximately 80
\emph{Kepler} asteroseismic targets that are also confirmed, validated
or candidate exoplanet host stars.

\begin{figure}
 \centerline {\epsfxsize=8.5cm\epsfbox{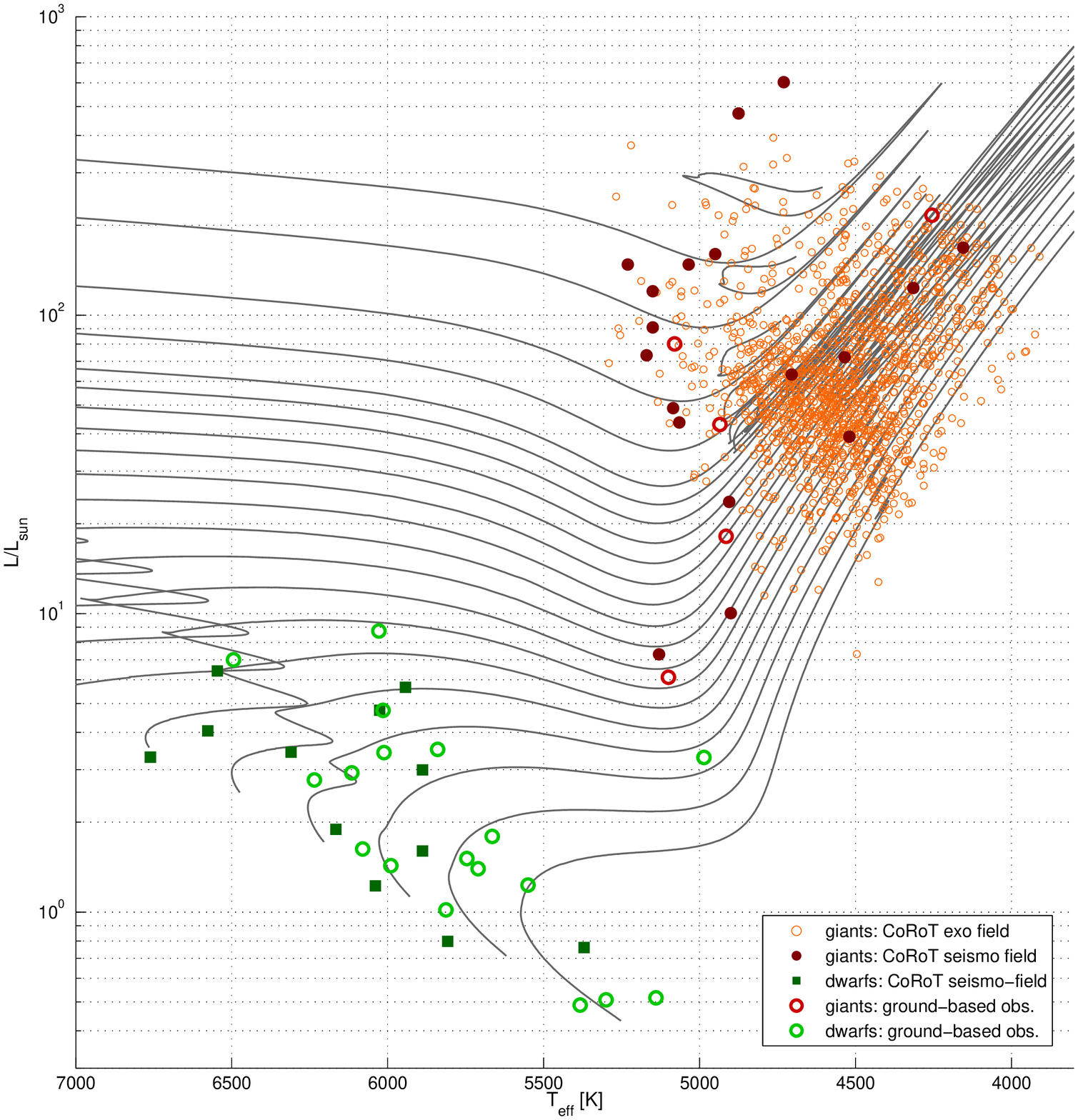}
              \epsfxsize=8.5cm\epsfbox{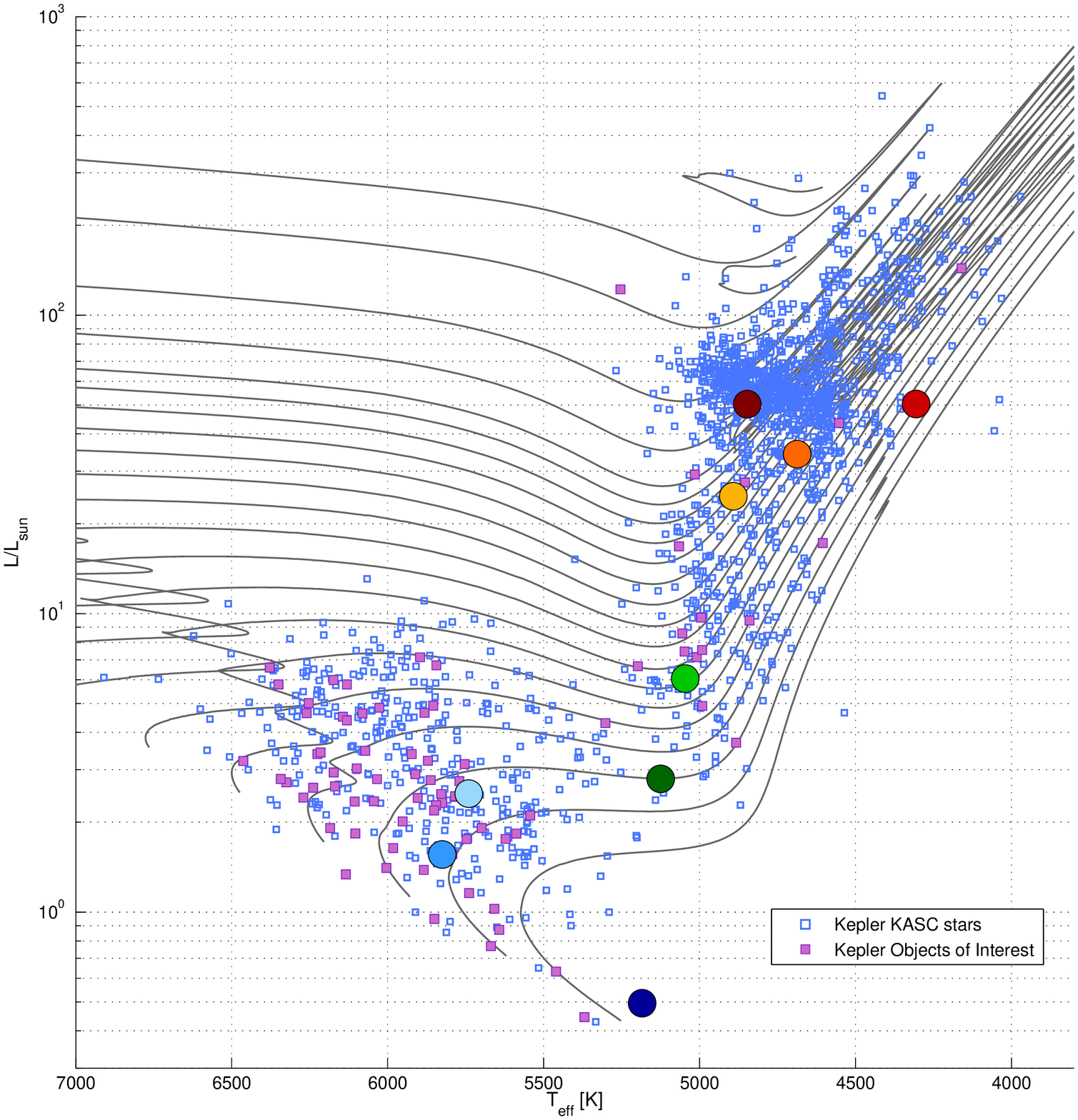}}

 \caption{Hertzsprung-Russell diagrams showing populations of stars
   with detected solar-like oscillations. Left-hand panel: Detections
   made by CoRoT and ground-based telescopes (see legend). Right-hand
   panel: Detections made by \emph{Kepler}, including \emph{Kepler}
   Objects of Interest (see legend). The large coloured circles mark
   the stars whose spectra are plotted in Figures~\ref{fig:spec}
   through~\ref{fig:stacked2}. Solid lines in both panels follow
   evolutionary tracks \citep{Ventura2008} computed assuming solar
   metallicity.}

 \label{fig:hr} 
\end{figure}

Figure~\ref{fig:hr} summarizes the asteroseismic data in-hand on
solar-like oscillators, plotting stars with detections on a
Hertzsprung-Russell diagram.

Developments in the analysis for CoRoT and \emph{Kepler} have built
naturally on the heritage and experience from two areas of the
discipline. The analysis of ground-based data on solar-like
oscillators offered considerable prior expertise in dealing with
moderate quality, low S/N data.  This provided the springboard for
many of the codes developed to extract average or global properties of
the oscillations spectra.  The analysis of Sun-as-a-star
helioseismology data was another obvious starting point from which to
develop techniques for application to main-sequence stars, most
notably for modelling and fitting asteroseismic parameters of
individual modes.

In preparation for \emph{Kepler}, and the expected large numbers of
asteroseismic targets, considerable effort was devoted to developing
and testing codes for automated detection of signatures of solar-like
oscillations (\citealt{Verner2011, Hekker2011a} and references therein),
and subsequent extraction of average asteroseismic parameters such as
$\left< \Delta\nu_{nl} \right>$ and the frequency of maximum
oscillations power, $\nu_{\rm max}$ (e.g., with hare-and-hounds
exercises using realistic, artificial asteroseimic data; see
\citealt{Stello2009}). The much more difficult task of extracting
accurate estimates of parameters of individual modes is colloquially
referred to as ``peak bagging'' (e.g., see
\citealt{Appourchaux2011}). Peak-bagging often proceeds by fitting
multi-parameter models to the observed oscillation spectra, with the
basic building-block being a Lorentzian-like function to describe each
resonant peak having an underlying maximum power spectral density or
height $H$ and width $\Gamma$, where $H \propto A^2/\Gamma$, $A$ being
the oscillation amplitude of the mode. In order to resolve the
Lorentzian, the length of the timeseries, $T$, must span several
amplitude e-folding lifetimes, $\tau$, where $\Gamma = 1/\left( \pi
\tau \right)$. The requirement $T/\tau \ge 10$ is a good guideline
threshold. When $T$ drops to only $2\tau$, so the observed profile
tends to a sinc-squared function (set by $T$), and a Lorentzian is no
longer the correct function to fit. Indeed, when $T \le 2\tau$, a
sine-wave fitting approach (sometimes, slightly misleadingly, called
``pre-whitening'') is the better option (i.e., one then also makes use
of the phase information).

Typical peak widths, $\Gamma$, in main-sequence stars are of the order
of one to a few $\rm \mu Hz$, which suggests that observations
spanning a month or so will just adequately resolve the peaks.  There
is however the added complexity of rotation (and magnetic fields). The
resulting frequency splittings between adjacent non-radial mode
components can range from a fraction to several $\rm \mu Hz$.  In
practice, multi-month datasets are required to get good constraints on
the peak widths and heights of radial modes in main-sequence stars
although it is safe to fit Lorentzian-like models to shorter
datasets; while datasets of a year to several years are needed to
also fully disentangle and so extract robust rotational splittings (of
which more in Section~\ref{sec:exo}), or asymmetries of those
splittings resulting from the near-surface magnetic fields.

Lessons learned from CoRoT's first solar-type targets, which were all
F-type stars, proved vital to the development of peak-bagging
methodology \citep{Appourchaux2008}.  The oscillation peaks in these F
stars turned out to be very wide in frequency (suggesting very heavy
damping of the oscillations) and modes adjacent in frequency were
extremely hard to resolve. This presented challenges not only to the
fitting but also, crucially, rendered visual inspection of the
spectrum useless as a means of tagging correctly the odd and even
angular-degree ridges (a problem that now appears to have been solved
by using the parameter $\epsilon$ from Equation~\ref{eq:asspall} as a
diagnostic for identifying the even-degree ridge; see
\citealt{White2012}).

The main lesson learned from CoRoT was that spectra of solar-type
stars were not always as straightforward to analyse as the Sun.
Further coordinated development and testing of peak-bagging codes, and
application to the more numerous \emph{Kepler} stars, has meant that
near automation of the peak-bagging is now possible for main-sequence
stars and also subgiants that have not yet evolved sufficiently to
show many modes of mixed character (see \citealt{Appourchaux2012}, and
references therein).  The initial step in the automated fitting
packages is to identify modes in the frequency power spectrum from
which a robust list of first-guess parameters then follows.
Identification often involves statistical testing, using false-alarm
or odds-ratio approaches (the latter involving the adoption of priors
on the expected mode signal; see \citealt{Appourchaux2011}).
Application of prior information to maximum-likelihood estimation is
now also common in peak-bagging, although this must be used with care;
and use of Markov Chain Monte Carlo (MCMC) methods is also very
fashionable, as in many other areas of astronomy
\citep{Gruberbauer2009, Benomar2009, Handberg2011}. While
computationally expensive, MCMC offers robust estimation of the
posterior distributions (and hence the confidence intervals) of the
best-fitting parameters. When the fitting is straightforward (high
S/N, with well-constrained fitting distributions) maximum-likeihood
estimators offer a more than adequate approach.

As stars evolve through the subgiant phase onto the RGB, the
appearance of modes of mixed p and g character offers new challenges,
and opportunities, for the fitting. The inertia and through it the
damping of the modes is affected by the coupling, so that the observed
linewidth $\Gamma$ of a mixed mode will be reduced by the ratio of the
(increased) inertia of the mixed mode and the inertia that a radial
mode would have at the same frequency. Provided the modes are well
resolved (see above), the mode height will \emph{not} be affected by
the coupling. However, when the mode peak is not resolved, the height
$H$ will be reduced by the same factor as $\Gamma$. The g-dominated
modes, which have very high inertia and hence very narrow peak
linewidths, are therefore much harder to detect than p-dominated modes
of lower inertia. Those lower-inertia modes that are observable will
still have the benefit of presenting narrower linewidths than pure p
modes at similar frequencies, helping estimation of frequencies and
frequency splittings (as we shall see in Section~\ref{sec:rot}).

The complicated spectra found in evolved solar-like oscillators makes
extraction of robust lists of identified modes and first-guess
parameters non-trivial, although enough is known already about the
pathology of the observed spectra to provide useful guidance
\citep[e.g., see][]{Huber2010, Mosser2011a}. Frequency spacings
between modes and frequency splittings due to rotation may be very
similar, making it hard to disentangle the observed signatures. Use of
reasonably strong priors is advisable when fitting. As we learn more
about how to handle evolved solar-like oscillators it should be
possible to ease these strong priors which in the short-term are
probably needed to extract usable frequencies.

Besides the frequencies and frequency splittings, other important data
products given by analysis of the oscillation spectra are mode powers
and linewidths. The long timeseries of \emph{Kepler} and CoRoT are
important for in principle providing unbiased estimates of both
parameters. Initial results have been obtained from peak-bagging
(e.g., \citealt{Baudin2011, Appourchaux2012}) and from other
techniques \citep{Huber2011, Mosser2012a} developed from ground-based
analysis methods \citep[e.g.,][]{Kjeldsen2008a}, including results on
stars in open clusters \citep{Stello2011a} and first theoretical
interpretation of the results \citep{Belkacem2012,Samadi2012}.

 \section{Asteroseismic inference on stellar properties}
 \label{sec:prop}

There is growing effort being devoted to the testing and development
of asteroseismic techniques for estimating fundamental stellar
properties. These techniques are now making it possible to estimate
precise and accurate properties of large numbers of field stars, on
which sparse inferences were only previously available.

The most accurate, and precise, stellar properties have come from
observations of stars in detached eclipsing binaries
\citep{Torres2010}. Other types of data on bright stars, e.g.,
trigonometric parallaxes, or interferometric radii, have also provided
precise and accurate (largely model-independent) properties. As we
shall see later, when asteroseismic observations are also available on
these stars it is possible to test the robustness of the asteroseismic
techniques. Moreover, excellent prior constraints on the stellar
properties allows asteroseismology to make unique tests of stellar
interiors physics, something we return to in later sections. The
excellent accuracy and precision achievable in asteroseismic estimates
of surface gravities is of considerable interest for helping to
calibrate spectroscopic data-reduction pipelines, in particular
automated pipelines for large-scale surveys (of which more later in
Sections~\ref{sec:exo} and~\ref{sec:conc}).

For many field stars we do not have the luxury of very accurate data
on fundamental properties. Stellar properties estimation has then
usually had to rely upon the use of spectroscopic or photometric
observations of basic surface properties, e.g., colours, gravities and
metallicities. Moreover, some uncertainty remains over the solar
photospheric composition \citep{Basu2008, Asplund2009}, which impacts
on the determination of absolute abundances in other stars.

Comparison of these observations with modelled observable surface
properties from stellar atmospheres and stellar evolution theory
yields ``best-fitting'' estimates of the fundamental properties. Such
data may provide little, if any, discrimination between stars with
different fundamental properties that share similar surface
properties. However, the asteroseismic observables -- most notably the
individual frequencies, which can already be measured to a precision
of better than 1 part in $10^4$ with \emph{Kepler} data -- provide the
means to discriminate in such cases. It is important to stress that
complementary, non-seismic inputs are required to optimize the
potential of the seismic data. At the very least one requires the
effective temperature, $T_{\rm eff}$, to obtain tight constraints on
the stellar radius and surface gravity; while in order to fully
constrain the mass and age, strong prior constraints on the
metallicity are essential \citep{Brown1994b, Basu2012}.

Later, we shall discuss the use of individual frequencies in stellar
properties estimation. But first, we shall look at the use of average
or global asteroseismic parameters, including application of the
asteroseismic scaling relations from Section~\ref{sec:theory}.

 \subsection{Use of average seismic parameters, and asteroseismic scaling relations}
 \label{sec:scaling}

When the S/N ratios in the asteroseismic data are insufficient to
allow robust fitting of individual mode frequencies, it is still
possible to extract average or global asteroseismic
parameters. Indeed, as noted previously, the automated analysis codes
developed for application to \emph{Kepler} and CoRoT data have enabled
efficient extraction of these parameters on large numbers of
stars. The main parameters are the average frequency separation
$\left< \Delta\nu_{nl} \right>$, and the frequency of maximum
oscillations power, $\nu_{\rm max}$. It may also be possible to
extract the average small frequency separations (see
Section~\ref{sec:theory}) from moderate-quality data.

Equations~\ref{eq:scalednu} and~\ref{eq:scalenumax} imply that if
estimates of $\left< \Delta\nu_{nl} \right>$ and $\nu_{\rm max}$ are
available, together with an independent estimate of $T_{\rm eff}$,
``direct'' estimation of the stellar radius, mass, mean density and
surface gravity is possible. This so-called direct method is
particularly attractive because it in principle provides estimates
that are independent of stellar evolutionary theory.

The most convenient practical application invokes the assumption that
for all evolutionary phases, from the main sequence to the red-giant
phase, it is safe to scale against precisely measured solar values of
the parameters. Re-arrangement of the scaling relation equations then
gives, for example:
 \begin{equation}
 \left(\frac{R}{\rm R_{\odot}}\right) \simeq 
 \left(\frac{\nu_{\rm max}}{\rm \nu_{max,\odot}}\right)
 \left(\frac{\left<\Delta\nu_{nl}\right>}{\left<\Delta\nu_{nl}\right>_{\odot}}\right)^{-2}
 \left(\frac{T_{\rm eff}}{\rm T_{eff,\odot}}\right)^{0.5},
 \label{eq:rest}
 \end{equation}
 \begin{equation}
 \left(\frac{M}{\rm M_{\odot}}\right) \simeq
 \left(\frac{\nu_{\rm max}}{\rm \nu_{max,\odot}}\right)^{3}
 \left(\frac{\left<\Delta\nu_{nl}\right>}{\left<\Delta\nu_{nl}\right>_{\odot}}\right)^{-4}
 \left(\frac{T_{\rm eff}}{\rm T_{eff,\odot}}\right)^{1.5},
 \label{eq:mest}
 \end{equation}
 \begin{equation}
 \left(\frac{\rho}{\rm \rho_{\odot}}\right) \simeq 
 \left(\frac{\left<\Delta\nu_{nl}\right>}{\left<\Delta\nu_{nl}\right>_{\odot}}\right)^{2}
 \label{eq:dest}
 \end{equation}
and
 \begin{equation}
 \left(\frac{g}{\rm g_{\odot}}\right) \simeq
 \left(\frac{\nu_{\rm max}}{\rm \nu_{max,\odot}}\right)
 \left(\frac{T_{\rm eff}}{\rm T_{eff,\odot}}\right)^{0.5}.
 \label{eq:gest}
 \end{equation}
Notice that the form of Equations~\ref{eq:rest} and~\ref{eq:mest}
imply that estimated masses are inherently more uncertain than radii.

One may also use $\left< \Delta\nu_{nl} \right> $, in addition to
$\nu_{\rm max}$, as input to so-called ``grid-based'' estimation of
the stellar properties \citep[e.g.,][]{Stello2009, Basu2010,
  Quirion2010, Gai2011}. This is essentially the well-used approach of
matching the observations to stellar evolutionary tracks, but with the
powerful diagnostic information contained in the seismic
$\left<\Delta\nu_{nl}\right>$ and $\nu_{\rm max}$ also brought to
bear. Properties of stars are determined by searching among a grid of
stellar evolutionary models to get a ``best match'' to the observed
set of input parameters, which should include $T_{\rm eff}$ and
[Fe/H]. While the direct method assumes that all values of temperature
are possible for a star of a given mass and radius, we know from
stellar evolution theory that only a narrow range of $T_{\rm eff}$ is
allowed for a given $M$ and $R$, assuming a known chemical composition
and given stellar interiors physics.  This prior information is
implicit in the grid-based approach, and means that estimated
uncertainties are typically lower than for the direct method because a
narrower range of outcomes is permitted. The model grids must be well
sampled in the various input parameters.  \citet{Bazot2012} have
recently discussed how use of MCMC techniques may help to mitigate
problems related to grid-sampling.

There is one more level of subtlety to consider in application of the
grid-based technique. It may rely wholly on the scaling relations, so
that the fundamental properties of the models (i.e., $R$, $M$, $T_{\rm
  eff}$) are used as inputs to compute values of
$\left<\Delta\nu_{nl}\right>$ and $\nu_{\rm max}$ for comparison with
the observations. Or one may instead compute theoretical oscillation
frequencies of each model, and from those compute a suitable average
$\left<\Delta\nu_{nl}\right>$ for comparison with the observations.
Another approach to testing the accuracy of the scaling relation for
$\left<\Delta\nu_{nl}\right>$ is therefore to compare the average
$\left<\Delta\nu_{nl}\right>$ implied by the $M$ and $R$ of each model
with the average $\left<\Delta\nu_{nl}\right>$ of the model's computed
oscillation frequencies. Comparisons of this type (e.g.,
\citealt{Ulrich1986}, \citealt{White2011}) suggest that for the
solar-type stars and RGB stars the relation is accurate to 2 to
3\,\%. Similar comparisons between models of red giants in different
evolutionary states show that relative differences in the
$\left<\Delta\nu_{nl}\right>$ scaling of the order few per cent are
expected between low-mass stars on the RGB and in the core-helium
burning RC phase \citep{Miglio2012b}. This is because low-mass stars
with same $M$ and $R$ can have significantly different sound speed
profiles, and hence different acoustic radii, in these two
evolutionary states. \citet{Mosser2013} have studied the impact of
including higher-order terms from asymptotic expressions describing the
frequencies (e.g., Equation~\ref{eq:asspall}), to estimate modified
average large separations for use with the scaling relations.

It is important to note that such comparisons of
$\left<\Delta\nu_{nl}\right>$ do not allow for the impact of poor
modelling of the near-surface layers and the effects of this on the
model-predicted values of the frequencies, and hence the average large
separations. At least in so far as the Sun is concerned, the effect on
$\left<\Delta\nu_{nl}\right>$ is relatively small. However, the issue
is much more of a concern when it comes to modelling of the individual
frequencies. We shall discuss these surface effects in more detail in
Section~\ref{sec:modelling} below.  Because we have much less
confidence in theoretical computations of $\nu_{\rm max}$ -- which
rely on the complicated excitation and damping processes -- than we do
in theoretical predictions of the oscillation frequencies,
model-computed $\nu_{\rm max}$ have so far not been used in stellar
properties estimation. Irrespective, it is clear that much more work
is needed to understand the diagnostic potential of $\nu_{\rm max}$,
and how far it can be pushed in terms of accuracy (in particular for
high-precision data).

When a grid-based approach is used, accuracy is of course demanded of
the stellar evolutionary models.  Those models must include all of the
requisite physics that we consider to be significant in determining
the evolutionary state of the star, and as a result its observable
properties. Extensive tests made using a variety of different stellar
evolutionary codes have shown that density, surface gravity and radius
(and through that luminosity) are largely model independent, and quite
insensitive to the input physics (see \citealt{Lebreton2008};
\citealt{Monteiro2009} and references therein). However, the masses
and ages are much more sensitive to choices made in construction of
the grid of stellar models, in particular the chemical composition,
and can be affected by, for example, the inclusion of the effects of
microscopic diffusion in the stellar evolutionary models. Overshoot at
convective boundaries can also be an issue in this regard. An obvious
way to capture the uncertainties in known modelling ingredients is to
use a variety of model grids, with different input physics, and then
include the resulting grid-to-grid scatter of the estimated properties
in the final, quoted uncertainties. This approach has been adopted in
the analysis of \emph{Kepler} stars (e.g., on the exoplanet host stars
discussed in Section~\ref{sec:exo}). Although, as noted above, the
uncertainties on the direct-method estimates are larger than for the
grid-based method, one might hope that they are expected to largely
capture any uncertainties or systematics in the scaling relations due
to, for example, metallicity effects.

How accurate are the scaling relations? Crucial to making fundamental
tests of the scaling relations is having independent and accurate
estimates of the stellar properties against which to compare the
asteroseismic values, on as wide a range of evolutionary states as
possible. Tests against independent mass estimates are limited to
cases of visual binaries, eclipsing binaries, high S/N exoplanet
transits, and to some extent stars in clusters. To test radii it
suffices to have accurate parallaxes, and interferometrically measured
angular radii.

\citet{Bruntt2010, Bedding2011b} and \citet{Miglio2012} compared
asteroseismic and independently determined properties (e.g., from
binaries) of a selection of bright solar-type stars and red giants,
which all showed solar-like oscillations in observations made from
either ground-based telescopes or CoRoT. The estimated properties were
found to agree at the level of precision of the uncertainties, i.e.,
to 10\,\% or better. The fundamental comparisons have recently been
extended to slightly fainter stars observed by
\emph{Kepler}. \citet{SilvaAguirre2012} used asteroseismic data on 22
of the brightest \emph{Kepler} targets with detected solar-like
oscillations, and found excellent agreement between stellar radii
inferred from the scaling relations and those inferred from using
Hipparcos parallaxes (at the level of a few percent). \citet{Huber2012}
combined interferometric observations of some of the brightest
\emph{Kepler} and CoRoT targets with Hipparcos parallaxes, and also
found excellent agreement with the scaling-relation inferred stellar
radii, at the 5\,\% level.

The analysis of high S/N exoplanet lightcurves provides accurate and
precise stellar densities independent of stellar evolutionary theory,
assuming the orbit is well constrained.  Four such examples, where
solar-like oscillations were also detected in the host star, are
TrES-2, HAT-P-7 and HAT-P-11 \citep{jcd2010a}, and HD~17156
\citep{Gilliland2011} (see Section~\ref{sec:exo}). As noted by
\citet{Southworth2011} the asteroseismic densities are in good
agreement with the lightcurve-derived densities for TrES-2 and
HD~17156, but not for HAT-P-7 and HAT-P-11.  The prospects for
expanding this very small sample look promising, given the healthy
yield of asteroseismic exoplanet host stars detected by \emph{Kepler}.

The detection by \emph{Kepler} of solar-like oscillations in red giant
members of open clusters has provided additional data for testing the
accuracy of the scaling relations, particularly when largely
model-independent constraints are available for the cluster members.
Independent radius estimates of stars in NGC\,6791 based on the
distance determination by \citet{Brogaard2011} were used by
\citet{Miglio2012b} to check the consistency of masses estimated from
the scaling relations (Equations~\ref{eq:rest}
and~\ref{eq:mest}). While no significant systematic effect was found
on the RGB, \citet{Miglio2012b} suggested that a relative correction
to the $\left<\Delta\nu_{nl}\right>$ scaling relation should be
considered between RC and RGB stars (as noted above). If scaling
relations are used, this could affect the mass determination of clump
stars at the $\sim$ 10\,\% level. By combining constraints from
near-turnoff eclipsing binaries and stellar models,
\citet{Brogaard2012} estimated masses of stars on the lower part of
the RGB in NGC\,6791. \citet{Basu2011} used asteroseismic techniques
to estimate an average RGB mass that agreed with \citet{Brogaard2012}
to within $\simeq 7\,\%$.  The difference is however significant given
the quoted uncertainties, and further work is needed to understand the
origin of the discrepancy.  Finally, asteroseismic masses of giants in
NGC\,6819 \citep{Basu2011, Miglio2012b} are in good agreement (at the
$\simeq 10\,\%$ level) with estimates given by isochrone fitting
\citep{Kalirai2004, Hole2009}. It should however be borne in mind that
uncertainties and model-dependences associated with the isochrone
method mean the resulting data do not provide stringest tests of the
asteroseismic masses. Results are however now coming available on
eclipsing binaries in NGC\,6819, which will allow tighter constraints
to be placed on the systematics (as discussed by
\citealt{Sandquist2013}).

Additional and more stringent tests of the asteroseismic scaling
relations will be possible in the near future using results from
\emph{Kepler} and CoRoT on solar-like oscillators that are members of
visual binaries and/or detached eclipsing binaries
\citep[e.g.,][]{Hekker2010}.

 \section{Inference from individual oscillation frequencies}
 \label{sec:nuf}

 \subsection{Estimation of stellar properties}
 \label{sec:modelling}

Use of individual frequencies increases the information content
provided by the seismic data for making inference on the stellar
properties. It is therefore possible to tighten constraints on those
properties, most notably constraints on the age. Asteroseismology not
only provides excellent precision in the age estimates -- with
realistic levels of 10 to 15\,\% achievable (e.g., see discussions in
\citealt{lebreton2009, lebreton2010, Soderblom2010}) -- but may also
be used to improve the accuracy of the age determinations through
tests of stellar models for systems with already well-constrained
properties.

The analysis again proceeds via a grid-based approach through
comparison of the observed frequencies with frequencies calculated for
the grid of stellar evolutionary models, with minimization of the
deviations between the observed and modelled parameters yielding the
best-fitting solutions.

A well-known problem in comparing observed and model-calculated
frequencies comes from the hard-to-model near-surface layers of
stars. Models usually employ simplified model atmospheres, and mixing
length theory is used to describe convection which leads to errors in
the structure of the superadiabatic region. Moreover, model
oscillation frequencies calculated in the usual adiabatic
approximation neglect the effects of turbulent pressure.  In the case
of the Sun, this has all been shown to lead to an offset (sometimes
called the ``surface term'') between observed p-mode frequencies and
the model-predicted p-mode frequencies (e.g., see \citealt{jcd1997},
and references therein), with the model frequencies being on average
too high by a few $\rm \mu Hz$ (giving a negative surface term).  The
offset is larger in modes at higher frequencies (i.e., the closer
$\nu_{nl}$ is to $\nu_{\rm ac}$; the lowest-frequency modes are
evanescent very close to the surface). As noted previously the average
large frequency separation is also affected by the surface term, by an
amount that depends on the gradient of the offset with radial order,
$n$. In the case of the Sun, the model-predicted $\left<
\Delta\nu_{nl} \right>$ is about $1\,\rm \mu Hz$ higher than the
observed $\left< \Delta\nu_{nl} \right>$ (fractionally an overestimate
of about $0.75\,\%$).

We should expect offsets between observations and predictions in other
stars. If the Sun is typical, the offsets for the frequencies will be
many times the sizes of the expected frequency uncertainties. The
achievable precision in the large separations is nowhere near as
good, and so there fractional offsets would in many cases be
comparable to the observed uncertainties. Failure to account for these
offsets will lead to errors in the estimated stellar properties. But
how important might those errors be?

Consider first a simple case, estimation of the mean stellar density
$\left< \rho \right>$, taking the Sun as an example. We assume that
for small fractional changes -- e.g., those describing the differences
between a reasonably good model and observations -- homologous scaling
of the frequencies holds to good approximation. This implies that
$\delta\nu_{nl}/\nu_{nl} \simeq \delta\left< \Delta\nu_{nl}
\right>/\left< \Delta\nu_{nl} \right> \simeq 1/2\,\delta\left< \rho
\right>/\left< \rho \right>$.  The fractional solar surface term
offsets in $\nu_{nl}$ (for frequencies close to $\nu_{\rm max}$) and
$\left< \Delta\nu_{nl} \right>$ quoted above imply inferred densities
would be in error by, respectively, a few tenths of a percent when
using frequencies, and 1 to 2\,\% when using the average large
separation. From Equations~\ref{eq:rest} and~\ref{eq:mest}, one might
naively expect fractional errors of roughly similar size and double
the size in radius and mass, respectively.

Provided the fractional offset in the stellar $\left< \Delta\nu_{nl}
\right>$ is similar to the fractional offset for the Sun, one can
actually suppress the effects of the surface term quite
straightforwardly when the large separation is used (and remove it
when analysing the Sun). When using the scaling relations, one should
scale the observed separation against the observed $\left<
\Delta\nu_{nl} \right>_{\odot}$.  When using stellar-model calculated
frequencies to estimate $\left< \Delta\nu_{nl} \right>$ in the
grid-based method, one should again scale the observed separation
against $\left< \Delta\nu_{nl} \right>_{\odot}$ whilst making sure
that each model-calculated average separation is scaled against a
solar-model average separation made from a properly calibrated solar
model having the same input physics as the model grid. Any residual
bias in the estimated properties will be due to fractional differences
between the solar and stellar surface term, all other things being
equal. Even with a surface term twice as large as solar, errors of
only a few percent would result. 

Given that the typical precision in direct-method and grid-based
estimated properties is at the several percent level, the surface term
effects are most likely not a significant cause for concern, with bias
from other ingredients of the modelling being more of an issue. What
of the individual frequencies? \citet{Kjeldsen2008} have proposed an
empirically motivated procedure to correct individual
frequencies. Since the solar surface term may be described fairly well
by a power law in frequency, they suggest in effect treating the
surface term of other stars as a homologously scaled version of the
solar offset. Application of the Kjeldsen et al. correction to
ground-based data on stars with similar $T_{\rm eff}$ and $\log\,g$
(i.e., similar $\nu_{\rm ac}$ and hence $\nu_{\rm max}$) to the Sun
has indicated similar surface-term properties in those stars, while on
a larger sample of 22 solar-type stars observed by \emph{Kepler} the
average size of the best-fitting surface term was found to increase in
magnitude with increasing $\log\,g$ \citep{Mathur2012}. Moreover,
further inspection of these data shows that the ratio of the absolute
size of the term to $\nu_{\rm max}$ is almost constant. It does
however remain to be seen whether this result instead carries
information related largely to other shortcomings of the stellar
models, not necessarily information associated with the surface term
(i.e., due to degeneracies from the modelling). It must be borne in
mind that different choices made in the construction of stellar models
can impact not only on the shape but possibly also on the sign of the
surface term. One needs only compare the surface term of different
standard solar models to see that for some the surface term can be
positive rather than negative up to frequencies close to $\nu_{\rm
  max\,\odot}$ (albeit at a level smaller than at higher frequencies).

Results from 3D numerical simulations of convection will provide
important input, e.g., as discussed by \citet{Goupil2011}. But what is
also needed are good asteroseismic targets with independently measured
accurate (and precise) properties, which would allow an exploration of
the model degeneracies associated with the surface term. 

Meanwhile, an obvious way to circumvent the problems presented by the
surface term is to instead employ frequency separation ratios in the
modelling, as first pointed out by \citet{Roxburgh2003}. Recall that
these ratios are somewhat independent of the structure of the
near-surface layers.

 \subsection{Use of signatures of modes of mixed character}
 \label{sec:mixed}

As a star evolves and its central regions contract the gravitational
acceleration near the core increases and so do the frequencies of
gravity modes (see Equations~\ref{eq:N}
to~\ref{eq:gspace}). Eventually, typically in the subgiant phase,
g-mode frequencies increase sufficiently to give interactions with
acoustic modes of the same angular degree. The resulting frequency
spectrum is thus not merely a simple superposition of p and g modes
but is determined by the coupling strength of the interacting
modes. When the observed mixed modes are mostly dominated by p-mode as
opposed to g-mode characteristics information on the interactions is
most easily recovered by studying departures from a constant frequency
spacing. Visually, one could inspect an {\'e}chelle diagram (see
Section~\ref{sec:theory}) folded by the average large frequency
separation (e.g., see \citealt{Metcalfe2010} as applied to the
\emph{Kepler} subgiant KIC\,11026764). When g-mode characteristics
dominate, it is better to fold in period, using the average period
spacing \citep{Bedding2011b}.

Mode bumping can affect the frequencies of several p (g) modes in
consecutive orders dependent on the coupling strength and the density
of modes in the spectrum. In favourable cases where several perturbed
p (or g) modes may be detected information on the coupling strength,
and on the underlying unperturbed p-mode and g-mode frequencies, may
be inferred by modelling the interaction between the g-mode and p-mode
cavities \citep{Unno1989, Bedding2011b, Deheuvels2011, jcd2011b,
  Mosser2012a, Mosser2012b}.  This approach, when applicable, provides
an invaluable tool to explore and exploit the diagnostic potential of
using modes of mixed character as probes of internal properties, such
as the behaviour of $N$ in the deep stellar interior, and the
characteristics of the evanescent region between the p- and g-mode
propagation cavities.  The full potential of \emph{Kepler} and CoRoT
data is still to be exploited. However, even from analyses of the
first few months of observations there were cases where the estimated
frequencies were already of sufficient quality to enable detailed
studies of the coupling strength of interacting modes, including the
frequencies of the unperturbed g modes in subgiants
\citep{Deheuvels2011, Benomar2012}.

Provided that the underlying coupling model is accurate, information
on the frequency of the underlying g modes, and on the coupling
itself, in principle allows robust inference to be made on the
fundamental stellar properties \citep{Deheuvels2011}. In subgiants and
low-luminosity giants of a given mass and chemical composition the
frequency of a g mode (or the period spacing of g modes, when
detectable) is a monotonic function of age. By fitting simultaneously
the frequency of the avoided crossing and the average large frequency
separation a precise value of the age and mass can be determined for a
given chemical composition and input physics, as has been shown for
the subgiant HD\,49385 observed by CoRoT \citep{Deheuvels2011,
  Deheuvels2012}. The coupling strength is also a promising proxy of
stellar mass during the subgiant phase, as demonstrated by
\citet{Benomar2012} for a selection of subgiants observed by
\emph{Kepler}, CoRoT and ground-based telescopes.

Following the subgiant phase, the period spacing of the g modes
decreases throughout evolution on the RGB as the stars develop very
compact helium cores. At the same time large frequency separations
decrease in response to significant expansion of the stellar radii,
hence the number of g modes per large separation increases markedly
leading to very dense spectra of modes.

In perhaps the most significant asteroseismic result of recent years,
\emph{Kepler} data were used to show that observed g-mode period
spacings are substantially higher for helium-core-burning stars in the
red clump (RC) than for stars ascending the RGB, providing a clear way
to discriminate stars that otherwise show very similar surface
properties \citep{Bedding2011}. Theoretical work \citep{Montalban2010}
had already indicated that the properties of the $l=1$ modes were
expected to be sensitive to the evolutionary state of giants.
Detection of the dense spectra of $l=1$ modes and measurement of the
period spacings \citep{Bedding2011, Beck2011} revealed a clear
division of the observed spacings and comparison with predictions from
stellar evolutionary models allowed the results to be properly
interpreted.  Similar results on the period spacings using CoRoT data
were subsequently obtained by \citet{Mosser2011}.

The typical period spacing in low-mass RGB stars (at the luminosity of
stars in the RC) is $\simeq 50$ to $70\,\rm s$, while in the RC the
average spacing increases to a few-hundred seconds. The differences
may be understood as follows: The helium core experiences a sudden
expansion following ignition of helium-burning
reactions. Re-adjustment of the stellar structure due to the increased
luminosity in the core results in a significant decrease of $N$ in the
central regions, giving rise to an increased period spacing. In
addition to the decreased core density, the release of energy from the
helium-burning nuclear reactions leads to the onset of convection in
the energy-generating core, contributing to an even larger increase of
the spacing. (See \citealt{jcd2011a} and \citealt{Montalban2013} for
further discussion.)

A firm identification of the evolutionary state of thousands of giants
has wide-ranging implications, from the study of stellar populations
(which we discuss in Section~\ref{sec:pop}), to providing a much
larger sample on which to test the efficiency of extra mixing
mechanisms in the giant phase
\citep[e.g.,][]{Charbonnel2005a}. Moreover, a carefully chosen sample
of RGB and RC stars could be used to quantify the effects of mass loss
on the RGB \citep{Mosser2011}.

Full exploration of the diagnostic potential of the period spacings --
including detailed comparison of observations and model predictions --
promises to shed light on the detailed core properties, including
helping to constrain the efficiency of mixing processes in low- and
intermediate-mass stars. As noticed in \citet{Montalban2013} the
average value of the period spacing in helium-burning
intermediate-mass stars is tightly correlated with the mass of the
helium core. An accurate calibration of the relation between
core-helium mass and stellar mass for different metallicities will
provide stringent constraints on the efficiency of extra mixing that
occurred in the near-core regions during the main sequence phase
\citep[e.g., see][]{Girardi1999}. Moreover, in core-helium burning
stars the period spacing is sensitive to the size of the adiabatically
stratified convective core, and to the temperature stratification in
the near-core regions. A detailed characterisation of such regions
promises to reduce current uncertainties on the required extra mixing,
and its efficiency during the core-helium burning phase (see
\citealt{Salaris2012} for a recent review). Such uncertainties have
implications for the total helium-burning lifetimes, the resulting
carbon-oxygen profile and subsequent evolutionary phases.

Finally, we note the exciting possibility of detecting stars in the
fast evolutionary phases between the occurrence of the first helium
flash and the quiet RC phase \citep{Bildsten2012}, where data on the
period spacings could add significant knowledge to our understanding
of the structural changes that follow helium ignition in low-mass
stars \citep{Weiss2012}.

 \subsection{Use of signatures of abrupt structural variation}
 \label{sec:abrupt}

Regions of stellar interiors where the structure changes abruptly,
such as the boundaries of convective regions, give rise to departures
from the regular frequency separations implied by an asymptotic
description. Careful measurement of these signatures -- sometimes
referred to as glitches -- not only has the potential to provide
additional information on the stellar properties, but also elucidates
other important parameters and physical properties of the stars. There
are signatures left by the ionization of helium in the near-surface
layers of the stars.  Measurements of these signatures should allow
tight constraints to be placed on the helium abundance, something that
would not otherwise be possible in such cool stars (because the
ionization temperatures are too high to yield usable photospheric
lines for spectroscopy). And as noted above, there are also signatures
left by the locations of convective boundaries. It is therefore
possible to pinpoint the lower boundaries of convective envelopes,
potentially important information for dynamo studies of cool
stars. Furthermore, it is also possible to estimate the sizes of
convective cores. Measurement of the sizes of these cores, and the
overshoot of the convective motions into the layers above, is
important because it can provide a more accurate calibration of the
ages of the affected stars.  The mixing implied by the convective
cores, and the possibility of mixing of fresh hydrogen fuel into the
nuclear burning cores -- courtesy of the regions of overshoot --
affects the main-sequence lifetimes.

 \subsubsection{Signatures from stellar envelopes}
 \label{sec:env}

The characteristics of the signatures imposed on the mode frequencies
depend on the properties and locations of the regions of abrupt
structural change. When the regions lie well within the mode cavities
a periodic component is manifest in the frequencies $\nu_{nl}$, which
is proportional to
 \begin{equation}
 \sin\left(4\pi\nu_{nl}\tau + \phi \right),
 \label{eq:glitch1}
 \end{equation}
where
 \begin{equation}
 \tau = \int_r^R \frac{{\rm d}r}{c}
 \label{eq:glitch1}
 \end{equation}
is the acoustic depth of the glitch feature, $r$ is the corresponding
location in radius, $c$ is the sound speed, and $\phi$ is a phase
offset \citep{Vorontsov1988, Gough1990b}. The ``period'' of the
signature induced in the frequencies, which equals $2\tau$, provides
information on the location of the region. The amplitude of the
signature provides a measure of the size of the structural
perturbation, while the decrease in amplitude with increasing
frequency gives information on the radial extent of the glitch. Two
signatures of this type have already been well studied in the solar
case: one due to the sharp variation in the gradient of the sound
speed at the base of the convective envelope; and another due to
changes in the adiabatic exponent in the near-surface helium
ionization zones \citep[see][ and references therein]{jcd2002}.

\begin{figure}

 \centerline {\epsfxsize=14.0cm\epsfbox{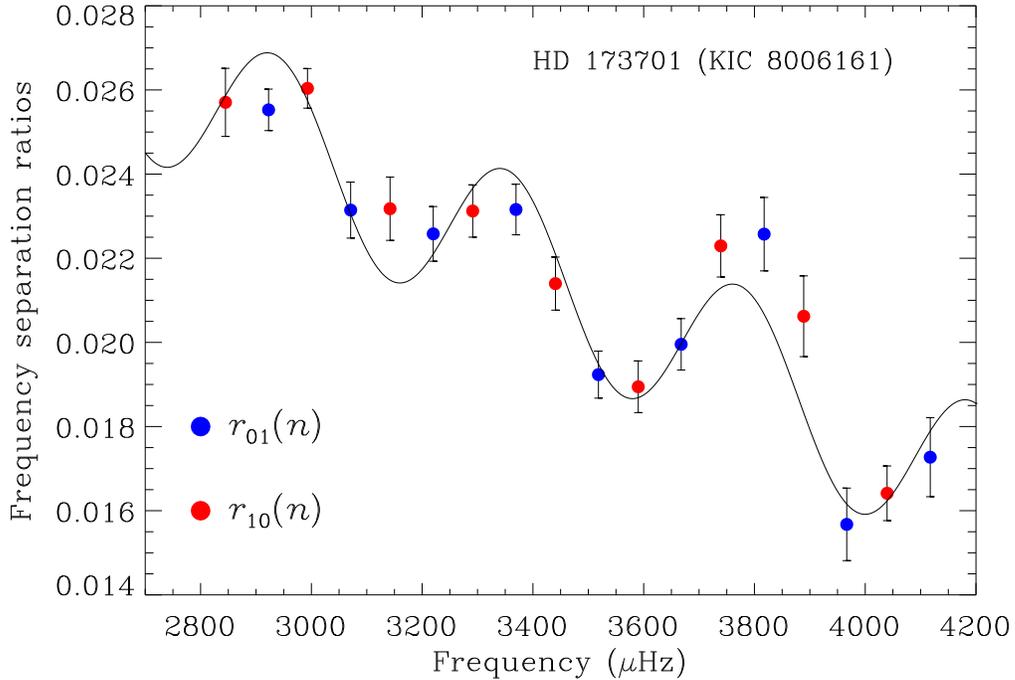}}

 \caption{Observed acoustic-glitch signature arising from the base of
   the convective envelope in the main-sequence \emph{Kepler} target
   HD\,173701 (KIC\,8006161; see also Figure~\ref{fig:stacked1}). The
   blue [red] points with error bars show frequency separation ratios
   $r_{01}(n)$ [$r_{10}(n)$] constructed from the estimated $l=0$ and
   $l=1$ frequencies of the star. The black solid line is a
   best-fitting model (sinusoid plus low-order polynomial) to help
   guide the eye. The best-fitting period of the sine wave implies
   that the base of the envelope lies at an acoustic radius of $t$
   just under $1200\,\rm sec$. Given that the acoustic radius of the
   star is $T_0 \simeq 3350\,\rm s$, this implies an acoustic depth
   $\tau$ of approximately $2150\,\rm s$, i.e., $\tau/T_0 \simeq
   0.65$.}

 \label{fig:glitch} 
\end{figure}

While the periodic signatures from the helium ionization zones may
already be readily apparent in the large frequency separations
$\Delta\nu_{nl}$, their signals may be better isolated by, for
example, taking second differences of frequencies of modes having the
same angular degree $l$, i.e., $\Delta_2\nu_{nl} = \nu_{n-1\,l} -
2\nu_{nl}+\nu_{n+1\,l}$, or by subtracting the frequencies from a
smoothly varying function in the overtone number, $n$. The signature
from the base of the convective envelope is also apparent in the
second differences, although at a reduced amplitude compared to the
helium signatures. Potentially better diagnostics from which to
extract the convective-envelope signature are the frequency separation
ratios constructed by using the $l=0$ and $l=1$ modes
\citep{Roxburgh2009}, i.e., $r_{01}(n) =
\delta\nu_{01}(n)/\Delta\nu_1(n)$ and $r_{10}(n) =
\delta\nu_{10}(n)/\Delta\nu_0(n+1)$.  Since the small and large
separations are affected in a similar way by the near-surface layers,
signatures from the near-surface ionization zones are largely
suppressed in the ratios, leaving the signature from the base of the
convective envelope. Use of the separation ratios gives a periodic
signal in the frequencies equal to twice the acoustic radius, $2t$,
not twice the acoustic depth, $2\tau$.  Acoustic radii, $t$, are
defined by:
 \begin{equation}
 t = T_0 - \tau,
 \label{eq:glitchr}
 \end{equation}
where the acoustic radius of the star, $T_0$, is given by
 \begin{equation}
 T_0 = \int_0^R \frac{{\rm d}r}{c} = 1/(2\Delta\nu)
                             \simeq 1/(2\left< \Delta\nu_{nl} \right>).
 \label{eq:taustar}
 \end{equation}
Figure~\ref{fig:glitch} shows the convection-zone-base glitch
signature of the main-sequence \emph{Kepler} target HD\,173701
(KIC\,8006161; see \citealt{Appourchaux2012, Mazumdar2012b}), as
observed in the frequency separation ratios made from the star's
estimated oscillation frequencies.

Since both $\tau$ and $T_0$ contain contributions from the
hard-to-model near-surface layers, translation to the radii $t$ when
using frequencies or second differences as observables will suppress
the surface contribution, in principle making comparisons with model
predictions more straightforward.  An important caveat worth adding is
that use of the different frequency diagnostics (or combinations)
outlined above will give small differences in the results, since the
exact properties of the glitches depend on the chosen diagnostic. For
example, explicit in the construction of the diagnostic used by
\citet{Houdek2007} is an assumed matching of the stellar interior to a
model atmosphere, where the upper reflecting layer for many modes
lies. Other methods may not take explicit account of the exact
location of the reflecting surface.  That said, these differences are
typically rather small, and do not compromise the precision of
inferences made at the few percent level.

The acoustic depth of the base of the solar convective envelope lies
at $\tau_{\rm BCZ} \approx 2300\,\rm s$, and produces a glitch signal
with an amplitude of approximately $0.1\,\rm \mu Hz$ in the
frequencies.  The solar He~II ionization zone produces a signal at
$\tau_{\rm HeII} \approx 700\,\rm s$, and a glitch-signal amplitude of
approximately $1\,\rm \mu Hz$.

The key to extracting the envelope glitch signatures is to have
sufficient precision in estimates of the frequencies, and data on a
sufficient number of overtones in the spectrum. Ideally, one typically
requires individual frequency uncertainties of the order of a few
tenths of a $\rm \mu Hz$ or lower, which demands multi-month
observations of the stars.  The resolution achievable in period will
correspond to $\simeq 1/\left( \Delta n \times \left< \Delta\nu_{nl}
\right> \right)$, i.e., the total range in frequency spanned by the
useable frequencies, where $\Delta n$ is the number of overtones
covered. When the S/N in the glitch signatures is good, the resolution
largely fixes the uncertainty in the estimated acoustic depth.

The range of depths accessible to the analysis is determined by the
large frequency separation of the star.  One may cast a discussion on
the limits in terms of the Nyquist-Shannon sampling theorem. When
individual frequencies are available on $l=0$ and $l=1$ modes the
diagnostics data are sampled in frequency at an approximately regular
interval of $\simeq \left< \Delta\nu_{nl} \right>/2$, implying a
Nyquist ``period'' of $1/\left< \Delta\nu_{nl} \right>$. This sets a
notional upper limit on measureable ``periods'' of $\simeq 2T_0 \simeq
1/\left< \Delta\nu_{nl} \right>$, and hence an upper limit on
measureable acoustic depths of $\simeq T_0 \simeq 1/(2\left<
\Delta\nu_{nl} \right>)$. In subgiants and red giants that show mixed
modes one may not have the luxury of being able to use non-radial
modes to construct glitch diagnostics since they will show the effects
of mode coupling. The sampling in frequency is then only $\simeq
\left< \Delta\nu_{nl} \right>$, which halves the upper-limit
measureable acoustic depth and it is then not possible to distinguish
between a structural glitch located at an acoustic depth $\tau$, and
one located at an acoustic depth $T_{0} - \tau$. We return later in
the section to consider the case of evolved stars.

In solar-type stars the base of the convective envelope is located at
an acoustic depth of roughly $ \approx T_0/2$ or deeper, and so $l=1$
modes are needed to avoid potential aliasing problems. Of all the
solar-type stars, early F-type stars have the shallowest convective
envelopes. Since the ratio of the envelope and He~II zone depths is
much smaller than in Sun-like analogues, it can be hard to disentangle
the signatures (most notably in stars with $M \ge 1.4\,\rm
M_{\odot}$). Even though the separation of the signals is in contrast
most pronounced in later-type, lower-mass dwarfs, the amplitude of the
convective envelope signature is then very weak and hence more
challenging to extract.

The first measurements of the envelope glitch signatures of a
solar-type star other than the Sun were made by
\citet{Mazumdar2012}. They extracted the envelope and He~II signatures
of the F-type star HD49933 using estimated frequencies obtained from
180\,days of data collected by CoRoT. The analysis proved challenging
for two reasons: first, due to the close correspondance of the glitch
periods (see above); and second, due to the large frequency
uncertainties from the heavy damping of the modes (a notable
characteristic of F-type stars). This resulted in large uncertainties
on the estimated envelope depth.

Thanks to \emph{Kepler}, more precise frequencies (from longer
datasets) are now available on a large ensemble of solar-type
stars. \citet{Mazumdar2012b} analyzed a representative sample of 19 of
these stars, and demonstrated that it is already possible to extract
very precise estimates of the glitch signature properties. Four
different techniques of analysis were applied to extract the glitch
signatures, using different combinations of the observed frequencies,
as based on the methods outlined in \citet{Houdek2007, Monteiro2000,
  Mazumdar2012} and \citet{Roxburgh2009}. Very good agreement was
found in the inferred glitch properties, giving confidence in the
extracted values. For example, acoustic convective envelope depths
were estimated to a typical precision of a few percent.

The results from \citet{Mazumdar2012b} have validated the robustness of
the analysis techniques, and indicate that we are now in a position to
apply them to, and exploit the results from, more than 100 solar-type
stars with multi-month \emph{Kepler} data. The \emph{Kepler} ensemble
will provide a comprehensive set of convective envelope depths for
testing stellar evolution theory, and for validating stellar dynamo
models. Results of those tests should also open the possibility to use
the estimated envelope and He~II depths as input to help constrain
both the gross stellar properties, and interiors structure
\citep{Mazumdar2005}. It will also be possible to use the extracted
envelope glitch signatures to place constraints on overshoot into the
radiative interiors (e.g., \citealt{Monteiro2000}).  Moreover, as
pointed out by \citet{Houdek2007} and \citet{Houdek2011}, careful
measurement and subsequent removal of glitch signatures from the mode
frequencies will in principle provide cleaner inference on the stellar
properties, notably age, when those frequencies are used to model
stars.

Extraction of the He~II signatures allows an estimate to be made of
the helium abundances in the stellar envelopes, as discussed by, for
example, \citet{Basu2004} and \citet{Monteiro2005}. If we assume that
the stellar mass and radius have already been determined to a
fractional precision of $\approx 10$\,\% and 5\,\%, respectively --
for example from the standard asteroseismic methods outlined in
previous sections -- it should be possible to use the measured
amplitude of the glitch signal to constrain the envelope helium
abundance to better than 10\,\%.

Evolved stars present different challenges for the analysis of glitch
signatures. First, as noted above, significant coupling may render
mixed modes poor diagnostics of the sought-for signatures. Exceptions
will be when the coupling with g modes is weak, e.g., in $l=1$ modes
in luminous RGB stars, or in $l=2$ modes. Second, the number of
acoustic modes trapped in the stellar interior decreases as the
stellar radius increases \citep[e.g., see][]{jcd2011a}. This is
because the frequency domain covered by the acoustic modes scales
approximately as $\nu_{\rm max}$, and since the modes are spaced
according to $\Delta\nu$ the number of overtones observed will scale
as $\left( \nu_{\rm max}/\Delta\nu \right) \propto
M^{1/2}R^{-1/2}T_{\rm eff}^{-1/2}$.  As stars evolve up the RGB, we
might therefore expect to observe fewer overtones than in
main-sequence stars, which makes extraction of periodic signatures in
frequency more challenging. It is worth adding that the S/N in those
modes that are detected will be higher (and peak linewidths narrower)
than in main-sequence stars on account of the higher mode amplitudes
(and lower damping rates) shown by evolved stars. Third, as a star
evolves off the main sequence and the convective envelope deepens, the
base of the envelope is displaced to greater acoustic depths
(typically to $\ge 0.9T_0$, depending on the mass, chemical
composition and evolutionary state).  In this respect the detection of
seismic signatures of helium ionisation is simpler for giants than for
solar-type stars, since the He signatures do not have to be
disentangled from the convective envelope signatures.

Glitch signatures have already been detected in red giant data
collected by CoRoT, the first published example being HR\,7349
\citep{Miglio2010}. Given the quality of the \emph{Kepler} data, we
may expect signatures to be detected and characterised precisely in a
large number of giants, opening the possibility to constrain the
envelope helium abundance in old stars (although further work is
needed to understand how parameters extracted on the signatures impact
on the accuracy and precision of estimated helium abundances).

 \subsubsection{Signatures from stellar cores}
 \label{sec:cores}

When regions of abrupt structural change do not lie well within the
mode cavities the signatures they leave in the mode frequencies are
more subtle. This is the case for the signatures left by convective
cores found in solar-type stars slightly more massive than the Sun.

Seismic diagnostics of convective cores are particularly important
\citep{Noels2010}. Several physical processes -- e.g., rotation,
overshooting, semiconvection and diffusion -- may alter the shape of
the chemical composition profile near the border of a convective core,
mainly through full or partial mixing of the radiatively stable layers
beyond the formal boundary of the convective region, as set by the
Schwarzschild criterion. Constraining the efficiency of such extra
mixing has wide relevance, most obviously since it would decrease
systematic uncertainties on the age determinations of the affected
main-sequence stars. There are significant uncertainties in the
modelling of transport processes in stellar interiors (e.g., due to
the lack of a satisfactory theory for convection); seismic constraints
on near-core mixing may be used to test such models, and to improve,
eventually, our understanding of stellar physics.

Comparisons of theoretical models and ``classical'' non-seismic
observations of stars show clearly that standard stellar models
underestimate the size of the centrally mixed region, for example from
stringent observational constraints provided by detached eclipsing
binaries and open clusters \citep[e.g., see][] {Andersen1990,
  Ribas2000}. While the need for extra mixing is generally accepted,
the calibration of its efficiency for stars of different mass and
chemical composition is very uncertain, in particular in the mass
range $1.1$ to $1.6\,\rm M_\odot$ \citep[e.g.,
  see][]{Demarque2004,VandenBerg2004,VandenBerg2006}. Moreover, there
is no clear consensus regarding the physical processes responsible for
the required extra-mixing that is missing in the standard
models. Possibilities include overshooting in its classical and
diffusive flavors \citep[e.g.][]{Maeder1975, Ventura1998}, microscopic
diffusion (Michaud et al. 2004), rotationally induced mixing
\citep[e.g., see][and references therein]{Maeder2000, Mathis2004}, and
mixing generated by propagation of internal waves
(e.g. \citealt{Young2003, Talon2005}).

A necessary condition for seismology to act as a diagnostic is that
the mixing process leaves a distinct signature in the chemical
composition profile, hence in the sound speed or $N$, which acoustic
and gravity modes are sensitive to. Several studies have proposed and
discussed the theoretical potential of diagnostics for solar-type
stars based upon use of the small frequency separations, frequency
separation ratios (see Section~\ref{sec:theory}), and other more
elaborate combinations of individual frequencies \citep{Popielski2005,
  Mazumdar2006, Cunha2007a, Cunha2011, SilvaAguirre2011}. The most
recent work has focussed in particular on developing diagostics based
on use of $l=0$ and $l=1$ modes. These modes have significantly higher
visibilities in the \emph{Kepler} and CoRoT data than do modes of
higher $l$. Diagnostics are now beginning to be applied to the
space-based data, which have the requisite frequency precision not
achieved from ground-based observations.  A possible exception
involving ground-based data is the well-studied binary $\alpha$\,Cen
where tight complementary constraints (including masses from the
analysis of the visual binary orbit) are available.
\citet{Demeulenaer2010} showed that the small frequency separations
$\delta\nu_{01}(n)$ extracted from ground-based data on
$\alpha$\,Cen~A were at odds with models computed using a substantial
amount of extra mixing in the core.

In solar-type stars, an asymptotic formalism of the frequencies based
on a description of the internal phase shifts in principle provides
the foundations for carrying out inverse analyses to infer the
structures of stellar cores, using only low-$l$ modes
\citep{Roxburgh2010}.  The observation of mixed modes would help to
dramatically improve the quality of the inversions, since they provide
kernels localised in the core of the star \citep{Basu2002}.

Constraints on the detailed properties of mixing in the near-core
regions may also be inferred from the analysis of data on more evolved
stars showing mixed modes. Mixed modes observed in subgiant stars
result from coupling of high-order p modes and low-order g modes. The
frequencies of the latter may present significant deviations from the
asymptotic limit and therefore bear information on abrupt structural
variations related to the chemical composition gradient in the
near-core regions (which controls $N$). Such a gradient is determined
primarily by the evolutionary state, but can also be modified by
different mixing processes taking place in the radiative interior. If
the profile of $N$ changes because of a different composition
gradient, then we can expect a signature of the different mixing
processes to be present in the frequencies of the mixed modes.

Seismic signatures of a smooth composition profile arising from
non-instantaneous mixing were investigated by \citet{Miglio2007},
while \citet{Deheuvels2011} have also highlighted the potential
effects of diffusion on the detailed behaviour of $N$ near the
core. These studies suggest that such subtle effects may be detectable
in stars that are evolved enough to present avoided crossings, yet
sufficiently close to the end of the main sequence that the mean
molecular weight profile has not yet been significantly modified by
nuclear reactions in a surrounding shell.

As discussed in Section~\ref{sec:mixed}, comparison of observed and
model-predicted frequencies of giants should provide information on
mixing processes occuring during the core-helium burning phase, which
can have a significant impact on later evolutionary phases
\citep[e.g., see][]{Straniero2003}.

Finally in this section we note that rotation may also act to induce
deep-seated mixing in stars with radiative
cores. \citet{Eggenberger2010} studied the effects of this by
comparing rotating and non-rotating models having the same fundamental
properties.  They noticed that rotational mixing increased the average
small separations and frequency separation ratios, and lead to a
slightly steeper slope of the small separations with frequency,
reflecting the impact of rotation on the chemical composition
gradients and in particular changes in the abundance of hydrogen in
the central parts of the star.

 \subsection{Inferences on internal rotation}
 \label{sec:rot}

Despite the importance of effects of rotation in the formation and
evolution of stars \citep[e.g., see][]{Pinsonneault1997, Maeder2000}
little is known about the efficiency of, and interplay between, the
physical processes regulating the transport of angular momentum in
stellar interiors.  Observational constraints on the impact of
rotation during stellar evolution are generally limited to surface
abundances signatures of deep mixing, and to measurements of the
surface rotation of stars in different evolutionary states.

A notable exception is of course the Sun, where helioseismic analysis
of the frequency splittings of non-radial p modes has provided a
detailed picture of rotation in the solar interior \citep[e.g.,
  see][]{Thompson2003, Howe2009}. Helioseismology revealed the solar
tachocline, a narrow region in the stably stratified layer just
beneath the base of the convective envelope which mediates the
transition from differential rotation in the envelope to a
solid-body-like profile in the radiative interior and is believed to
play an integral r\^ole in dynamo action
\citep{Ossen2003}. Helioseismic inversions showed that the
solid-body-like profile persists down to at least $R \sim 0.2\,\rm
R_{\odot}$ (deeper down inferences are rather uncertain), results that
present severe challenges to models of angular momentum transfer in
the solar interior \citep[e.g., see][]{Pinsonneault1989}.

Long, continuous datasets are needed to extract accurate and precise
estimates of the rotational frequency splittings of low-$l$ p modes in
main-sequence stars. The frequency splittings, $\delta\nu_{nlm}$, will
typically vary from a few $\rm \mu Hz$ (in more massive, and/or very
young solar-type stars) down to a fraction of a $\rm \mu Hz$ (in less
massive, and/or more mature solar-type stars; see inset to
Figure~\ref{fig:spec}). The splittings may often be of comparable size
to the linewidths of the damped modes in the frequency-power spectrum,
which will in some cases prevent robust extraction of the splittings
and in others means that care is needed to deal with potential
biases. Extensive testing with Sun-as-a-star and artificial data mean
that these problems are well understood \citep{Chaplin2006}.

The robust detection of variations of p-mode splittings with
frequency, angular degree and azimuthal order will be challenging, but
promises to provide information on near-surface magnetic fields (which
can contribute to the splittings) as well as internal differential
rotation (much harder). A first obvious exercise using the
\emph{Kepler} ensemble will be to compare the average frequency
splittings with measures of surface rotation periods.

While the frequency splittings of p modes observed in main-sequence
stars are largely determined by the rotation profile in the stellar
envelope (as in the case of the Sun), the situation is different in
the case of evolved stars where modes of mixed character are sensitive
to the rotation in deeper-lying layers. As noted in
Section~\ref{sec:obs}, the fact that these modes are also less heavily
damped than pure p modes means they present very narrow peaks in the
frequency-power spectrum. This makes extraction of the observed
frequency splittings in principle more straightforward than in
main-sequence stars. Long datasets are nevertheless still required,
and during some evolutionary epochs (e.g., giants in the RC) the
complicated appearance of the oscillation spectra can present
additional challenges for disentangling the splittings.

Recently, \citet{Beck2012} measured rotational splittings of mixed
modes in three low-luminosity red-giant stars observed with
\emph{Kepler}. Mixed modes were identified in each power spectrum as
dense clusters of $l=1$ modes. The mode at the centre of each cluster
is dominated by p-mode characteristics, and is hence most sensitive to
the external layers. Adjacent modes are more g-mode-like in character,
and so their rotational kernels (Equation~\ref{eq:rot}) are localised
predominantly in the central regions of the star. Through a first
comparison with rotational kernels representative of the modes
observed in one of the giants (KIC~8366239), the authors found
evidence for a core rotating at least ten-times faster than the
surface.

\citet{Eggenberger2012} compared the observed splittings with those
predicted by models including meridional circulation and shear
instability. They concluded that these processes alone produce an
insufficient coupling to account for the rotational splittings
observed in KIC 8366239, and that an additional mechanism for the
transport of angular momentum must operate in stellar interiors during
post-main sequence evolution. Moreover, by comparing the ratio of
splittings for p- and g-dominated modes, they estimated the efficiency
of this additional physical process.

\citet{Deheuvels2012} also recently reported the detection of
rotationally split modes in HIP~92775 (KIC~7341231), a halo star at
the base of the RGB. By applying various inversion techniques and
assumptions on the functional form of the rotation profile, they could
set quantitative constraints on the internal rotation profile of the
star. This led to a robust inference on the rotation of the core, and
an upper limit to the surface rotation, establishing that in HIP~92775
the core rotates at least five-times faster than the surface.  As
pointed out by the authors, while the rotation period of the core
could be robustly derived, weaker constraints were obtained on the
surface rotation, due to the fact that rotational splittings of
dipolar modes detected in these low-luminosity giants are
significantly contaminated by the core rotation. In the future the
detection of rotationally split $l=2$ modes, and splittings in stars
with different trapping properties, should provide information on
rotation profiles in the stellar envelopes (including potentially
differential rotation).

\citet{Mosser2012c} estimated average rotational frequency splittings
for around 300 giants, providing direct constraints on the evolution
of the mean core rotation for stars in the red-giant phase. Comparison
of the measured core rotation rates for RGB and RC stars provides
evidence for core spin-down in the final phases of evolution on the
RGB.

Detailed comparisons with theoretical predictions of internal
rotational profiles at different evolutionary states are needed to
quantify the effects of additional processes transporting angular
momentum in radiative \citep{Charbonnel2005, Gough1998, Mathis2009}
and convective regions \citep{Palacios2012}.  Moreover, in order to
achieve a fully consistent physical picture of transport of chemicals
and angular momentum in stars of different masses, chemical
compositions, and evolutionary stages, models will have to reproduce
not only seismic constraints on the internal rotational profile but
also chemical signatures of deep mixing obtained via spectroscopic
methods \citep[e.g., see][ and references therein]{Smiljanic2009}.

 \section{Asteroseismology, exoplanets, and stellar activity studies}
 \label{sec:exo}

Asteroseismology can be particularly powerful when it is applied to
stars that are exoplanet hosts. It can provide the accurate and
precise estimates of the stellar properties (i.e., density, surface
gravity, mass, radius and age) that are needed to make robust
inference on the properties of the planets, and information on the
internal rotation and stellar angle of inclination to help better
understand the evolutionary dynamics of the systems. Moreover,
asteroseismology can be used to probe levels of near-surface magnetic
activity, interior-atmosphere linkages, and stellar activity cycles,
all relevant to understanding the influence that stars have on their
local environments, where planets are found.

The first asteroseismic studies of exoplanet hosts showing solar-like
oscillations used ground-based Doppler velocity observations of
$\mu$\,Arae \citep{Bouchy2005, Bazot2005, Soriano2010} and
$\iota$\,Hor \citep{Vauclair2008}. The Hubble Space Telescope fine
guidance sensor provided several days of asteroseismic data on the
solar-type host HD~17156 \citep{Gilliland2011}. Studies of four other
known exoplanet hosts were made possible by early prioritisation of
targets observed by CoRoT (HD52265; see \citealt{Ballot2011,
  Escobar2012}) and by \emph{Kepler} (HAT-P-7, HAT-P-11 and TrES-2;
see \citealt{jcd2010a}).

\emph{Kepler} has further opened the possibilities to combine
exoplanet studies and asteroseismology, with (at the time of writing)
solar-like oscillations detected in around 80 stars which also have
single or multiple candidate, validated or confirmed transiting
exoplanets \citep{Huber2013}. These asteroseismic \emph{Kepler}
Objects of Interest (KOIs) span the \emph{Kepler} apparent magnitude
range $7.4 \le K_p \le 13.5$, the peak of the sample lying around $K_p
\simeq 12$. This 80-strong sample of asteroseismic KOIs provides a
unique ensemble with extremely well-constrained stellar properties.

\emph{Kepler}'s transit observations provide a direct estimate of
$R_{\rm p}/R$, i.e., the ratio of the radii of the planet and star,
hence accurate and precise radii from asteroseismology allow tight
constraints to be placed on the absolute sizes of the planets. The
stellar radius is also required to fix the stellar luminosity and
hence the location of the habitable zone around the star. There are
already several examples in the literature where asteroseismically
estimated stellar radii put tight constraints on the radii of small
planets, e.g., Kepler-10b, \emph{Kepler}'s first rocky exoplanet
\citep{Batalha2011}, and Kepler-21b \citep{Howell2012}. Particularly
noteworthy was Kepler-22b \citep{Borucki2012}, \emph{Kepler}'s first
validated planet lying in the habitable zone of its host star. This
Sun-like analogue is quite faint for asteroseismology, and provided an
excellent example of how a detection of just the large frequency
separation was sufficient to get a good estimate of the stellar radius
(the modes were too weak to estimate individual frequencies).

In systems that are bright enough for follow-up radial velocity
observations those velocity data may be combined with the transit data
to estimate planetary masses, $M_{\rm p}$. An accurate estimate of the
stellar mass, $M$, is required (which asteroseismology can again
provide) with the inferred planetary mass scaling with the stellar
mass according to $M_{\rm p} \propto M^{2/3}$. A recent example is the
Kepler-68 system \citep{Gilliland2013}.

When usable radial velocity data cannot be obtained (due to a
combination of the faintness of the star, and small sizes of the
planets) \emph{Kepler} has shown how detected transit timing
variations (TTVs) in multi-planet systems may be used to help
constrain the planetary properties. These TTVs correspond to
deviations from strictly periodic transit intervals, and arise due to
the mutual gravitational interactions of the planets. So-called
photodynamical models combine information from the observed transits
-- including data on the TTVs -- with the estimated properties of the
host star.  Here, the inferred planetary masses depend on the stellar
mass, via $M_{\rm p} \propto M$.  \citet{Carter2012} presented the
first example of combining TTVs and asteroseismology, the Kepler-36
system containing two planets. The small uncertainties on the stellar
mass and radius proved crucial to constraining the planet masses and
radii to better than 8\,\% and 3\,\% respectively, helping to confirm
the unusual nature of the system (planets having markedly different
densities lying in closely spaced orbits).

Stellar ages estimated from asteroseismology of course provide
upper-limit age estimates for the planets. This information is of
particular interest for cases where planets are potentially habitable,
and for those developing planet evolutionary models, but is also
relevant to calculations of the long-term stability of discovered
systems, e.g., for confirming that inferred configurations are stable
on the required timescales (as in the case of Kepler-36) or for ruling
out potential configurations that may be unstable in systems where
there are degeneracies in the inferred best-fitting solutions.

Asteroseismology may also be used to improve the accuracy of
spectroscopic estimates of temperature and metallicity, which are
needed for analysis of the host stars. Spectroscopic $\log\,g$ can be
very uncertain and inaccuracies can lead to bias in estimates of
$T_{\rm eff}$ and [Fe/H] due to strong correlations in the
analysis. In the case of the asteroseismic KOIs
\citep[e.g.,][]{Borucki2012, Carter2012, Huber2013}, estimates of
$\log\,g$ from asteroseismology (which used initial estimates of
$T_{\rm eff}$ and [Fe/H] from spectroscopy as inputs) were used as a
strong prior on $\log\,g$ in re-analysis of the spectroscopic
data. Only one iteration was needed to reach convergence in the
asteroseismically estimated stellar properties. A similar approach has
been adopted using stellar densities measured from transit lightcurves
\citep{Torres2012}.

One of the most exciting areas where asteroseismology may be brought
to bear is to provide information on the spin-orbit alignments of
exoplanet systems. Asteroseismology may be applied to estimate the
angle $i$ between the stellar rotation axis and the
line-of-sight. When applied to transiting systems, where the orbital
plane of the planets must be close to edge-on, an estimated $i$
significantly different from 90\,degrees -- such that the rotation
axis is well inclined relative to the plane of the sky -- will imply a
spin-orbit misalignment.

Asteroseismic estimation of $i$ rests on our ability to resolve and
extract signatures of rotation in the non-radial modes of the
oscillation spectrum. The mode patterns of the non-radial modes are
non spherically symmetric, and hence the observed amplitudes of the
modes depend on the viewing angle. For the slow to moderate rates of
rotation expected in solar-like oscillators, measurement of the
observed relative power of the different azimuthal components in each
non-radial mode provides a direct estimate of $i$ (for a detailed
discussion of the method, see \citealt{Gizon2003, Ballot2006,
  Ballot2008, Chaplin2013}). In the best cases, asteroseismology can
constrain $i$ to an uncertainty of just a few degrees.  The analysis
requires bright targets and long-duration timeseries to give the
requisite signal-to-noise and frequency resolution for extracting
clear signatures of rotation in the oscillation spectrum.

When peak-bagging models are fitted to the modes to extract the
required information, the maximized likelihoods exhibit strong
correlations between the rotational frequency splittings
$\delta\nu_{\rm s}$ and $\sin\,i$. This means that when the S/N in the
modes or the length of the dataset is insufficient to extract a unique
solution for the splitting and angle it is still often possible to
estimate the product $\delta\nu_{\rm s}\sin\,i$. \emph{Kepler} data
may also provide estimates of the surface rotation period, $P_{\rm
  rot}$, from rotational modulation of starspots and active
regions. Hence, even when a unique asteroseismic solution for $i$ is
unavailable, the observed seismic product $\delta \nu_{\rm s}\sin\,i$
can be combined with the observed $P_{\rm rot}$ to estimate the sine
of the angle, i.e., from $\sin\,i = (\delta \nu_{\rm s}\sin\,i) P_{\rm
  rot}$. This is similar to the method that combines $P_{\rm rot}$
with $v\sin\,i$ from spectroscopy (e.g., see \citealt{Hirano2012}
for examples using \emph{Kepler} data). Implicit is the assumption
that the internal rotation rates probed by the seismic splittings are
similar to the surface rotation, which appears to be reasonably valid
for moderately slowly rotating main-sequence stars, where the
splittings are most sensitive to the rotation in the envelope.

The asteroseismic technique provides a useful complement to other
existing methods. Observations of the Rossiter-McLaughlin (RM) effect
and use of spot-crossing events during exoplanet transits have
provided estimates of the sky-projected angle between the planetary
orbital axis and the stellar rotation axis in more than 50 systems
\citep{Albrecht2012}. The RM observations need bright targets and are
extremely challenging for small planets (e.g. super-Earth-sized
objects). Thanks to \emph{Kepler} the spot-crossing method can be
extended to much fainter targets (e.g., see
\citealt{Sanchis-Ojeda2012}). However, it also works best on systems
with large planets, and on active stars with large spots (to give the
required SNR in the spot-crossing events).

In contrast, the asteroseismic method gives information independent of
the planet properties, and is therefore useful for investigating
alignments in systems with very small planets. However, the
limitations of having information on $i$ only must be borne in mind:
it is possible for the difference between the orbital and spin angles
to be small even when the system is misaligned (e.g., see
\citealt{Hirano2012}). Dependent on the system properties, statistical
arguments can however be employed to help discuss the likely
correlation of the orbital and spin axes; and the full
three-dimensional configuration may of course be recovered when the
sky-projected angle and the stellar inclination angle are both
available.

Theories which propose that strong misalignments are the result of
dynamical interactions (e.g., see \citealt{Winn2010}) are supported by
the well-ordered alignments found in the multi-planet system Kepler-30
studied by \citet{Sanchis-Ojeda2012} using spot-crossing data; and two
multiple-systems with solar-type host stars investigated by
\citet{Chaplin2013} with asteroseismology. Application of seismology
to other \emph{Kepler} systems with RGB hosts, or subgiants showing
many mixed modes, will be particularly attractive since the narrow
mode peaks are particularly conducive to accurate measurement of the
splittings and inclination.

We finish with some remarks on stellar activity and activity cycles.
The asteroseismic \emph{Kepler} ensemble offers the prospect of being
able to select with high precision and accuracy ``evolutionary
sequences'', i.e., stars of very similar mass and chemical composition
spanning different evolutionary epochs \citep{SilvaAguirre2011a}, from
the main sequence potentially all the way through to the AGB, which
may be combined with ground-based observations of stellar activity
(e.g. Ca H+K) to give a powerful diagnostic of how activity evolves in
solar-type stars, and to better understand the influence that stars
have on their local environments (with the obvious implications for
exoplanet habitability).

The availability of long timeseries data is now making it possible to
``sound'' stellar cycles with asteroseismology. The prospects for such
studies have been considered in some depth (e.g., see
\citealt{Karoff2009} and references therein). The first convincing
results on stellar-cycle variations of the p-mode frequencies of a
solar-type star (the $F$-type star HD\,49933) were reported by
\citet{Garcia2010}, from observations made by CoRoT. This result is
important for two reasons: first, the obvious one of demonstrating the
feasibility of such studies; and second, the period of the stellar
cycle was evidently significantly shorter than the 11-yr period of the
Sun (probably between 1 and 2\,yr). 

The results on HD\,49933 are interesting when set against the paradigm
(e.g., see \citealt{Bohm-Vitense2007}) that stars showing cycle
periods divide activity-wise into two groups, with stars in each group
displaying a similar number of rotation periods per cycle period. The
implication is that stars with short rotation periods -- HD\,49933 has
a surface rotation period of about 3\,days -- tend to have short cycle
periods. If other similar stars show similar short-length cycles,
there is the prospect of being able to track asteroseismically two or
more complete cycles of such stars with the extended \emph{Kepler}
mission.  We note that \citet{Metcalfe2010a} recently used
chromospheric Ca H \& K data to detect a short (1.6\,yr) cycle period
in another F-type star, the aforementioned exoplanet host
$\iota$\,Hor.

\emph{Kepler} will also make it possible to detect full swings in
activity in stars with cycles having periods up to approximately the
length of the solar cycle.  Detailed analysis of the \emph{Kepler}
asteroseismic ensemble is now underway to search for stellar-cycle
signatures. Finally, we should not forget the prospects for detecting
seismic signatures of stellar cycles from suitably separated (in
time) episodic campaigns on ground-based telescopes.

 \section{Asteroseismology and stellar populations studies}
 \label{sec:pop}

\begin{figure}
 \centerline {\epsfxsize=14.0cm\epsfbox{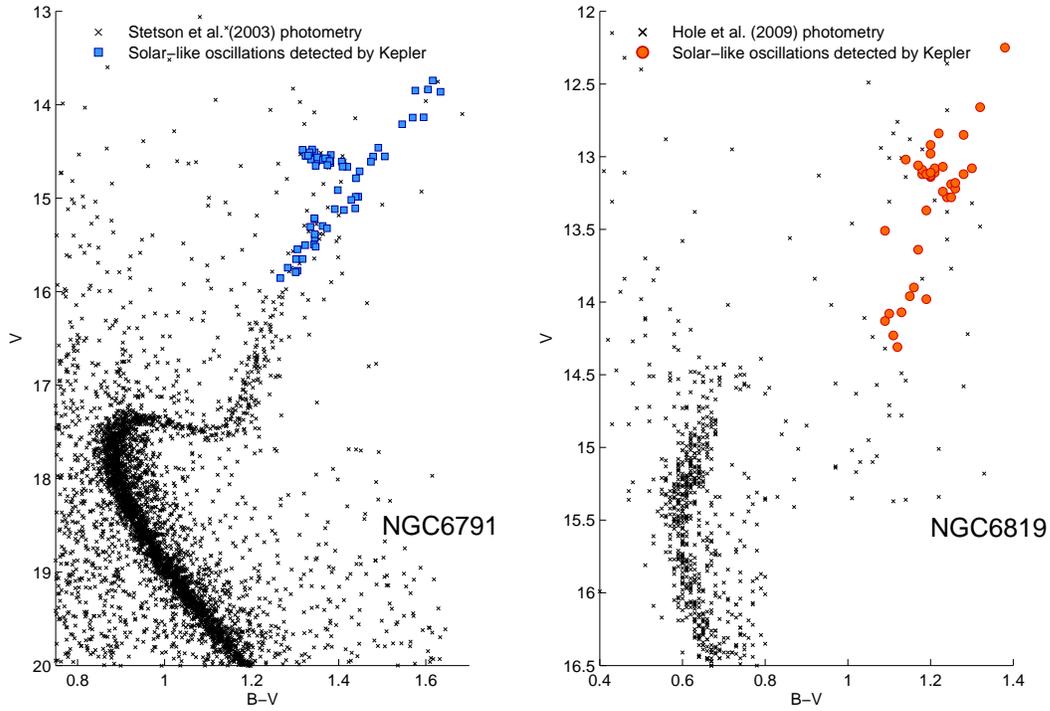}}

 \caption{Colour-magnitude diagrams ($B-V$ versus apparent visual
   magnitude, $V$) of members of the open clusters NGC\,6791 and
   NGC\,6819. Coloured symbols mark the locations of stars with
   \emph{Kepler} detections of solar-like oscillations (e.g., see
   \citealt{Stello2011}).}

 \label{fig:clust} 
\end{figure}

Undoubtedly one the highlights of the \emph{Kepler} asteroseismology
programme has been the detection of oscillations in giants belonging
to the open clusters NGC\,6791, NGC\,6819 (see
Figure~\ref{fig:clust}), and NGC\,6811 (which has far fewer
detections).  An obvious advantage of modelling stars belonging to a
supposedly simple population is to work under the assumption that all
stars have the same age, initial chemical composition, and lie at
essentially the same distance. These strong priors reduce the number
of free parameters during modelling, allowing for stringent tests of
stellar evolutionary theory.

While the analysis and interpretation of asteroseismic data in cluster
members is still in its infancy, several promising results have
already emerged.  \citet{Basu2011} estimated the age, mass and
distance of RGB stars belonging to the old-open clusters NGC\,6791 and
NGC\,6819 through application of a grid-based method (see
Section~\ref{sec:scaling}) using the observed $\left< \Delta\nu_{nl}
\right>$, $\nu_{\rm max}$, photometrically derived $T_{\rm eff}$, and
metallicities from spectroscopic analyses. Given the assumption of a
common age and distance, and provided that systematic uncertainties
have been accounted for, this approach leads to very precise and
potentially very accurate estimates of the cluster properties. Such
estimates may be then compared against those obtained using different
techniques and observational constraints (e.g., isochrone fitting,
eclipsing binaries). These comparisons, as discussed in
Section~\ref{sec:scaling}, are crucial to test the methods used to
estimate radii, masses, and ages of giants, since those methods are
also applied to single field stars belonging to composite populations
(see below).  \citet{Stello2011} compared the observed and expected
$\left< \Delta\nu_{nl} \right>$ and $\nu_{\rm max}$ to discriminate
between asteroseismic cluster members and likely non-members, while
\citet{Miglio2012b} estimated the integrated RGB mass loss in NGC6791
by comparing the average masses of stars in the RC and on the RGB.
The availability of seismic constraints beyond the average seismic
parameters, and eventually the comparison between observed and
theoretically predicted frequencies of individual modes, will turn
these targets into astrophysical laboratories to test models of the
internal structure of giants (see Section~\ref{sec:mixed}).

Along with studies of simple stellar populations, the detection by
CoRoT and \emph{Kepler} of oscillations in thousands of field stars
has opened the door to detailed studies of stellar populations
belonging to the Milky Way, which can be used to inform models of the
Galaxy.  The mechanisms of the formation and evolution of the Milky
Way are encoded in the kinematics, chemistry, locations and ages of
stars. Particularly important observational constraints are relations
linking velocity dispersion and metallicity to age in different parts
of the Galaxy, as well as spatial gradients of metallicity and key
abundance ratios at different ages \citep[e.g., for reviews
  see][]{Freeman2002,Chiappini2012}.  The difficulties associated with
estimating distances and, to an even greater extent, ages
\citep{Soderblom2010} of individual field stars has been a major
obstacle to discriminating between different scenarios of formation
and evolution of the major components of the Milky Way.  In this
context, solar-like oscillators represent key tracers of the
properties of stellar populations.

Once data from the first CoRoT observational run had been analysed,
and solar-like oscillations had been detected in thousands of red
giant stars \citep{Hekker2009}, it became clear that the newly
available observational constraints would allow novel approaches to
the study of Galactic stellar populations.  \citet{Miglio2009}
presented a first comparison between observed and predicted seismic
properties of giants observed in the first CoRoT field, which
highlighted the expected signatures of RC stars in both
distributions. As discussed in Section~\ref{sec:scaling}, we may
estimate radii and masses for all stars with detected oscillations by
combining average and global asteroseismic parameters with estimates
of surface temperature.

There are several important reasons why asteroseismic data on red
giants offer huge potential for populations studies. First, G and K
giants are numerous: They are therefore substantial contributors by
number to magnitude-limited surveys of stars, such as CoRoT and
\emph{Kepler}. Moreover, the large intrinsic oscillation amplitudes
and long oscillation periods mean that oscillations may be detected in
faint targets observed in the long-cadence modes of CoRoT and
\emph{Kepler} (see Section~\ref{sec:obs}). Second, with asteroseismic
data in hand red giants may be used as accurate distance indicators
probing regions out to about $10\,\rm kpc$: As in the case of
eclipsing binaries the distance to each red giant may be estimated
from the absolute luminosity, which is obtained from the
asteroseismically determined radius and $T_{\rm eff}$. This differs
from the approach adopted to exploit pulsational information from
classical pulsators, notably Cepheid variables, where the observed
pulsation frequency leads to an estimate of the mean density only and
hence additional calibration and assumptions are needed to yield an
estimated distance.  Giants observed by CoRoT and \emph{Kepler} may be
used as distance indicators, mapping regions at different
Galactocentric radii and, in the case of \emph{Kepler}, exploring
regions where thick-disc and halo giants are expected.

Third, seismic data on RGB stars in principle provide robust ages that
probe a wide age range: Once a star has evolved to the RGB its age is
determined to a first approximation by the time spent in the
core-hydrogen burning phase, which is predominantly a function of mass
and metallicity. Hence, the estimated masses of red giants provide
important constraints on age. The CoRoT and \emph{Kepler} giants cover
a mass range from $\simeq 0.9$ to $\simeq 3\,\rm M_\odot$, which in
turn maps to an age range spanning $\simeq 0.3$ to $\simeq 12\,\rm
Gyr$, i.e., the entire Galactic history.  As a word of caution it is
worth remembering that estimation of the ages is inherently model
dependent. Systematic uncertainties from predictions of main-sequence
lifetimes need to be taken into account.

Consider, for example, the impact on RGB ages of uncertainties in
predictions of the size of the central, fully-mixed region in
main-sequence stars.  We take the example of a model of mass $1.4\,\rm
M_\odot$. The difference between the main-sequence lifetime of a model
with and without overshooting\footnote{We assume an extension of the
  overshooting region equal to $0.2\,H_p$, where $H_p$ is the pressure
  scale height at the boundary of the convective core, as defined by
  the Schwarzschild criterion.} from the core is of the order of
$20\,\%$. However, once the model reaches the giant phase, this
difference is reduced to about $5\,\%$. Low-mass models with a larger
centrally mixed region experience a significantly shorter subgiant
phase, the reason being that they end the main sequence with an
isothermal helium core which is closer to the Sh\"onberg-Chandrasekar
limit \citep[see][]{Maeder1975}, hence partially offsetting the impact
of a longer main-sequence lifetime.  On the other hand, the effect of
core overshooting on the age of RGB stars is more pronounced when the
mass of the He core at the end of the main sequence is close to (or
even larger then) the Sh\"onberg-Chandrasekar limit (e.g., in the case
of higher-mass stars, or in models computed assuming large
overshooting parameters).

When considering RC stars, an additional complication in the age
determination arises from the rather uncertain mass-loss rates
occurring during the RGB phase \citep[e.g.,][]{Catelan2009}. In this
case the characterisation of populations of giants will benefit
greatly from estimation of the period spacings of the observed g
modes, which we have seen allows a clear distinction to be made
between RGB and RC stars (see Section~\ref{sec:mixed}).

\emph{Kepler} is contributing significantly to the characterisation
not only of red-giant populations, but has also opened the way for
``ensemble asteroseismology'' of solar-type stars. The detection of
solar-like oscillations in about 500 F and G-type dwarfs allowed
\citet{Chaplin2011} to perform a first quantitative comparison between
the distributions of observed masses and radii of these stars with
predictions from models of synthetic populations in the Galaxy.  This
first comparison showed intriguing differences in the distribution of
mass which will need to be addressed with more detail. This sample of
stars should provide a gold standard for the age determination of
field dwarfs, and further information on age-metallicity relations
albeit out to a more limited distance than the giants, of a few
hundreds parsecs from the Sun \citep{SilvaAguirre2012}.

 \section{Concluding Remarks}
 \label{sec:conc}

We finish by looking ahead, with a few remarks on future possibilities
to supplement those made already throughout the review. In the short
to medium term, there are exciting prospects for fully exploiting the
\emph{Kepler} and CoRoT data. The long \emph{Kepler} datasets provide
a unique opportunity to make exquisite tests of stellar interiors
physics, to probe the internal rotation and dynamics of evolutionary
sequences of low-mass stars, and to elucidate our understanding of
stellar activity and stellar dynamos by detecting seismic signatures
of stellar cycles and applying diagnostics of the internal
dynamics. The sample of asteroseismic planet-hosting stars will
increase in size as more data are collected and the analysis of the
existing raw \emph{Kepler} data continues to evolve. A significant
number of systems with red-giant hosts is likely to be added. This
will provide crucial results to address how the dynamics of systems,
and planet-star interactions, evolve over time (including engulfment
of planets).

There is the potential for combining asteroseismic results from
\emph{Kepler} and CoRoT with results from large spectroscopic surveys,
e.g., APOGEE \citep{Majewski2010}, HERMES \citep{Barden2010}, and the
GAIA ESO Public Spectroscopic Survey \citep{Gilmore2012} (with CoRoT
providing results on red giants in different fields of the
Galaxy). Since the asteroseismic data provide a very accurate way to
determine surface gravities, they can also play an important r\^ole in
helping to calibrate the spectroscopic analyses. A formal
collaboration (APOKASC) has already been established between APOGEE
and the \emph{Kepler} Asteroseismic Science Consortium. These
collaborations open the possibility to provide strong constraints on
age-metallicity and age-velocity relations in different parts of the
Milky Way.

With regards to future requirements on the asteroseismic observations,
there is an obvious need for oscillations data on solar-type stars in
clusters (the open clusters observed by \emph{Kepler} are too faint to
yield detections in cool main-sequence stars). Such data would allow
further tests of stellar interiors physics in main-sequence and
subgiant stars and help to calibrate the age ladder. Long timeseries
data are also needed on bright stars in the solar
neighbourhood. Photometric observations of bright stars of the type
proposed for the ESA and NASA candidate missions PLATO and TESS would
provide data for asteroseismology limited by intrinsic stellar noise,
not photon counting noise. PLATO and TESS would also observe targets
in different fields.

There are also exciting developments to come for ground-based
asteroseismology. The Stellar Observations Network Group (SONG) of 1-m
telescopes is now being deployed \citep{Grundahl2011}, with one of the
main goals being the dedicated study of solar-like oscillators. SONG
will provide Doppler velocity data of unprecedented quality on a
selection of bright, nearby stars, including data on the
lowest-frequency modes not accessible to the photometric observations.

 \section{Acknowledgements}
 \label{sec:acknowledgements}

This review has benefited significantly from numerous and varied
discussions with many colleagues. We would in particular like to thank
Sarbani Basu, Yvonne Elsworth, Ron Gilliland and Arlette Noels for
providing detailed feedback on earlier drafts; and Guy Davies and
G\"unter Houdek for their expert help with figures.  The authors
acknowledge financial support from the UK Science and Technology
Facilities Council (STFC). We also acknowledge use of \emph{Kepler}
data for some of the figures. These data were made available through
the KASC website, and we thank R.~Garc\'ia who prepared the data for
asteroseismic analysis.

%%%%%%%%%%%%%%%%%%%%%%%%%%%%%%%%%%%%%%%%%%%%%%%%%%%%%%%%%%%%%%%%%%%%%%%

%% Figure 1:

%%%%%%%%%%%%%%%%%%%%%%%%%%%%%%%%%%%%%%%%%%%%%%%%%%%%%%%%%%%%%%%%%%%%%%%

%%%%%%%%%%%%%%%%%%%%%%%%%%%%%%%%%%%%%%%%%%%%%%%%%%%%%%%%%%%%%%%%%%%%%%%

%% Figure 2:

%%%%%%%%%%%%%%%%%%%%%%%%%%%%%%%%%%%%%%%%%%%%%%%%%%%%%%%%%%%%%%%%%%%%%%%

%% Figure 3:

%%%%%%%%%%%%%%%%%%%%%%%%%%%%%%%%%%%%%%%%%%%%%%%%%%%%%%%%%%%%%%%%%%%%%%%

%% Figure 4:

%%%%%%%%%%%%%%%%%%%%%%%%%%%%%%%%%%%%%%%%%%%%%%%%%%%%%%%%%%%%%%%%%%%%%%%

%% Figure 5:

%%%%%%%%%%%%%%%%%%%%%%%%%%%%%%%%%%%%%%%%%%%%%%%%%%%%%%%%%%%%%%%%%%%%%%%

%% Figure 6:

%%%%%%%%%%%%%%%%%%%%%%%%%%%%%%%%%%%%%%%%%%%%%%%%%%%%%%%%%%%%%%%%%%%%%%%

%% Figure 7:

%%%%%%%%%%%%%%%%%%%%%%%%%%%%%%%%%%%%%%%%%%%%%%%%%%%%%%%%%%%%%%%%%%%%%%%

%\bibliographystyle{Astronomy}
\bibliography{araa1}

\begin{thebibliography}{}
\expandafter\ifx\csname natexlab\endcsname\relax\def\natexlab#1{#1}\fi

\bibitem[{{Aerts}, {Christensen-Dalsgaard} \& {Kurtz}(2010)}]{Aerts2010}
{Aerts} C, {Christensen-Dalsgaard} J, {Kurtz} DW. 2010.
\newblock \textit{{Asteroseismology}}.
\newblock Springer Science+Business Media B.V.

\bibitem[{{Aizenman}, {Smeyers} \& {Weigert}(1977)}]{Aizenman1977}
{Aizenman} M, {Smeyers} P, {Weigert} A. 1977.
\newblock \textit{\aap} 58:41

\bibitem[{{Albrecht} et~al.(2012){Albrecht}, {Winn}, {Johnson}, {Howard},
  {Marcy} et~al.}]{Albrecht2012}
{Albrecht} S, {Winn} JN, {Johnson} JA, {Howard} AW, {Marcy} GW, et~al. 2012.
\newblock \textit{\apj} 757:18

\bibitem[{Andersen, Clausen \& Nordstrom(1990)}]{Andersen1990}
Andersen J, Clausen JV, Nordstrom B. 1990.
\newblock \textit{\apjl} 363:L33--L36

\bibitem[{{Appourchaux}(2011)}]{Appourchaux2011}
{Appourchaux} T. 2011.
\newblock In \textit{Asteroseismology}, ed. P~{Pall{\'e}}, vol.~23 of
  \textit{Canary Islands Winter School of Astrophysics}. Cambridge University
  Press.
\newblock In the press (arXiv:1103.5352)

\bibitem[{{Appourchaux} et~al.(2012){Appourchaux}, {Chaplin}, {Garc{\'{\i}}a},
  {Gruberbauer}, {Verner} et~al.}]{Appourchaux2012}
{Appourchaux} T, {Chaplin} WJ, {Garc{\'{\i}}a} RA, {Gruberbauer} M, {Verner}
  GA, et~al. 2012.
\newblock \textit{\aap} 543:A54

\bibitem[{{Appourchaux} et~al.(2008){Appourchaux}, {Michel}, {Auvergne},
  {Baglin}, {Toutain} et~al.}]{Appourchaux2008}
{Appourchaux} T, {Michel} E, {Auvergne} M, {Baglin} A, {Toutain} T, et~al.
  2008.
\newblock \textit{\aap} 488:705--714

\bibitem[{{Arentoft} et~al.(2008){Arentoft}, {Kjeldsen}, {Bedding}, {Bazot},
  {Christensen-Dalsgaard} et~al.}]{Arentoft2008}
{Arentoft} T, {Kjeldsen} H, {Bedding} TR, {Bazot} M, {Christensen-Dalsgaard} J,
  et~al. 2008.
\newblock \textit{\apj} 687:1180--1190

\bibitem[{{Asplund} et~al.(2009){Asplund}, {Grevesse}, {Sauval} \&
  {Scott}}]{Asplund2009}
{Asplund} M, {Grevesse} N, {Sauval} AJ, {Scott} P. 2009.
\newblock \textit{\araa} 47:481--522

\bibitem[{{Ballot}(2010)}]{Ballot2010}
{Ballot} J. 2010.
\newblock \textit{Astronomische Nachrichten} 331:933

\bibitem[{{Ballot} et~al.(2008){Ballot}, {Appourchaux}, {Toutain} \&
  {Guittet}}]{Ballot2008}
{Ballot} J, {Appourchaux} T, {Toutain} T, {Guittet} M. 2008.
\newblock \textit{\aap} 486:867--875

\bibitem[{{Ballot}, {Garc{\'{\i}}a} \& {Lambert}(2006)}]{Ballot2006}
{Ballot} J, {Garc{\'{\i}}a} RA, {Lambert} P. 2006.
\newblock \textit{\mnras} 369:1281--1286

\bibitem[{{Ballot} et~al.(2011){Ballot}, {Gizon}, {Samadi}, {Vauclair},
  {Benomar} et~al.}]{Ballot2011}
{Ballot} J, {Gizon} L, {Samadi} R, {Vauclair} G, {Benomar} O, et~al. 2011.
\newblock \textit{\aap} 530:A97

\bibitem[{{Barden} et~al.(2010){Barden}, {Jones}, {Barnes}, {Heijmans}, {Heng}
  et~al.}]{Barden2010}
{Barden} SC, {Jones} DJ, {Barnes} SI, {Heijmans} J, {Heng} A, et~al. 2010.
\newblock In \textit{Society of Photo-Optical Instrumentation Engineers (SPIE)
  Conference Series}, vol. 7735 of \textit{Society of Photo-Optical
  Instrumentation Engineers (SPIE) Conference Series}

\bibitem[{{Basu} \& {Antia}(2008)}]{Basu2008}
{Basu} S, {Antia} HM. 2008.
\newblock \textit{\physrep} 457:217--283

\bibitem[{{Basu}, {Chaplin} \& {Elsworth}(2010)}]{Basu2010}
{Basu} S, {Chaplin} WJ, {Elsworth} Y. 2010.
\newblock \textit{\apj} 710:1596--1609

\bibitem[{{Basu}, {Christensen-Dalsgaard} \& {Thompson}(2002)}]{Basu2002}
{Basu} S, {Christensen-Dalsgaard} J, {Thompson} MJ. 2002.
\newblock In \textit{Stellar Structure and Habitable Planet Finding}, eds.
  B~{Battrick}, F~{Favata}, IW~{Roxburgh}, D~{Galadi}, vol. 485 of \textit{ESA
  Special Publication}

\bibitem[{{Basu} et~al.(2011){Basu}, {Grundahl}, {Stello}, {Kallinger},
  {Hekker} et~al.}]{Basu2011}
{Basu} S, {Grundahl} F, {Stello} D, {Kallinger} T, {Hekker} S, et~al. 2011.
\newblock \textit{\apjl} 729:L10

\bibitem[{Basu et~al.(2004)Basu, Mazumdar, Antia \& Demarque}]{Basu2004}
Basu S, Mazumdar A, Antia HM, Demarque P. 2004.
\newblock \textit{\mnras} 350:277--286

\bibitem[{{Basu} et~al.(2012){Basu}, {Verner}, {Chaplin} \&
  {Elsworth}}]{Basu2012}
{Basu} S, {Verner} GA, {Chaplin} WJ, {Elsworth} Y. 2012.
\newblock \textit{\apj} 746:76

\bibitem[{{Batalha} et~al.(2011){Batalha}, {Borucki}, {Bryson}, {Buchhave},
  {Caldwell} et~al.}]{Batalha2011}
{Batalha} NM, {Borucki} WJ, {Bryson} ST, {Buchhave} LA, {Caldwell} DA, et~al.
  2011.
\newblock \textit{\apj} 729:27

\bibitem[{{Baudin} et~al.(2011){Baudin}, {Barban}, {Belkacem}, {Hekker},
  {Morel} et~al.}]{Baudin2011}
{Baudin} F, {Barban} C, {Belkacem} K, {Hekker} S, {Morel} T, et~al. 2011.
\newblock \textit{\aap} 529:A84

\bibitem[{{Bazot}, {Bourguignon} \& {Christensen-Dalsgaard}(2012)}]{Bazot2012}
{Bazot} M, {Bourguignon} S, {Christensen-Dalsgaard} J. 2012.
\newblock \textit{\mnras} 427:1847--1866

\bibitem[{{Bazot} et~al.(2005){Bazot}, {Vauclair}, {Bouchy} \&
  {Santos}}]{Bazot2005}
{Bazot} M, {Vauclair} S, {Bouchy} F, {Santos} NC. 2005.
\newblock \textit{\aap} 440:615--621

\bibitem[{{Beck} et~al.(2011){Beck}, {Bedding}, {Mosser}, {Stello}, {Garcia}
  et~al.}]{Beck2011}
{Beck} PG, {Bedding} TR, {Mosser} B, {Stello} D, {Garcia} RA, et~al. 2011.
\newblock \textit{Science} 332:205

\bibitem[{{Beck} et~al.(2012){Beck}, {Montalban}, {Kallinger}, {De Ridder},
  {Aerts} et~al.}]{Beck2012}
{Beck} PG, {Montalban} J, {Kallinger} T, {De Ridder} J, {Aerts} C, et~al. 2012.
\newblock \textit{\nat} 481:55--57

\bibitem[{{Bedding}(2011)}]{Bedding2011b}
{Bedding} TR. 2011.
\newblock In \textit{Asteroseismology}, ed. P~{Pall{\'e}}, vol.~23 of
  \textit{Canary Islands Winter School of Astrophysics}. Cambridge University
  Press.
\newblock In the press (arXiv:1107.1723)

\bibitem[{{Bedding} et~al.(2010){Bedding}, {Huber}, {Stello}, {Elsworth},
  {Hekker} et~al.}]{Bedding2010}
{Bedding} TR, {Huber} D, {Stello} D, {Elsworth} YP, {Hekker} S, et~al. 2010.
\newblock \textit{\apjl} 713:L176--L181

\bibitem[{{Bedding} et~al.(2007){Bedding}, {Kjeldsen}, {Arentoft}, {Bouchy},
  {Brandbyge} et~al.}]{Bedding2007}
{Bedding} TR, {Kjeldsen} H, {Arentoft} T, {Bouchy} F, {Brandbyge} J, et~al.
  2007.
\newblock \textit{\apj} 663:1315--1324

\bibitem[{{Bedding} et~al.(2011){Bedding}, {Mosser}, {Huber}, {Montalb{\'a}n},
  {Beck} et~al.}]{Bedding2011}
{Bedding} TR, {Mosser} B, {Huber} D, {Montalb{\'a}n} J, {Beck} P, et~al. 2011.
\newblock \textit{\nat} 471:608--611

\bibitem[{{Belkacem} et~al.(2012){Belkacem}, {Dupret}, {Baudin}, {Appourchaux},
  {Marques} \& {Samadi}}]{Belkacem2012}
{Belkacem} K, {Dupret} MA, {Baudin} F, {Appourchaux} T, {Marques} JP, {Samadi}
  R. 2012.
\newblock \textit{\aap} 540:L7

\bibitem[{{Belkacem} et~al.(2011){Belkacem}, {Goupil}, {Dupret}, {Samadi},
  {Baudin} et~al.}]{Belkacem2011}
{Belkacem} K, {Goupil} MJ, {Dupret} MA, {Samadi} R, {Baudin} F, et~al. 2011.
\newblock \textit{\aap} 530:A142

\bibitem[{{Benomar}, {Appourchaux} \& {Baudin}(2009)}]{Benomar2009}
{Benomar} O, {Appourchaux} T, {Baudin} F. 2009.
\newblock \textit{\aap} 506:15--32

\bibitem[{{Benomar} et~al.(2012){Benomar}, {Bedding}, {Stello}, {Deheuvels},
  {White} \& {Christensen-Dalsgaard}}]{Benomar2012}
{Benomar} O, {Bedding} TR, {Stello} D, {Deheuvels} S, {White} TR,
  {Christensen-Dalsgaard} J. 2012.
\newblock \textit{\apjl} 745:L33

\bibitem[{{Bildsten} et~al.(2012){Bildsten}, {Paxton}, {Moore} \&
  {Macias}}]{Bildsten2012}
{Bildsten} L, {Paxton} B, {Moore} K, {Macias} PJ. 2012.
\newblock \textit{\apjl} 744:L6

\bibitem[{{B{\"o}hm-Vitense}(2007)}]{Bohm-Vitense2007}
{B{\"o}hm-Vitense} E. 2007.
\newblock \textit{\apj} 657:486--493

\bibitem[{{Borucki} et~al.(2012){Borucki}, {Koch}, {Batalha}, {Bryson}, {Rowe}
  et~al.}]{Borucki2012}
{Borucki} WJ, {Koch} DG, {Batalha} N, {Bryson} ST, {Rowe} J, et~al. 2012.
\newblock \textit{\apj} 745:120

\bibitem[{{Bouchy} et~al.(2005){Bouchy}, {Bazot}, {Santos}, {Vauclair} \&
  {Sosnowska}}]{Bouchy2005}
{Bouchy} F, {Bazot} M, {Santos} NC, {Vauclair} S, {Sosnowska} D. 2005.
\newblock \textit{\aap} 440:609--614

\bibitem[{{Brogaard} et~al.(2011){Brogaard}, {Bruntt}, {Grundahl}, {Clausen},
  {Frandsen} et~al.}]{Brogaard2011}
{Brogaard} K, {Bruntt} H, {Grundahl} F, {Clausen} JV, {Frandsen} S, et~al.
  2011.
\newblock \textit{\aap} 525:A2

\bibitem[{{Brogaard} et~al.(2012){Brogaard}, {VandenBerg}, {Bruntt},
  {Grundahl}, {Frandsen} et~al.}]{Brogaard2012}
{Brogaard} K, {VandenBerg} DA, {Bruntt} H, {Grundahl} F, {Frandsen} S, et~al.
  2012.
\newblock \textit{\aap} 543:A106

\bibitem[{{Brown} et~al.(1994){Brown}, {Christensen-Dalsgaard},
  {Weibel-Mihalas} \& {Gilliland}}]{Brown1994b}
{Brown} TM, {Christensen-Dalsgaard} J, {Weibel-Mihalas} B, {Gilliland} RL.
  1994.
\newblock \textit{\apj} 427:1013--1034

\bibitem[{{Brown} \& {Gilliland}(1994)}]{Brown1994a}
{Brown} TM, {Gilliland} RL. 1994.
\newblock \textit{\araa} 32:37--82

\bibitem[{Brown et~al.(1991)Brown, Gilliland, Noyes \& Ramsey}]{Brown1991}
Brown TM, Gilliland RL, Noyes RW, Ramsey LW. 1991.
\newblock \textit{\apj} 368:599--609

\bibitem[{{Bruntt} et~al.(2010){Bruntt}, {Bedding}, {Quirion}, {Lo Curto},
  {Carrier} et~al.}]{Bruntt2010}
{Bruntt} H, {Bedding} TR, {Quirion} PO, {Lo Curto} G, {Carrier} F, et~al. 2010.
\newblock \textit{\mnras} 405:1907--1923

\bibitem[{{Carter} et~al.(2012){Carter}, {Agol}, {Chaplin}, {Basu}, {Bedding}
  et~al.}]{Carter2012}
{Carter} JA, {Agol} E, {Chaplin} WJ, {Basu} S, {Bedding} TR, et~al. 2012.
\newblock \textit{Science} 337:556--

\bibitem[{Catelan(2009)}]{Catelan2009}
Catelan M. 2009.
\newblock \textit{\apss} 320:261--309

\bibitem[{{Chaplin} et~al.(2006){Chaplin}, {Appourchaux}, {Baudin}, {Boumier},
  {Elsworth} et~al.}]{Chaplin2006}
{Chaplin} WJ, {Appourchaux} T, {Baudin} F, {Boumier} P, {Elsworth} Y, et~al.
  2006.
\newblock \textit{\mnras} 369:985--996

\bibitem[{{Chaplin} et~al.(2010){Chaplin}, {Appourchaux}, {Elsworth},
  {Garc{\'{\i}}a}, {Houdek} et~al.}]{Chaplin2010}
{Chaplin} WJ, {Appourchaux} T, {Elsworth} Y, {Garc{\'{\i}}a} RA, {Houdek} G,
  et~al. 2010.
\newblock \textit{\apjl} 713:L169--L175

\bibitem[{{Chaplin} et~al.(2005){Chaplin}, {Houdek}, {Elsworth}, {Gough},
  {Isaak} \& {New}}]{Chaplin2005}
{Chaplin} WJ, {Houdek} G, {Elsworth} Y, {Gough} DO, {Isaak} GR, {New} R. 2005.
\newblock \textit{\mnras} 360:859--868

\bibitem[{{Chaplin} et~al.(2011){Chaplin}, {Kjeldsen}, {Christensen-Dalsgaard},
  {Basu}, {Miglio} et~al.}]{Chaplin2011}
{Chaplin} WJ, {Kjeldsen} H, {Christensen-Dalsgaard} J, {Basu} S, {Miglio} A,
  et~al. 2011.
\newblock \textit{Science} 332:213--216

\bibitem[{{Chaplin} et~al.(2013){Chaplin}, {Sanchis-Ojeda}, {Campante},
  {Handberg}, {Stello} et~al.}]{Chaplin2013}
{Chaplin} WJ, {Sanchis-Ojeda} R, {Campante} TL, {Handberg} R, {Stello} D,
  et~al. 2013.
\newblock \textit{\apj} 766:101

\bibitem[{{Charbonnel}(2005)}]{Charbonnel2005a}
{Charbonnel} C. 2005.
\newblock In \textit{Cosmic Abundances as Records of Stellar Evolution and
  Nucleosynthesis}, eds. TG~{Barnes} III, FN~{Bash}, vol. 336 of
  \textit{Astronomical Society of the Pacific Conference Series}

\bibitem[{{Charbonnel} \& {Talon}(2005)}]{Charbonnel2005}
{Charbonnel} C, {Talon} S. 2005.
\newblock \textit{Science} 309:2189--2191

\bibitem[{{Chiappini}(2012)}]{Chiappini2012}
{Chiappini} C. 2012.
\newblock In \textit{Red Giants as Probes of the Structure and Evolution of the
  Milky Way}, eds. A~{Miglio}, J~{Montalb{\'a}n}, A~{Noels}

\bibitem[{{Christensen-Dalsgaard}(2002)}]{jcd2002}
{Christensen-Dalsgaard} J. 2002.
\newblock \textit{Reviews of Modern Physics} 74:1073--1129

\bibitem[{{Christensen-Dalsgaard}(2011)}]{jcd2011a}
{Christensen-Dalsgaard} J. 2011.
\newblock In \textit{Asteroseismology}, ed. P~{Pall{\'e}}, vol.~23 of
  \textit{Canary Islands Winter School of Astrophysics}. Cambridge University
  Press.
\newblock In the press (arXiv:1106.5946)

\bibitem[{{Christensen-Dalsgaard}, {Bedding} \& {Kjeldsen}(1995)}]{jcd1995}
{Christensen-Dalsgaard} J, {Bedding} TR, {Kjeldsen} H. 1995.
\newblock \textit{\apjl} 443:L29--L32

\bibitem[{{Christensen-Dalsgaard} \& {Houdek}(2010)}]{jcd2010}
{Christensen-Dalsgaard} J, {Houdek} G. 2010.
\newblock \textit{\apss} 328:51--66

\bibitem[{{Christensen-Dalsgaard} et~al.(2010){Christensen-Dalsgaard},
  {Kjeldsen}, {Brown}, {Gilliland}, {Arentoft} et~al.}]{jcd2010a}
{Christensen-Dalsgaard} J, {Kjeldsen} H, {Brown} TM, {Gilliland} RL, {Arentoft}
  T, et~al. 2010.
\newblock \textit{\apjl} 713:L164--L168

\bibitem[{{Christensen-Dalsgaard} \& {Thompson}(1997)}]{jcd1997}
{Christensen-Dalsgaard} J, {Thompson} MJ. 1997.
\newblock \textit{\mnras} 284:527--540

\bibitem[{{Christensen-Dalsgaard} \& {Thompson}(2011)}]{jcd2011b}
{Christensen-Dalsgaard} J, {Thompson} MJ. 2011.
\newblock In \textit{IAU Symposium}, eds. NH~{Brummell}, AS~{Brun},
  MS~{Miesch}, Y~{Ponty}, vol. 271 of \textit{IAU Symposium}

\bibitem[{{Cunha} et~al.(2007){Cunha}, {Aerts}, {Christensen-Dalsgaard},
  {Baglin}, {Bigot} et~al.}]{Cunha2007}
{Cunha} MS, {Aerts} C, {Christensen-Dalsgaard} J, {Baglin} A, {Bigot} L, et~al.
  2007.
\newblock \textit{\aapr} 14:217--360

\bibitem[{{Cunha} \& {Brand{\~a}o}(2011)}]{Cunha2011}
{Cunha} MS, {Brand{\~a}o} IM. 2011.
\newblock \textit{\aap} 529:A10

\bibitem[{{Cunha} \& {Metcalfe}(2007)}]{Cunha2007a}
{Cunha} MS, {Metcalfe} TS. 2007.
\newblock \textit{\apj} 666:413--422

\bibitem[{{de Meulenaer} et~al.(2010){de Meulenaer}, {Carrier}, {Miglio},
  {Bedding}, {Campante} et~al.}]{Demeulenaer2010}
{de Meulenaer} P, {Carrier} F, {Miglio} A, {Bedding} TR, {Campante} TL, et~al.
  2010.
\newblock \textit{\aap} 523:A54

\bibitem[{{De Ridder} et~al.(2009){De Ridder}, Barban, Baudin, Carrier, Hatzes
  et~al.}]{DeRidder2009}
{De Ridder} J, Barban C, Baudin F, Carrier F, Hatzes AP, et~al. 2009.
\newblock \textit{\nat} 459:398--400

\bibitem[{{Deheuvels} et~al.(2012){Deheuvels}, {Garc{\'{\i}}a}, {Chaplin},
  {Basu}, {Antia} et~al.}]{Deheuvels2012}
{Deheuvels} S, {Garc{\'{\i}}a} RA, {Chaplin} WJ, {Basu} S, {Antia} HM, et~al.
  2012.
\newblock \textit{\apj} 756:19

\bibitem[{{Deheuvels} \& {Michel}(2011)}]{Deheuvels2011}
{Deheuvels} S, {Michel} E. 2011.
\newblock \textit{\aap} 535:A91

\bibitem[{{Demarque} et~al.(2004){Demarque}, {Woo}, {Kim} \&
  {Yi}}]{Demarque2004}
{Demarque} P, {Woo} JH, {Kim} YC, {Yi} SK. 2004.
\newblock \textit{\apjs} 155:667--674

\bibitem[{Dupret et~al.(2009)Dupret, Belkacem, Samadi, Montalban, Moreira
  et~al.}]{Dupret2009}
Dupret MA, Belkacem K, Samadi R, Montalban J, Moreira O, et~al. 2009.
\newblock \textit{\aap} 506:57--67

\bibitem[{{Eggenberger} et~al.(2010){Eggenberger}, {Meynet}, {Maeder},
  {Miglio}, {Montalban} et~al.}]{Eggenberger2010}
{Eggenberger} P, {Meynet} G, {Maeder} A, {Miglio} A, {Montalban} J, et~al.
  2010.
\newblock \textit{\aap} 519:A116

\bibitem[{{Eggenberger}, {Montalb{\'a}n} \& {Miglio}(2012)}]{Eggenberger2012}
{Eggenberger} P, {Montalb{\'a}n} J, {Miglio} A. 2012.
\newblock \textit{\aap} 544:L4

\bibitem[{{Escobar} et~al.(2012){Escobar}, {Th{\'e}ado}, {Vauclair}, {Ballot},
  {Charpinet} et~al.}]{Escobar2012}
{Escobar} ME, {Th{\'e}ado} S, {Vauclair} S, {Ballot} J, {Charpinet} S, et~al.
  2012.
\newblock \textit{\aap} 543:A96

\bibitem[{{Freeman} \& {Bland-Hawthorn}(2002)}]{Freeman2002}
{Freeman} K, {Bland-Hawthorn} J. 2002.
\newblock \textit{\araa} 40:487--537

\bibitem[{{Gai} et~al.(2011){Gai}, {Basu}, {Chaplin} \& {Elsworth}}]{Gai2011}
{Gai} N, {Basu} S, {Chaplin} WJ, {Elsworth} Y. 2011.
\newblock \textit{\apj} 730:63

\bibitem[{{Garc{\'{\i}}a} et~al.(2010){Garc{\'{\i}}a}, {Mathur}, {Salabert},
  {Ballot}, {R{\'e}gulo} et~al.}]{Garcia2010}
{Garc{\'{\i}}a} RA, {Mathur} S, {Salabert} D, {Ballot} J, {R{\'e}gulo} C,
  et~al. 2010.
\newblock \textit{Science} 329:1032--

\bibitem[{{Gautschy} \& {Saio}(1995)}]{gautschy95}
{Gautschy} A, {Saio} H. 1995.
\newblock \textit{\araa} 33:75--114

\bibitem[{{Gautschy} \& {Saio}(1996)}]{gautschy96}
{Gautschy} A, {Saio} H. 1996.
\newblock \textit{\araa} 34:551--606

\bibitem[{{Gilliland} et~al.(2010){Gilliland}, {Brown},
  {Christensen-Dalsgaard}, {Kjeldsen}, {Aerts} et~al.}]{Gilliland2010}
{Gilliland} RL, {Brown} TM, {Christensen-Dalsgaard} J, {Kjeldsen} H, {Aerts} C,
  et~al. 2010.
\newblock \textit{\pasp} 122:131--143

\bibitem[{{Gilliland} et~al.(2013){Gilliland}, {Marcy}, {Rowe}, {Rogers},
  {Torres} et~al.}]{Gilliland2013}
{Gilliland} RL, {Marcy} GW, {Rowe} JF, {Rogers} L, {Torres} G, et~al. 2013.
\newblock \textit{\apj} 766:40

\bibitem[{{Gilliland} et~al.(2011){Gilliland}, {McCullough}, {Nelan}, {Brown},
  {Charbonneau} et~al.}]{Gilliland2011}
{Gilliland} RL, {McCullough} PR, {Nelan} EP, {Brown} TM, {Charbonneau} D,
  et~al. 2011.
\newblock \textit{\apj} 726:2

\bibitem[{{Gilmore} et~al.(2012){Gilmore}, {Randich}, {Asplund}, {Binney},
  {Bonifacio} et~al.}]{Gilmore2012}
{Gilmore} G, {Randich} S, {Asplund} M, {Binney} J, {Bonifacio} P, et~al. 2012.
\newblock \textit{The Messenger} 147:25--31

\bibitem[{Girardi(1999)}]{Girardi1999}
Girardi L. 1999.
\newblock \textit{\mnras} 308:818--832

\bibitem[{{Gizon} \& {Solanki}(2003)}]{Gizon2003}
{Gizon} L, {Solanki} SK. 2003.
\newblock \textit{\apj} 589:1009--1019

\bibitem[{{Gough}(1986)}]{Gough1986}
{Gough} DO. 1986.
\newblock In \textit{Hydrodynamic and Magnetodynamic Problems in the Sun and
  Stars}, ed. Y~{Osaki}

\bibitem[{{Gough}(1990)}]{Gough1990b}
{Gough} DO. 1990.
\newblock In \textit{Progress of Seismology of the Sun and Stars}, eds.
  Y~{Osaki}, H~{Shibahashi}, vol. 367 of \textit{Lecture Notes in Physics,
  Berlin Springer Verlag}

\bibitem[{{Gough} \& {McIntyre}(1998)}]{Gough1998}
{Gough} DO, {McIntyre} ME. 1998.
\newblock \textit{\nat} 394:755--757

\bibitem[{{Gough} \& {Thompson}(1990)}]{Gough1990}
{Gough} DO, {Thompson} MJ. 1990.
\newblock \textit{\mnras} 242:25--55

\bibitem[{{Goupil} et~al.(2011){Goupil}, {Lebreton}, {Marques}, {Samadi} \&
  {Baudin}}]{Goupil2011}
{Goupil} MJ, {Lebreton} Y, {Marques} JP, {Samadi} R, {Baudin} F. 2011.
\newblock \textit{Journal of Physics Conference Series} 271:012031

\bibitem[{{Gruberbauer} et~al.(2009){Gruberbauer}, {Kallinger}, {Weiss} \&
  {Guenther}}]{Gruberbauer2009}
{Gruberbauer} M, {Kallinger} T, {Weiss} WW, {Guenther} DB. 2009.
\newblock \textit{\aap} 506:1043--1053

\bibitem[{{Grundahl} et~al.(2011){Grundahl}, {Christensen-Dalsgaard}, {Gr{\aa}e
  J{\o}rgensen}, {Frandsen}, {Kjeldsen} \& {Kj{\ae}rgaard
  Rasmussen}}]{Grundahl2011}
{Grundahl} F, {Christensen-Dalsgaard} J, {Gr{\aa}e J{\o}rgensen} U, {Frandsen}
  S, {Kjeldsen} H, {Kj{\ae}rgaard Rasmussen} P. 2011.
\newblock \textit{Journal of Physics Conference Series} 271:012083

\bibitem[{{Handberg} \& {Campante}(2011)}]{Handberg2011}
{Handberg} R, {Campante} TL. 2011.
\newblock \textit{\aap} 527:A56

\bibitem[{{Hekker} et~al.(2010){Hekker}, {Debosscher}, {Huber}, {Hidas}, {De
  Ridder} et~al.}]{Hekker2010}
{Hekker} S, {Debosscher} J, {Huber} D, {Hidas} MG, {De Ridder} J, et~al. 2010.
\newblock \textit{\apjl} 713:L187--L191

\bibitem[{{Hekker} et~al.(2011{\natexlab{a}}){Hekker}, {Elsworth}, {De Ridder},
  {Mosser}, {Garc{\'{\i}}a} et~al.}]{Hekker2011a}
{Hekker} S, {Elsworth} Y, {De Ridder} J, {Mosser} B, {Garc{\'{\i}}a} RA, et~al.
  2011{\natexlab{a}}.
\newblock \textit{\aap} 525:A131

\bibitem[{{Hekker} et~al.(2011{\natexlab{b}}){Hekker}, {Gilliland}, {Elsworth},
  {Chaplin}, {De Ridder} et~al.}]{Hekker2011}
{Hekker} S, {Gilliland} RL, {Elsworth} Y, {Chaplin} WJ, {De Ridder} J, et~al.
  2011{\natexlab{b}}.
\newblock \textit{\mnras} 414:2594--2601

\bibitem[{{Hekker} et~al.(2009){Hekker}, {Kallinger}, {Baudin}, {De Ridder},
  {Barban} et~al.}]{Hekker2009}
{Hekker} S, {Kallinger} T, {Baudin} F, {De Ridder} J, {Barban} C, et~al. 2009.
\newblock \textit{\aap} 506:465--469

\bibitem[{{Hirano} et~al.(2012){Hirano}, {Sanchis-Ojeda}, {Takeda}, {Narita},
  {Winn} et~al.}]{Hirano2012}
{Hirano} T, {Sanchis-Ojeda} R, {Takeda} Y, {Narita} N, {Winn} JN, et~al. 2012.
\newblock \textit{\apj} 756:66

\bibitem[{{Hole} et~al.(2009){Hole}, {Geller}, {Mathieu}, {Platais}, {Meibom}
  \& {Latham}}]{Hole2009}
{Hole} KT, {Geller} AM, {Mathieu} RD, {Platais} I, {Meibom} S, {Latham} DW.
  2009.
\newblock \textit{\aj} 138:159--168

\bibitem[{{Houdek}(2010)}]{Houdek2010}
{Houdek} G. 2010.
\newblock \textit{\apss} 328:237--244

\bibitem[{{Houdek} et~al.(1999){Houdek}, {Balmforth}, {Christensen-Dalsgaard}
  \& {Gough}}]{Houdek1999}
{Houdek} G, {Balmforth} NJ, {Christensen-Dalsgaard} J, {Gough} DO. 1999.
\newblock \textit{\aap} 351:582--596

\bibitem[{Houdek \& Gough(2007)}]{Houdek2007}
Houdek G, Gough DO. 2007.
\newblock \textit{\mnras} 375:861--880

\bibitem[{{Houdek} \& {Gough}(2011)}]{Houdek2011}
{Houdek} G, {Gough} DO. 2011.
\newblock \textit{\mnras} 418:1217--1230

\bibitem[{{Howe}(2009)}]{Howe2009}
{Howe} R. 2009.
\newblock \textit{Living Reviews in Solar Physics} 6:1

\bibitem[{{Howell} et~al.(2012){Howell}, {Rowe}, {Bryson}, {Quinn}, {Marcy}
  et~al.}]{Howell2012}
{Howell} SB, {Rowe} JF, {Bryson} ST, {Quinn} SN, {Marcy} GW, et~al. 2012.
\newblock \textit{\apj} 746:123

\bibitem[{{Huber} et~al.(2011{\natexlab{a}}){Huber}, {Bedding}, {Arentoft},
  {Gruberbauer}, {Guenther} et~al.}]{Huber2011a}
{Huber} D, {Bedding} TR, {Arentoft} T, {Gruberbauer} M, {Guenther} DB, et~al.
  2011{\natexlab{a}}.
\newblock \textit{\apj} 731:94

\bibitem[{{Huber} et~al.(2011{\natexlab{b}}){Huber}, {Bedding}, {Stello},
  {Hekker}, {Mathur} et~al.}]{Huber2011}
{Huber} D, {Bedding} TR, {Stello} D, {Hekker} S, {Mathur} S, et~al.
  2011{\natexlab{b}}.
\newblock \textit{\apj} 743:143

\bibitem[{{Huber} et~al.(2010){Huber}, {Bedding}, {Stello}, {Mosser}, {Mathur}
  et~al.}]{Huber2010}
{Huber} D, {Bedding} TR, {Stello} D, {Mosser} B, {Mathur} S, et~al. 2010.
\newblock \textit{\apj} 723:1607--1617

\bibitem[{{Huber} et~al.(2013){Huber}, {Chaplin}, {Christensen-Dalsgaard},
  {Gilliland}, {Kjeldsen} et~al.}]{Huber2013}
{Huber} D, {Chaplin} WJ, {Christensen-Dalsgaard} J, {Gilliland} RL, {Kjeldsen}
  H, et~al. 2013.
\newblock \textit{\apj} 767:127

\bibitem[{{Huber} et~al.(2012){Huber}, {Ireland}, {Bedding}, {Brand{\~a}o},
  {Piau} et~al.}]{Huber2012}
{Huber} D, {Ireland} MJ, {Bedding} TR, {Brand{\~a}o} IM, {Piau} L, et~al. 2012.
\newblock \textit{\apj} 760:32

\bibitem[{{Kalirai} \& {Tosi}(2004)}]{Kalirai2004}
{Kalirai} JS, {Tosi} M. 2004.
\newblock \textit{\mnras} 351:649--662

\bibitem[{{Kallinger} et~al.(2010){Kallinger}, {Mosser}, {Hekker}, {Huber},
  {Stello} et~al.}]{Kallinger2010}
{Kallinger} T, {Mosser} B, {Hekker} S, {Huber} D, {Stello} D, et~al. 2010.
\newblock \textit{\aap} 522:A1

\bibitem[{{Karoff} et~al.(2009){Karoff}, {Metcalfe}, {Chaplin}, {Elsworth},
  {Kjeldsen} et~al.}]{Karoff2009}
{Karoff} C, {Metcalfe} TS, {Chaplin} WJ, {Elsworth} Y, {Kjeldsen} H, et~al.
  2009.
\newblock \textit{\mnras} 399:914--923

\bibitem[{{Kjeldsen} \& {Bedding}(1995)}]{Kjeldsen1995a}
{Kjeldsen} H, {Bedding} TR. 1995.
\newblock \textit{\aap} 293:87--106

\bibitem[{{Kjeldsen} et~al.(2008){Kjeldsen}, {Bedding}, {Arentoft}, {Butler},
  {Dall} et~al.}]{Kjeldsen2008a}
{Kjeldsen} H, {Bedding} TR, {Arentoft} T, {Butler} RP, {Dall} TH, et~al. 2008.
\newblock \textit{\apj} 682:1370--1375

\bibitem[{{Kjeldsen}, {Bedding} \&
  {Christensen-Dalsgaard}(2008)}]{Kjeldsen2008}
{Kjeldsen} H, {Bedding} TR, {Christensen-Dalsgaard} J. 2008.
\newblock \textit{\apjl} 683:L175--L178

\bibitem[{{Kjeldsen} et~al.(1995){Kjeldsen}, {Bedding}, {Viskum} \&
  {Frandsen}}]{Kjeldsen1995b}
{Kjeldsen} H, {Bedding} TR, {Viskum} M, {Frandsen} S. 1995.
\newblock \textit{\aj} 109:1313--1319

\bibitem[{{Kjeldsen} et~al.(2010){Kjeldsen}, {Christensen-Dalsgaard},
  {Handberg}, {Brown}, {Gilliland} et~al.}]{Kjeldsen2010}
{Kjeldsen} H, {Christensen-Dalsgaard} J, {Handberg} R, {Brown} TM, {Gilliland}
  RL, et~al. 2010.
\newblock \textit{Astronomische Nachrichten} 331:966

\bibitem[{{Lebreton} \& {Montalb{\'a}n}(2009)}]{lebreton2009}
{Lebreton} Y, {Montalb{\'a}n} J. 2009.
\newblock In \textit{IAU Symposium}, eds. EE~{Mamajek}, DR~{Soderblom}, RFG
  {Wyse}, vol. 258 of \textit{IAU Symposium}

\bibitem[{{Lebreton} \& {Montalb{\'a}n}(2010)}]{lebreton2010}
{Lebreton} Y, {Montalb{\'a}n} J. 2010.
\newblock \textit{\apss} 328:29--38

\bibitem[{{Lebreton} et~al.(2008){Lebreton}, {Montalb{\'a}n},
  {Christensen-Dalsgaard}, {Roxburgh} \& {Weiss}}]{Lebreton2008}
{Lebreton} Y, {Montalb{\'a}n} J, {Christensen-Dalsgaard} J, {Roxburgh} IW,
  {Weiss} A. 2008.
\newblock \textit{\apss} 316:187--213

\bibitem[{{Maeder}(1975)}]{Maeder1975}
{Maeder} A. 1975.
\newblock \textit{\aap} 43:61--69

\bibitem[{Maeder \& Meynet(2000)}]{Maeder2000}
Maeder A, Meynet G. 2000.
\newblock \textit{\araa} 38:143--190

\bibitem[{{Majewski} et~al.(2010){Majewski}, {Wilson}, {Hearty}, {Schiavon} \&
  {Skrutskie}}]{Majewski2010}
{Majewski} SR, {Wilson} JC, {Hearty} F, {Schiavon} RR, {Skrutskie} MF. 2010.
\newblock In \textit{IAU Symposium}, eds. K~{Cunha}, M~{Spite}, B~{Barbuy},
  vol. 265 of \textit{IAU Symposium}

\bibitem[{{Mathis}(2009)}]{Mathis2009}
{Mathis} S. 2009.
\newblock \textit{\aap} 506:811--828

\bibitem[{Mathis, Palacios \& Zahn(2004)}]{Mathis2004}
Mathis S, Palacios A, Zahn JP. 2004.
\newblock \textit{\aap} 425:243--247

\bibitem[{{Mathur} et~al.(2012){Mathur}, {Metcalfe}, {Woitaszek}, {Bruntt},
  {Verner} et~al.}]{Mathur2012}
{Mathur} S, {Metcalfe} TS, {Woitaszek} M, {Bruntt} H, {Verner} GA, et~al. 2012.
\newblock \textit{\apj} 749:152

\bibitem[{{Mazumdar}(2005)}]{Mazumdar2005}
{Mazumdar} A. 2005.
\newblock \textit{\aap} 441:1079--1086

\bibitem[{{Mazumdar} et~al.(2006){Mazumdar}, {Basu}, {Collier} \&
  {Demarque}}]{Mazumdar2006}
{Mazumdar} A, {Basu} S, {Collier} BL, {Demarque} P. 2006.
\newblock \textit{\mnras} 372:949--958

\bibitem[{{Mazumdar} et~al.(2012{\natexlab{a}}){Mazumdar}, {Michel}, {Antia} \&
  {Deheuvels}}]{Mazumdar2012}
{Mazumdar} A, {Michel} E, {Antia} HM, {Deheuvels} S. 2012{\natexlab{a}}.
\newblock \textit{\aap} 540:A31

\bibitem[{{Mazumdar} et~al.(2012{\natexlab{b}}){Mazumdar}, {Monteiro},
  {Ballot}, {Antia}, {Basu} et~al.}]{Mazumdar2012b}
{Mazumdar} A, {Monteiro} MJPFG, {Ballot} J, {Antia} HM, {Basu} S, et~al.
  2012{\natexlab{b}}.
\newblock \textit{Astronomische Nachrichten} 333:1040--1043

\bibitem[{{Metcalfe} et~al.(2010{\natexlab{a}}){Metcalfe}, {Basu}, {Henry},
  {Soderblom}, {Judge} et~al.}]{Metcalfe2010a}
{Metcalfe} TS, {Basu} S, {Henry} TJ, {Soderblom} DR, {Judge} PG, et~al.
  2010{\natexlab{a}}.
\newblock \textit{\apjl} 723:L213--L217

\bibitem[{{Metcalfe} et~al.(2012){Metcalfe}, {Chaplin}, {Appourchaux},
  {Garc{\'{\i}}a}, {Basu} et~al.}]{Metcalfe2012}
{Metcalfe} TS, {Chaplin} WJ, {Appourchaux} T, {Garc{\'{\i}}a} RA, {Basu} S,
  et~al. 2012.
\newblock \textit{\apjl} 748:L10

\bibitem[{{Metcalfe} et~al.(2010{\natexlab{b}}){Metcalfe}, {Monteiro},
  {Thompson}, {Molenda-{\.Z}akowicz}, {Appourchaux} et~al.}]{Metcalfe2010}
{Metcalfe} TS, {Monteiro} MJPFG, {Thompson} MJ, {Molenda-{\.Z}akowicz} J,
  {Appourchaux} T, et~al. 2010{\natexlab{b}}.
\newblock \textit{\apj} 723:1583--1598

\bibitem[{{Michel} \& {Baglin}(2012)}]{Michel2012}
{Michel} E, {Baglin} A. 2012.
\newblock In \textit{Second CoRoT Symposium: Transiting planets, vibrating
  stars and their connection}, eds. A~{Baglin}, M~{Deleuil}, E~{Michel},
  C~{Moutou}.
\newblock In the press (arXiv:1202.1422)

\bibitem[{{Michel} et~al.(2008){Michel}, {Baglin}, {Auvergne}, {Catala},
  {Samadi} et~al.}]{Michel2008}
{Michel} E, {Baglin} A, {Auvergne} M, {Catala} C, {Samadi} R, et~al. 2008.
\newblock \textit{Science} 322:558--

\bibitem[{{Miglio}(2012)}]{Miglio2012}
{Miglio} A. 2012.
\newblock In \textit{Red Giants as Probes of the Structure and Evolution of the
  Milky Way}, eds. A~Miglio, J~Montalb{\'a}n, A~Noels, ApSS Proceedings

\bibitem[{{Miglio} et~al.(2012){Miglio}, {Brogaard}, {Stello}, {Chaplin},
  {D'Antona} et~al.}]{Miglio2012b}
{Miglio} A, {Brogaard} K, {Stello} D, {Chaplin} WJ, {D'Antona} F, et~al. 2012.
\newblock \textit{\mnras} 419:2077--2088

\bibitem[{{Miglio} et~al.(2009){Miglio}, {Montalb{\'a}n}, {Baudin},
  {Eggenberger}, {Noels} et~al.}]{Miglio2009}
{Miglio} A, {Montalb{\'a}n} J, {Baudin} F, {Eggenberger} P, {Noels} A, et~al.
  2009.
\newblock \textit{\aap} 503:L21--L24

\bibitem[{Miglio et~al.(2010)Miglio, Montalb\'{a}n, Carrier, {De Ridder},
  Mosser et~al.}]{Miglio2010}
Miglio A, Montalb\'{a}n J, Carrier F, {De Ridder} J, Mosser B, et~al. 2010.
\newblock \textit{\aap} 520:L6

\bibitem[{Miglio, Montalb\'{a}n \& Maceroni(2007)}]{Miglio2007}
Miglio A, Montalb\'{a}n J, Maceroni C. 2007.
\newblock \textit{\mnras} 377:373--382

\bibitem[{{Montalb{\'a}n} et~al.(2013){Montalb{\'a}n}, {Miglio}, {Noels},
  {Dupret}, {Scuflaire} \& {Ventura}}]{Montalban2013}
{Montalb{\'a}n} J, {Miglio} A, {Noels} A, {Dupret} MA, {Scuflaire} R, {Ventura}
  P. 2013.
\newblock \textit{\apj} 766:118

\bibitem[{{Montalb{\'a}n} et~al.(2010){Montalb{\'a}n}, {Miglio}, {Noels},
  {Scuflaire} \& {Ventura}}]{Montalban2010}
{Montalb{\'a}n} J, {Miglio} A, {Noels} A, {Scuflaire} R, {Ventura} P. 2010.
\newblock \textit{\apjl} 721:L182--L188

\bibitem[{{Monteiro}(2009)}]{Monteiro2009}
{Monteiro} MJPFG. 2009.
\newblock \textit{{Evolution and Seismic Tools for Stellar Astrophysics}}.
\newblock Springer

\bibitem[{Monteiro, Christensen-Dalsgaard \& Thompson(2000)}]{Monteiro2000}
Monteiro MJPFG, Christensen-Dalsgaard J, Thompson MJ. 2000.
\newblock \textit{\mnras} 316:165--172

\bibitem[{{Monteiro} \& {Thompson}(2005)}]{Monteiro2005}
{Monteiro} MJPFG, {Thompson} MJ. 2005.
\newblock \textit{\mnras} 361:1187--1196

\bibitem[{{Mosser} et~al.(2011{\natexlab{a}}){Mosser}, {Barban},
  {Montalb{\'a}n}, {Beck}, {Miglio} et~al.}]{Mosser2011}
{Mosser} B, {Barban} C, {Montalb{\'a}n} J, {Beck} PG, {Miglio} A, et~al.
  2011{\natexlab{a}}.
\newblock \textit{\aap} 532:A86

\bibitem[{{Mosser} et~al.(2011{\natexlab{b}}){Mosser}, {Belkacem}, {Goupil},
  {Michel}, {Elsworth} et~al.}]{Mosser2011a}
{Mosser} B, {Belkacem} K, {Goupil} MJ, {Michel} E, {Elsworth} Y, et~al.
  2011{\natexlab{b}}.
\newblock \textit{\aap} 525:L9

\bibitem[{{Mosser} et~al.(2012{\natexlab{a}}){Mosser}, {Elsworth}, {Hekker},
  {Huber}, {Kallinger} et~al.}]{Mosser2012a}
{Mosser} B, {Elsworth} Y, {Hekker} S, {Huber} D, {Kallinger} T, et~al.
  2012{\natexlab{a}}.
\newblock \textit{\aap} 537:A30

\bibitem[{{Mosser} et~al.(2012{\natexlab{b}}){Mosser}, {Goupil}, {Belkacem},
  {Marques}, {Beck} et~al.}]{Mosser2012c}
{Mosser} B, {Goupil} MJ, {Belkacem} K, {Marques} JP, {Beck} PG, et~al.
  2012{\natexlab{b}}.
\newblock \textit{\aap} 548:A10

\bibitem[{{Mosser} et~al.(2012{\natexlab{c}}){Mosser}, {Goupil}, {Belkacem},
  {Michel}, {Stello} et~al.}]{Mosser2012b}
{Mosser} B, {Goupil} MJ, {Belkacem} K, {Michel} E, {Stello} D, et~al.
  2012{\natexlab{c}}.
\newblock \textit{\aap} 540:A143

\bibitem[{{Mosser} et~al.(2013){Mosser}, {Michel}, {Belkacem}, {Goupil},
  {Baglin} et~al.}]{Mosser2013}
{Mosser} B, {Michel} E, {Belkacem} K, {Goupil} MJ, {Baglin} A, et~al. 2013.
\newblock \textit{\aap} 550:A126

\bibitem[{{Noels} et~al.(2010){Noels}, {Montalban}, {Miglio}, {Godart} \&
  {Ventura}}]{Noels2010}
{Noels} A, {Montalban} J, {Miglio} A, {Godart} M, {Ventura} P. 2010.
\newblock \textit{\apss} 328:227--236

\bibitem[{{Ossendrijver}(2003)}]{Ossen2003}
{Ossendrijver} M. 2003.
\newblock \textit{\aapr} 11:287--367

\bibitem[{{Palacios}(2012)}]{Palacios2012}
{Palacios} A. 2012.
\newblock In \textit{Red Giants as Probes of the Structure and Evolution of the
  Milky Way}, eds. A~{Miglio}, J~{Montalb{\'a}n}, A~{Noels}

\bibitem[{{Pinsonneault}(1997)}]{Pinsonneault1997}
{Pinsonneault} M. 1997.
\newblock \textit{\araa} 35:557--605

\bibitem[{{Pinsonneault} et~al.(1989){Pinsonneault}, {Kawaler}, {Sofia} \&
  {Demarque}}]{Pinsonneault1989}
{Pinsonneault} MH, {Kawaler} SD, {Sofia} S, {Demarque} P. 1989.
\newblock \textit{\apj} 338:424--452

\bibitem[{Popielski \& Dziembowski(2005)}]{Popielski2005}
Popielski BL, Dziembowski WA. 2005.
\newblock \textit{Acta Astronomica} 55:177--193

\bibitem[{{Quirion}, {Christensen-Dalsgaard} \& {Arentoft}(2010)}]{Quirion2010}
{Quirion} PO, {Christensen-Dalsgaard} J, {Arentoft} T. 2010.
\newblock \textit{\apj} 725:2176--2189

\bibitem[{{Reese}, {Ligni{\`e}res} \& {Rieutord}(2006)}]{Reese2006}
{Reese} D, {Ligni{\`e}res} F, {Rieutord} M. 2006.
\newblock \textit{\aap} 455:621--637

\bibitem[{Ribas, Jordi \& Gim\'{e}nez(2000)}]{Ribas2000}
Ribas I, Jordi C, Gim\'{e}nez A. 2000.
\newblock \textit{\mnras} 318:L55--L59

\bibitem[{{Roxburgh}(2009)}]{Roxburgh2009}
{Roxburgh} IW. 2009.
\newblock \textit{\aap} 493:185--191

\bibitem[{{Roxburgh}(2010)}]{Roxburgh2010}
{Roxburgh} IW. 2010.
\newblock \textit{\apss} 328:3--11

\bibitem[{{Roxburgh} \& {Vorontsov}(1997)}]{Roxburgh1997}
{Roxburgh} IW, {Vorontsov} SV. 1997.
\newblock \textit{\mnras} 292:L33--L36

\bibitem[{Roxburgh \& Vorontsov(2003)}]{Roxburgh2003}
Roxburgh IW, Vorontsov SV. 2003.
\newblock \textit{\aap} 411:215--220

\bibitem[{{Salaris}(2012)}]{Salaris2012}
{Salaris} M. 2012.
\newblock In \textit{Red Giants as Probes of the Structure and Evolution of the
  Milky Way}, eds. A~{Miglio}, J~{Montalb{\'a}n}, A~{Noels}

\bibitem[{{Samadi} et~al.(2012){Samadi}, {Belkacem}, {Dupret}, {Ludwig},
  {Baudin} et~al.}]{Samadi2012}
{Samadi} R, {Belkacem} K, {Dupret} MA, {Ludwig} HG, {Baudin} F, et~al. 2012.
\newblock \textit{\aap} 543:A120

\bibitem[{{Samadi} et~al.(2007){Samadi}, {Georgobiani}, {Trampedach}, {Goupil},
  {Stein} \& {Nordlund}}]{Samadi2007}
{Samadi} R, {Georgobiani} D, {Trampedach} R, {Goupil} MJ, {Stein} RF,
  {Nordlund} {\AA}. 2007.
\newblock \textit{\aap} 463:297--308

\bibitem[{{Samadi} \& {Goupil}(2001)}]{Samadi2001}
{Samadi} R, {Goupil} MJ. 2001.
\newblock \textit{\aap} 370:136--146

\bibitem[{{Sanchis-Ojeda} et~al.(2012){Sanchis-Ojeda}, {Fabrycky}, {Winn},
  {Barclay}, {Clarke} et~al.}]{Sanchis-Ojeda2012}
{Sanchis-Ojeda} R, {Fabrycky} DC, {Winn} JN, {Barclay} T, {Clarke} BD, et~al.
  2012.
\newblock \textit{\nat} 487:449--453

\bibitem[{{Sandquist} et~al.(2013){Sandquist}, {Mathieu}, {Brogaard}, {Meibom},
  {Geller} et~al.}]{Sandquist2013}
{Sandquist} EL, {Mathieu} RD, {Brogaard} K, {Meibom} S, {Geller} AM, et~al.
  2013.
\newblock \textit{\apj} 762:58

\bibitem[{{Silva Aguirre} et~al.(2011{\natexlab{a}}){Silva Aguirre}, {Ballot},
  {Serenelli} \& {Weiss}}]{SilvaAguirre2011}
{Silva Aguirre} V, {Ballot} J, {Serenelli} AM, {Weiss} A. 2011{\natexlab{a}}.
\newblock \textit{\aap} 529:A63

\bibitem[{{Silva Aguirre} et~al.(2012){Silva Aguirre}, {Casagrande}, {Basu},
  {Campante}, {Chaplin} et~al.}]{SilvaAguirre2012}
{Silva Aguirre} V, {Casagrande} L, {Basu} S, {Campante} TL, {Chaplin} WJ,
  et~al. 2012.
\newblock \textit{\apj} 757:99

\bibitem[{{Silva Aguirre} et~al.(2011{\natexlab{b}}){Silva Aguirre}, {Chaplin},
  {Ballot}, {Basu}, {Bedding} et~al.}]{SilvaAguirre2011a}
{Silva Aguirre} V, {Chaplin} WJ, {Ballot} J, {Basu} S, {Bedding} TR, et~al.
  2011{\natexlab{b}}.
\newblock \textit{\apjl} 740:L2

\bibitem[{{Smiljanic} et~al.(2009){Smiljanic}, {Gauderon}, {North}, {Barbuy},
  {Charbonnel} \& {Mowlavi}}]{Smiljanic2009}
{Smiljanic} R, {Gauderon} R, {North} P, {Barbuy} B, {Charbonnel} C, {Mowlavi}
  N. 2009.
\newblock \textit{\aap} 502:267--282

\bibitem[{Soderblom(2010)}]{Soderblom2010}
Soderblom DR. 2010.
\newblock \textit{Annual Review of \aap} 48:581--629

\bibitem[{{Soriano} \& {Vauclair}(2010)}]{Soriano2010}
{Soriano} M, {Vauclair} S. 2010.
\newblock \textit{\aap} 513:A49

\bibitem[{{Southworth}(2011)}]{Southworth2011}
{Southworth} J. 2011.
\newblock \textit{\mnras} 417:2166--2196

\bibitem[{{Stello} et~al.(2009){Stello}, {Chaplin}, {Bruntt}, {Creevey},
  {Garc{\'{\i}}a-Hern{\'a}ndez} et~al.}]{Stello2009}
{Stello} D, {Chaplin} WJ, {Bruntt} H, {Creevey} OL,
  {Garc{\'{\i}}a-Hern{\'a}ndez} A, et~al. 2009.
\newblock \textit{\apj} 700:1589--1602

\bibitem[{{Stello} et~al.(2011{\natexlab{a}}){Stello}, {Huber}, {Kallinger},
  {Basu}, {Mosser} et~al.}]{Stello2011a}
{Stello} D, {Huber} D, {Kallinger} T, {Basu} S, {Mosser} B, et~al.
  2011{\natexlab{a}}.
\newblock \textit{\apjl} 737:L10

\bibitem[{{Stello} et~al.(2011{\natexlab{b}}){Stello}, {Meibom}, {Gilliland},
  {Grundahl}, {Hekker} et~al.}]{Stello2011}
{Stello} D, {Meibom} S, {Gilliland} RL, {Grundahl} F, {Hekker} S, et~al.
  2011{\natexlab{b}}.
\newblock \textit{\apj} 739:13

\bibitem[{{Straniero} et~al.(2003){Straniero}, {Dom{\'{\i}}nguez}, {Imbriani}
  \& {Piersanti}}]{Straniero2003}
{Straniero} O, {Dom{\'{\i}}nguez} I, {Imbriani} G, {Piersanti} L. 2003.
\newblock \textit{\apj} 583:878--884

\bibitem[{{Talon} \& {Charbonnel}(2005)}]{Talon2005}
{Talon} S, {Charbonnel} C. 2005.
\newblock \textit{\aap} 440:981--994

\bibitem[{{Tassoul}(1980)}]{Tassoul1980}
{Tassoul} M. 1980.
\newblock \textit{\apjs} 43:469--490

\bibitem[{{Thompson} et~al.(2003){Thompson}, {Christensen-Dalsgaard}, {Miesch}
  \& {Toomre}}]{Thompson2003}
{Thompson} MJ, {Christensen-Dalsgaard} J, {Miesch} MS, {Toomre} J. 2003.
\newblock \textit{\araa} 41:599--643

\bibitem[{{Torres}, {Andersen} \& {Gim{\'e}nez}(2010)}]{Torres2010}
{Torres} G, {Andersen} J, {Gim{\'e}nez} A. 2010.
\newblock \textit{\aapr} 18:67--126

\bibitem[{{Torres} et~al.(2012){Torres}, {Fischer}, {Sozzetti}, {Buchhave},
  {Winn} et~al.}]{Torres2012}
{Torres} G, {Fischer} DA, {Sozzetti} A, {Buchhave} LA, {Winn} JN, et~al. 2012.
\newblock \textit{\apj} 757:161

\bibitem[{{Ulrich}(1986)}]{Ulrich1986}
{Ulrich} RK. 1986.
\newblock \textit{\apjl} 306:L37--L40

\bibitem[{Unno et~al.(1989)Unno, Osaki, Ando, Saio \& Shibahashi}]{Unno1989}
Unno W, Osaki Y, Ando H, Saio H, Shibahashi H. 1989.
\newblock \textit{{Nonradial oscillations of stars}}.
\newblock Nonradial oscillations of stars, Tokyo: University of Tokyo Press,
  1989, 2nd ed.

\bibitem[{VandenBerg, Bergbusch \& Dowler(2006)}]{VandenBerg2006}
VandenBerg DA, Bergbusch PA, Dowler PD. 2006.
\newblock \textit{\apjs} 162:375--387

\bibitem[{VandenBerg \& Stetson(2004)}]{VandenBerg2004}
VandenBerg DA, Stetson PB. 2004.
\newblock \textit{\pasp} 116:997--1011

\bibitem[{{Vauclair} et~al.(2008){Vauclair}, {Laymand}, {Bouchy}, {Vauclair},
  {Hui Bon Hoa} et~al.}]{Vauclair2008}
{Vauclair} S, {Laymand} M, {Bouchy} F, {Vauclair} G, {Hui Bon Hoa} A, et~al.
  2008.
\newblock \textit{\aap} 482:L5--L8

\bibitem[{{Ventura}, {D'Antona} \& {Mazzitelli}(2008)}]{Ventura2008}
{Ventura} P, {D'Antona} F, {Mazzitelli} I. 2008.
\newblock \textit{\apss} 316:93--98

\bibitem[{Ventura et~al.(1998)Ventura, Zeppieri, Mazzitelli \&
  D'Antona}]{Ventura1998}
Ventura P, Zeppieri A, Mazzitelli I, D'Antona F. 1998.
\newblock \textit{\aap} 334:953--968

\bibitem[{{Verner} et~al.(2011){Verner}, {Elsworth}, {Chaplin}, {Campante},
  {Corsaro} et~al.}]{Verner2011}
{Verner} GA, {Elsworth} Y, {Chaplin} WJ, {Campante} TL, {Corsaro} E, et~al.
  2011.
\newblock \textit{\mnras} 415:3539--3551

\bibitem[{{Vorontsov}(1988)}]{Vorontsov1988}
{Vorontsov} SV. 1988.
\newblock In \textit{Advances in Helio- and Asteroseismology}, eds.
  J~{Christensen-Dalsgaard}, S~{Frandsen}, vol. 123 of \textit{IAU Symposium}

\bibitem[{{Weiss}(2012)}]{Weiss2012}
{Weiss} A. 2012.
\newblock In \textit{Red Giants as Probes of the Structure and Evolution of the
  Milky Way}, eds. A~{Miglio}, J~{Montalb{\'a}n}, A~{Noels}

\bibitem[{{White} et~al.(2012){White}, {Bedding}, {Gruberbauer}, {Benomar},
  {Stello} et~al.}]{White2012}
{White} TR, {Bedding} TR, {Gruberbauer} M, {Benomar} O, {Stello} D, et~al.
  2012.
\newblock \textit{\apjl} 751:L36

\bibitem[{{White} et~al.(2011){White}, {Bedding}, {Stello},
  {Christensen-Dalsgaard}, {Huber} \& {Kjeldsen}}]{White2011}
{White} TR, {Bedding} TR, {Stello} D, {Christensen-Dalsgaard} J, {Huber} D,
  {Kjeldsen} H. 2011.
\newblock \textit{\apj} 743:161

\bibitem[{{Winn} et~al.(2010){Winn}, {Fabrycky}, {Albrecht} \&
  {Johnson}}]{Winn2010}
{Winn} JN, {Fabrycky} D, {Albrecht} S, {Johnson} JA. 2010.
\newblock \textit{\apjl} 718:L145--L149

\bibitem[{{Young} et~al.(2003){Young}, {Knierman}, {Rigby} \&
  {Arnett}}]{Young2003}
{Young} PA, {Knierman} KA, {Rigby} JR, {Arnett} D. 2003.
\newblock \textit{\apj} 595:1114--1123

\end{thebibliography}

\end{document}